\documentclass[a4paper,12pt]{article}
\usepackage{natbib}
\usepackage[colorlinks,citecolor=blue,urlcolor=blue]{hyperref}
\usepackage{epsf}
\usepackage{epsfig}
\usepackage{multirow}
\usepackage[english]{babel}
\usepackage{graphicx}
\usepackage{geometry}
\usepackage{natbib}
\usepackage{amssymb}
\usepackage{amsmath}
\usepackage{mathtools}
\mathtoolsset{showonlyrefs=true}
\usepackage{color}
\usepackage{enumerate}
\usepackage{authblk}

\usepackage{caption}
\usepackage[labelformat=simple]{subcaption}
\usepackage{bm}

\newcommand{\specificthanks}[1]{\@fnsymbol{#1}}

\title{A Latent Shrinkage Position Model \\for Binary and Count Network Data}
\author[]{Xian Yao Gwee\thanks{xian-yao.gwee@ucdconnect.ie} }
\author[]{Isobel Claire Gormley\thanks{claire.gormley@ucd.ie}}
\author[]{Michael Fop\thanks{michael.fop@ucd.ie}}
\affil[]{School of Mathematics and Statistics,\protect \\ University College Dublin,\protect \\ Ireland.}

\date{\today}

\begin{document}

\maketitle

\begin{abstract}
Interactions between actors are frequently represented using a network. The latent position model is widely used for analysing network data, whereby each actor is positioned in a latent space. Inferring the dimension of this space is challenging. Often, for simplicity, two dimensions are used or model selection criteria are employed to select the dimension, but this requires choosing a criterion and the computational expense of fitting multiple models. Here the latent shrinkage position model (LSPM) is proposed which intrinsically infers the effective dimension of the latent space. The LSPM employs a Bayesian nonparametric multiplicative truncated gamma process prior that ensures shrinkage of the variance of the latent positions across higher dimensions. Dimensions with non-negligible variance are deemed most useful to describe the observed network, inducing automatic inference on the latent space dimension. While the LSPM is applicable to many network types, logistic and Poisson LSPMs are developed here for binary and count networks respectively. Inference proceeds via a Markov chain Monte Carlo algorithm, where novel surrogate proposal distributions reduce the computational burden. The LSPM’s properties are assessed through simulation studies, and its utility is illustrated through application to real network datasets. Open source software assists wider implementation of the LSPM.
\end{abstract}

\section{Introduction}

Network data are a collection of interconnected objects which are usually represented formally using graph theory. The objects are often called nodes while their connections to each other are called edges. Living in an increasingly connected world, network data have garnered increased interest in recent years. Social network data are a well known example where friendships \citep{Liu_Chen_2021}, competitions \citep{DAngelo_Murphy_Alfo_2019}, companies \citep{Ryan_Wyse_Friel_2017}, households \citep{Fosdick_McCormick_Murphy_Ng_Westling_2019}, and other social relations are exhibited between individuals (or actors) and can be modelled to understand how people interact in different situations. Beyond that, the network-based perspective has found useful applications in complex systems across various fields including computational biology such as protein-protein interactions \citep{Chen_Liao_Chen_Wang_Chen_2019}; neuroscience on the topological properties of brain connectomes \citep{yang_priebe_park_marchette_2020}; economy on understanding trades between countries \citep{Sajedianfard_2021}; education on online problem based learning \citep{Saqr_Alamro_2019}; and public health on studying the spread of infectious disease \citep{Jo_Chang_You_Ghim_2021}.

There are many types of models for network data. Most are variants of the random graph model \citep{erdos1959, Gilbert_1959}, the stochastic blockmodel \citep{Holland_Laskey_Leinhardt_1983}, or the latent position model \citep[LPM,][]{hoff2002}. The LPM has received much attention as it provides a meaningful visualisation of the data and rich qualitative information \citep{JMLR:v21:17-470_ma,Tafakori_Pourkhanali_Rastelli_2021}. The LPM postulates that the nodes have positions in a latent space, and that the observed edge formation process is explained through the distance between the nodes' latent positions. In this way, important features such as transitivity, reciprocity, and homophily are easily accounted for in the model, which captures local and global structures \citep{Rastelli_Friel_Raftery_2016,Kim_Lee_Xue_Niu_2018, Smith_Asta_Calder_2019}. 
%The LPM postulates that the nodes have positions in a latent space, and that the observed edge formation process is explained through the distance between the nodes' latent positions. 

The number of dimensions of the latent space, $p$, in the LPM is usually unknown and needs to be inferred from the data. Typically, $p$ is fixed at two for easy visualisation and interpretation \citep{Sewell_Chen_2016, pmlr-v119-zhang20aa, Liu_Chen_2021}. However, this is an arbitrary choice and may result in an incomplete or overly complex description of the network. The estimation of $p$ has received some attention in the literature. Attempts to tackle the dimension selection issue have used model selection tools such as a variant of the well-known Bayesian information criterion \citep[BIC,][]{handcock2007}, the Akaike information criterion \citep{gormley_murphy_2010,sewell:2021}, the deviance information criterion \citep{friel_ras2016}, and the Watanabe-Akaike information
criterion \citep{Ng_Murphy_Westling_McCormick_Fosdick_2021,Sosa_Betancourt_2022}. Models with different numbers of dimensions are fitted and then compared using the chosen model selection criterion. However, the selected criterion may not be formally correct for choosing $p$ \citep{handcock2007}. Furthermore, fitting a range of models, each with a different value of $p$, can incur a large computational cost and restrict the set of models considered. 

An alternative approach to the dimension selection problem is to have an automatic process that infers the optimal dimension of the latent space from the data. This type of process has been used in areas such as factor analysis  using shrinkage priors \citep{bhattacharya_dunson_2011, durante_2017, Murphy_Viroli_Gormley_2020}, or spike and slab distributions \citep{legramanti_durante_dunson_2020}. In network analysis, within a stochastic blockmodel setting, automatic dimension estimation has been used in \cite{yang_priebe_park_marchette_2020} in a frequentist framework while \cite{passino_heard_2020} used a Bayesian framework. 
% \cite{yang_priebe_park_marchette_2020} extend the adjacency spectral embedding to have a distributional model that contains an informative part and a redundant part that holds the remaining dimensions which can be truncated. \cite{passino_heard_2020} utilise asymptotic results for the leading components of spectral embeddings while making realistic assumptions on the remaining components. The leading components will contain information for generating the latent positions while the remaining components are the non-relevant part that do not contribute to the network. Both of these models deliver an unsatisfactory performance if the within-community degree distributions are not homogeneous. \cite{Passino_Heard_Rubin-Delanchy_2021} tackle this unsatisfactory performance by using spherical coordinates under the degree-corrected stochastic blockmodel. Nevertheless, these models allow the estimation of the number of dimensions and communities simultaneously but in a stochastic blockmodel setting.

In the context of the LPM, a number of proposals for automatic inference of the dimension of the latent space using shrinkage priors have been suggested. \cite{rastelli2018} develops an approach to automatically determine the number of dimensions by creating a finite mixture of unidimensional LPMs and utilising a Dirichlet shrinkage prior on the mixing proportions. The key idea is to intentionally overfit the number of dimensions by considering a relatively large number of mixture components and then empty the superfluous components using the shrinkage prior during inference. The approach shows good performance in recovering the true dimension with small networks but tends to overestimate the number of dimensions when the number of nodes is large. Also, the approach is less interpretable than the original LPM as the Euclidean distance between 
nodes across dimensions has no relevant meaning.
% any dimensional coordinate vectors has no relevant meaning but rather one should study the distances between nodes for each latent dimension separately. Moreover, the realisation of any edges between two nodes only takes information from one dimension despite potentially having latent positions in more than one dimension. 
\cite{Durante_Dunson_2014} employ the shrinkage prior of \cite{bhattacharya_dunson_2011} when considering the dimension of the latent space in the projection model formulation of the LPM for a dynamic binary network; the distance model formulation of the LPM is considered here. \cite{Durante_Dunson_Vogelstein_2017} use a similar shrinkage approach but in the context of a population of networks.

% In the latent space model setting, automatic inference of dimension was done in \cite{Durante_Dunson_2014} and \cite{Durante_Dunson_Vogelstein_2017} with the former having the closest similarity with our proposing method where the shrinkage process prior is placed directly on the variance of the latent positions. However, the formulation for determining the edge formation probability in \cite{Durante_Dunson_2014} is more similar to the projection model while the proposing work here will be more similar to the distance model first introduced in \cite{hoff2002}. The shrinkage prior formulation was done for a dynamic binary network in \cite{Durante_Dunson_2014} while the work here will be using the generalised linear model to drive the formulation for networks that are of binary or count edges. In \cite{Durante_Dunson_Vogelstein_2017}, similar technique is used but now the latent positions have a dot product representation and the focus was on the multiple network nature instead of the dimension selection problem despite having good performance overall.

% The latent shrinkage position model (LSPM) is proposed here
%%% TODO
Here the {\em latent shrinkage position model} (LSPM), which facilitates automatic inference on the dimensionality of the latent space, is introduced. The LSPM is a Bayesian nonparametric model that theoretically allows infinitely many dimensions. This is established through a shrinkage process prior inspired by \cite{bhattacharya_dunson_2011}. The key idea is that the variance of the latent positions on each dimension becomes increasingly small as the number of dimensions grows. The informative, effective dimensions are those that have non-negligible variance in the latent positions while the uninformative, negligible dimensions will have latent position variances that tend to zero. With the LSPM, the need to select a model selection criterion is obviated, only a single model needs to be fitted and the LPM's ease of interpretation is retained. In addition, posterior uncertainty concerning the latent space dimension is automatically accounted for. The LSPM builds on \cite{Durante_Dunson_2014} and \cite{Durante_Dunson_Vogelstein_2017} by (a) considering binary and count valued networks with the development of the logistic LSPM and the Poisson LSPM respectively, (b) introducing a truncated version of the \cite{bhattacharya_dunson_2011} multiplicative gamma process prior and (c) proposing a computationally efficient Markov chain Monte Carlo inference scheme through the use of surrogate proposal distributions.

The remainder of this article is structured as follows: Section \ref{sec:2} describes the LPM and the proposed LSPM. Section \ref{sec:3} outlines the inferential process. Section \ref{sec:4} contains simulation studies exploring the performance of the LSPM in terms of inferring the number of effective dimensions and estimating parameters across realistic settings. Section \ref{sec:5} applies the proposed LSPM to a range of networks, having different edge types and characteristics. Section \ref{sec:6} concludes the article and discusses some potential extensions. R \citep{Rcore2022} code with which all results presented herein were produced is freely available from the \href{https://gitlab.com/gwee95/lspm}{\texttt{lspm}} GitLab repository.

% An R package, LSPM, is freely available from \hyperlink{https://www.r-project.org}{https://www.r-project.org} \citep{Rcore} to assist widespread use of the method, and with which all results presented herein were produced.

\section{The latent shrinkage position model}
\label{sec:2}

\subsection{The latent position model} 
\label{ssec:lpm}
Network data typically take the form of an $n \times n$ adjacency matrix, $\mathbf{Y}$, where $n$ is the number of observed nodes, with entries $y_{i,j}$ denoting the relationship between node $i$ and node $j$. To probabilistically model the presence or absence of an edge between two nodes, the latent position model \citep[LPM,][]{hoff2002} assumes that the observed edge formation process can be explained in terms of the nodes’ latent positions in a $p$-dimensional latent space. Under the LPM, edges are assumed independent, conditional on the latent positions of the nodes. Self-loops are not permitted and thus the diagonal elements of $\mathbf{Y}$ are zero. The sampling distribution is

$$
    \mathbb{P}(\mathbf{Y}\mid\alpha, \mathbf{Z}) = \prod_{i \neq j} \mathbb{P}(y_{i,j}\mid\alpha, \mathbf{z}_i,\mathbf{z}_j)
$$
where $\mathbf{Z}$ is the $n \times p$ matrix of latent positions with $\mathbf{z}_i$ denoting the latent position of node $i$, while $\alpha$ is a global parameter that captures the overall connectivity level in the network. A generalised linear model \citep{McCullagh_Nelder_1998} can be used to model the probability of a range of edge types between nodes as a function of their latent positions. \cite{hoff2002} propose the distance model formulation of the LPM by considering the Euclidean distance between $\mathbf{z}_i$ and $\mathbf{z}_j$, the latent locations of nodes $i$ and $j$ respectively, i.e.,
\begin{equation} 
\label{eqn:binprob}
g[\mathbb{E}(y_{i,j})] 
= \alpha - \Vert \mathbf{z}_i-\mathbf{z}_j\Vert^2 _2 \\
\end{equation}
where $g$ is an appropriate link function. As in \cite{gollini2016} and \cite{DAngelo_Murphy_Alfo_2019}, here the distance is taken to be the squared Euclidean distance. This model is particularly suitable for undirected networks or directed networks that exhibit strong reciprocity. The projection model formulation is an alternative LPM that considers the angle between nodes; \cite{Salter-Townshend_White_Gollini_Murphy_2012} provide a comprehensive review. 

The LPM has been predominantly used to model binary valued networks via the use of a logit link function
\begin{equation}\label{eq:logitprob} 
\log \frac{q_{i,j}}{1-q_{i,j}} = \alpha - \Vert \mathbf{z}_i-\mathbf{z}_j \Vert^2_2, \end{equation} 
where $q_{i,j}$ is the probability of forming an edge between node $i$ and node $j$. The probability of forming an edge between nodes when the distance between them is zero is determined by $\alpha$. There have been various extensions
to this model, for example, by including covariates \citep{hoff2002} and random effects \citep{Hoff03randomeffects}, sender and receiver effects \citep{Hoff_2005}, social reach \citep{sewell2015}, and stochastic blockmodels \citep{Ng_Murphy_Westling_McCormick_Fosdick_2021}. While feasible, such extensions are not considered here in the interest of simplicity. 

% This model has a simple interpretation where a large $\alpha$ reflects a network with a high level of connection between nodes, while increasing the distance between latent positions will decrease the level of connection between nodes. 

In addition to binary valued networks, various link functions within the generalised linear model framework can be utilised to model alternative edge types. This includes, but is not limited to, the use of a Poisson distribution \citep{Hoff_2005} for modelling count edges, and a truncated normal distribution \citep{Sewell_Chen_2016} and an exponential distribution \citep{rastelli2018} for non-negative continuous edges. In addition to binary edges, the LPM for count edges is considered here where
$$\log (\lambda_{i,j}) = \alpha - \Vert \mathbf{z}_i-\mathbf{z}_j \Vert^2_2,$$
where $\lambda_{i,j}$ is the rate parameter of a Poisson distribution between node $i$ and $j$. A similar interpretation to the binary case applies here where $\alpha$ is the rate for forming an edge between a pair of nodes when the distance between them is zero.

Beyond different edge types, the LPM has been extended, for example, to facilitate clustering of nodes in a network \citep{handcock2007, gormley_murphy_2010}, to model dynamic networks \citep{sewell2015}, and to model multiple networks \citep{gollini2016,salter:2017,DAngelo_Murphy_Alfo_2019}. While extending the LPM has received much attention, attempts to objectively infer the latent space dimension have been few; subjectively choosing $p=2$ or choosing between a set of possible dimensions via a selected model selection criterion are typical approaches. Here, the LSPM is proposed to address the inference of $p$ in a Bayesian nonparametric framework using a shrinkage prior.

\subsection{The latent shrinkage position model} \label{ssec:lspm}

The latent shrinkage position model (LSPM) is a LPM which employs a multiplicative truncated gamma process (MTGP) prior allowing the model to intrinsically infer the number of effective dimensions from the network data, where ``effective dimensions'' means the dimensions necessary to fully describe the network. The MTGP prior builds on the multiplicative gamma process (MGP) prior that originates from the infinite factor model literature \citep{bhattacharya_dunson_2011}, where the MGP prior allows an infinite number of factors whose loadings are shrunk towards zero as the factor number increases. \cite{legramanti_durante_dunson_2020} modified the MGP by employing a spike and slab distribution structure that increases the prior mass on the spike for later factors. Other popular shrinkage priors include the Dirichlet-Laplace \citep{bhattacharya_pati_pillai_dunson_2015}, generalized double Pareto \citep{armagan_dunson_lee_2013}, and the horseshoe \citep{carvalho_polson_scott_2010} priors.

Under the LPM, the prior on the latent positions is typically assumed to be a zero centred Gaussian with equal variance across each of the $p$ dimensions. Under the LSPM, the latent positions are assumed to have a zero centred Gaussian distribution with diagonal precision matrix $\mathbf{\Omega}$, whose entries $\omega_{\ell}$ denote the precision of the latent positions in dimension $\ell$, for $\ell=1, \ldots, \infty$. The LSPM employs a MTGP prior on the precision parameters: the latent dimension $h$ has an associated shrinkage strength parameter $\delta_h$, where the cumulative product of $\delta_1$ to $\delta_{\ell}$ gives the precision $\omega_{\ell}$. An unconstrained gamma prior is assumed for the shrinkage strength parameter for the first dimension, while a truncated gamma distribution is assumed for the remaining dimensions to ensure shrinkage. Specifically, for $i = 1, \ldots, n$
\begin{equation}
\begin{aligned}  
\label{eq:z_mvn}
    \mathbf{z}_{i} \sim \text{MVN}(\mathbf{0}, \mathbf{\Omega}^{-1}) \qquad
    \omega_{\ell} = \prod_{h=1}^{\ell} \delta_{h}  \quad \mbox{ for } \ell = 1, \, \ldots, \, \infty \\
\end{aligned}
\end{equation}
\begin{equation}
\begin{aligned}  
\notag
    \delta_1 \sim \text{Gam}(a_1, b_1=1) \qquad
    \delta_{h} \sim \text{Gam}^{\text{T}}(a_2, b_2=1, c_2 = 1) \quad \mbox{ for }h  > 1. \\
\end{aligned}
\end{equation}
Here $a_1$ and $b_1$ are the shape and rate parameters of the gamma prior on the shrinkage parameter of the first dimension, while $a_2$ is the shape parameter, $b_2$ is the rate parameter, and $c_2$ is the left truncation point (here set to 1) of the truncated gamma prior for dimensions $h > 1$. This MTGP prior results in an increasing precision and therefore a shrinking variance in the higher dimensions of the latent space.

While \cite{bhattacharya_dunson_2011} state that $\omega_{\ell}$ stochastically increases under the restriction $a_2 > 1$, \cite{durante_2017} shows that this does not guarantee shrinkage, and indeed that shrinkage does not occur when $a_1 = 1$ and $a_2 = 1.1$. While the reader is referred to \cite{durante_2017} for guidelines to study the  behavior of the MGP under all possible combinations of $a_1$ and $a_2$, \cite{durante_2017} theoretically and empirically suggests that using $a_1 = 2$ and $a_2 = 3$ induces posterior
distributions with the desired characteristics, with improved performance when the true number of dimensions is low. Although the $\omega_{\ell}$ will be stochastically increasing under such hyperparameter settings, this sometimes poses difficulties in the MCMC algorithm used for inference. In particular, $\delta_h < 1$ can be accepted in some cases, which increases the latent positions' variance on dimension $h$ rather than decreasing it. To tackle this, here a truncated gamma distribution bounded between 1 and $\infty$ is assumed for the second and higher dimensions to ensure shrinkage across these dimensions. Nevertheless, the shrinkage parameter on the first dimension does not require the use of a truncated gamma distribution allowing it to have an unconstrained range and allowing the first dimension to  encode as much information as required. The variance of the first dimension therefore determines the maximum possible variance for higher dimensions.

Empirically (see supplementary material, Appendix \ref{app:MGPvsMTGP}), while the use of the MGP rather than the MTGP prior tends to make little difference in terms of inferring the posterior mode of the number of dimensions, the MGP does tend to result in more diffuse posteriors, with the MTGP being more precise in its inference of the dimension and the variance parameters. Moreover, the MTGP prior is faithful to the inherent model principle of shrinking variance at higher dimensions. Furthermore, the MTGP prior induces an intuitively appealing decrease in the order of importance of subsequent dimensions, as it is typical of dimension reduction methods.

Under this MTGP prior, the LSPM is nonparametric with infinitely many dimensions, where unnecessary higher dimensions' variances are increasingly shrunk towards zero. Dimensions that have variance very close to zero will then have little to no meaningful information encoded in them as the distances between nodes will be close to zero. Thus, the effective dimensions are those in which the variance is non-negligible.

% This is evident with $\delta_{h} \geq 1$ which causes $\omega_{\ell}^{-1}$ to always shrink in value, thus changing from a stochastic shrinking nature to a deterministic case.

% Meaningless dimensions can be truncated and reduced into a finite dimension model with less complexity while still being able to capture most of the network structure. However, converting from an infinite model directly into a finite model will result in the $\alpha$ parameter to be bias since the $\alpha$ is estimated for the infinite model. This effect is explored further in Section \hyperlink{section.4}{4}.

% Although the MTGP prior is nonparametric, it is not implementable in practice without setting an upper truncation level, $p$, on the number of dimensions fitted. Given the context, it is intuitive to set $p < n$; further, as every edge formation considers a pair of nodes, the truncation level can reasonably be taken to be half of the total number of nodes, i.e. $p < \frac{n}{2}$ is considered. Nevertheless, choosing a suitable value for the truncation level in the LSPM is arguably much less impactful than choosing the correct dimension in the LPM, as long as the true number of effective dimensions is lower than the truncation level. 

\subsection{Properties of the latent shrinkage position model}
Given the use of the multiplicative truncated gamma process prior, it is of interest to explore the behaviour of distances in the latent space imposed by the LSPM. The full derivations for this section are given in the supplementary material in Appendix \ref{app:properties}. 

\begin{figure}[tb]
     \centering
     \begin{subfigure}[b]{0.495\linewidth}
         \centering
         \includegraphics[width=\linewidth]{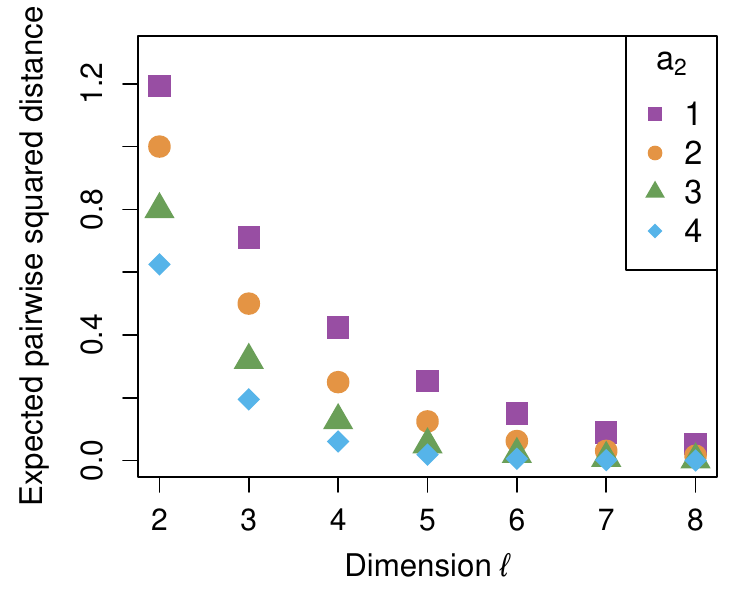}
         \caption{}
         \label{fig:DistWithinDim}
     \end{subfigure}
     \hfill
     \begin{subfigure}[b]{0.495\linewidth}
         \centering
         \includegraphics[width=\linewidth]{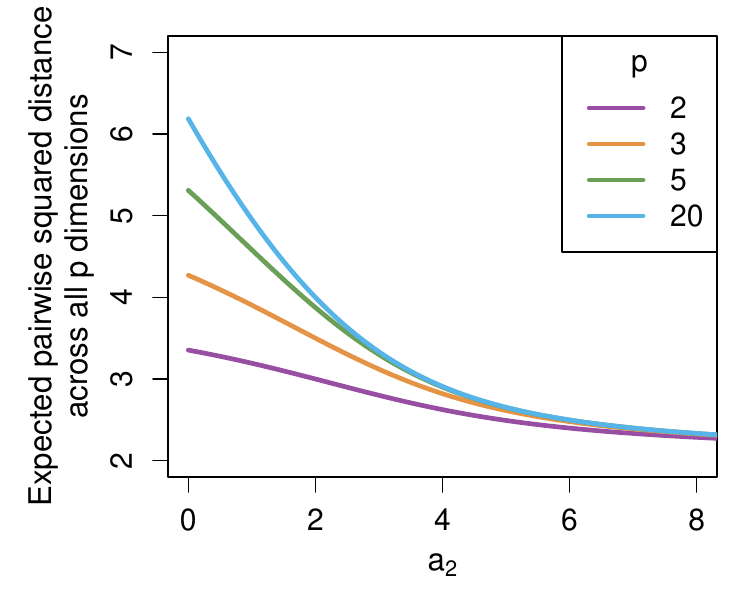}
         \caption{}
         \label{fig:DistAcrossDim}
     \end{subfigure}
        \caption{(a) The expected pairwise squared distance for each dimension $\ell$ for different values of $a_2$; (b) The expected pairwise squared distance in the latent space for different values of $a_2$ and different truncation levels $p$. }
        \label{fig:ExpectedDistance}
\end{figure}

The expected squared distance between nodes $i$ and $j$ within the $\ell$-th dimension given its precision $\omega_{\ell}$ is
\begin{equation}     \label{eq:dist}
    \mathbb{E}[(z_{i\ell}-z_{j\ell})^2 \mid \omega_{\ell}]  =  2\left(\frac{b_1}{a_1-1}\right)\left[\frac{ \Gamma(a_2-1,1)}{\Gamma(a_2,1)}\right]^{\ell-1} ,
\end{equation}
where $\Gamma$ is the upper incomplete gamma function. Since $\Gamma(a_2-1,1) < \Gamma(a_2,1)$, any dimension higher than $\ell=1$ is expected to have a smaller distance. Increasing $a_1$ will result in the biggest decrease in all of the expected distances for $\ell = 1, \ldots, \infty$ as later dimensions are based on previous dimension(s). Figure \ref{fig:ExpectedDistance}, in which $a_1$ is set to 2, illustrates the typical behaviour of the expected squared distance between a pair of nodes. Increasing $a_2$ will decrease the gamma function ratio in \eqref{eq:dist} and thus decrease the expected squared distance as shown in Figure \ref{fig:DistWithinDim}. For fixed $a_1$ and $b_1$, the minimum decrease in expected squared distance is $\lim_{a_2 \to 0} \frac{ \Gamma(a_2-1,1)}{\Gamma(a_2,1)} = 0.68$ which means the expected distance contribution from each higher dimension will be at least 32\% less than the previous dimension. 

% However, depending on the actual value sampled from the truncated gamma distribution, the actual distance contribution from the higher dimension may not always be lesser than the previous dimension.

The expected squared distance between nodes $i$ and $j$ in the latent space is the geometric sum of (\ref{eq:dist}) across all dimensions $\ell = 1, \ldots, p < \infty$, i.e.
\begin{equation*} \label{eq:distsum}
    \mathbb{E}[(\mathbf{z}_{i}-\mathbf{z}_{j})^2 \mid \omega_{1}, \ldots , \omega_{p}] = 2\left(\frac{b_1}{a_1-1}\right)\left[\frac{1 - \left( \frac{ \Gamma(a_2-1,1)}{\Gamma(a_2,1)} \right) ^{p}}{ 1 - \frac{b_2 \Gamma(a_2-1,1)}{\Gamma(a_2,1)}}\right].
\end{equation*}
Figure \ref{fig:DistAcrossDim} shows the behaviour of $ \mathbb{E}[(\mathbf{z}_{i}-\mathbf{z}_{j})^2 \mid \omega_{1}, \ldots , \omega_{p}] $ under different truncation levels while varying the hyperparameter $a_2$. As $a_2$ increases, the expected squared distance contribution from the higher dimensions gets increasingly small and encourages a smaller number of effective dimensions, which is notable at $a_2 > 6$, showing very small distance contributions from higher latent dimensions. The positive limit as $a_2 \rightarrow \infty$, alluded to in Figure \ref{fig:DistAcrossDim}, is due to the contribution of the expected squared distance from the first dimension.

%Small values of $a_2$ mean higher dimensions have an expected pairwise squared distance which is relatively similar to the previous dimension. 

% Networks where $n>100$ are robust with respect to hyperparameters initialisation; further, $a_2$ can also be learnt from the data if required. In general, the value of $a_2$ notably influences the result only when $n$ is small. 

\section{Inference}
\label{sec:3}

The joint posterior distribution of the LSPM is
$$
\mathbb{P}(\alpha, \mathbf{Z}, \bm{\delta} \mid \mathbf{Y})  \propto \mathbb{P}(\mathbf{Y}\mid\alpha, \mathbf{Z}) \mathbb{P}(\alpha) \mathbb{P}(\mathbf{Z} \mid \bm{\delta}) \mathbb{P}(\bm{\delta})
$$
where $\mathbb{P}(\mathbf{Z} \mid \bm{\delta})$ and $\mathbb{P}(\bm{\delta})$ are the prior distributions outlined in Section \ref{ssec:lspm}; for $\alpha$, a $N(\mu_{\alpha} = 0,\sigma^2_{\alpha} = 9)$ non-informative prior is assumed throughout, where $\mu_{\alpha}$ is the mean and $\sigma^2_{\alpha}$ is the variance. 
% For the MTGP priors, as suggested in \cite{durante_2017}, the hyperparameters are set here as $a_1 = 2 $ and $a_2 = 3$, which have been found to work well in practice. These hyperparameters, particularly $a_2$ have a considerable effect on the adaptive sampler; more details in Subsection \ref{ssec:adapt}.
Markov chain Monte Carlo (MCMC) is employed here to draw samples from the joint posterior distribution. Although the MTGP prior is nonparametric, it is not implementable in practice without setting a truncation level, $p$, on the number of dimensions fitted. Thus, an adaptive Metropolis-within-Gibbs sampler is used to dynamically shrink or augment the truncation level (see Section \ref{ssec:adapt}). Details of the derivations of the full conditional distributions of the latent positions and parameters are given in the supplementary material, in Appendix \ref{app:fullconditionals}.

The MCMC algorithm proceeds by iterating the following steps for $s=1, \ldots, S$ where $s$ denotes the current iteration and $S$ is the total number of iterations.

\begin{enumerate}
    \item Sample a value $\check{\mathbf{Z}}$ from the MVN$(\mathbf{Z}^{(s)}, k\mathbf{\Omega}^{-1(s)})$ proposal distribution where $k$ is a step size factor. Accept $\check{\mathbf{Z}}$ as $\mathbf{Z}^{(s+1)}$ with probability \\ $\frac{\mathbb{P}(\mathbf{Y}\mid\check{\mathbf{Z}},\alpha^{(s)})}{\mathbb{P}(\mathbf{Y}\mid\mathbf{Z}^{(s)},\alpha^{(s)})} \frac{\text{MVN}(\check{\mathbf{Z}}, k\mathbf{\Omega}^{-1(s)})}{\text{MVN}(\mathbf{Z}^{(s)}, k\mathbf{\Omega}^{-1(s)})}$, otherwise set $\mathbf{Z}^{(s+1)} = \mathbf{Z}^{(s)}$.
        
    \item Sample a value $\check{\alpha}$ from an informed Gaussian proposal distribution and accept $\check{\alpha}$ as $\alpha^{(s+1)}$ following the Metropolis-Hastings (M-H) acceptance ratio. Details of this informed proposal distribution are given in Section \ref{ssec:informA}.
    \item Sample $\delta_1^{(s+1)}$ from \\ $\text{Gam}\left(\frac{np}{2}+a_1, 
                \quad b_1 + \frac{1}{2}\sum_{i=1}^{n}\sum_{\ell=1}^{p} \prod_{m=2}^{\ell} \delta_{m}^{(s)} \left[z_{i\ell}^{(s+1)} \right]^2  \right) $, \\
                where $z_{i\ell}$ is the latent position of node $i$ in dimension $\ell$.
    \item Sample $\delta_h^{(s+1)}$ for $h=2, \ldots, p$ from \\
    $\text{Gam}^{\text{T}}\left(\frac{n(p-h+1)}{2}+a_2, \hspace{1pt} b_2 + \frac{1}{2}\sum_{i=1}^{n}\sum_{\ell=h}^{p} \prod_{m=1, m \neq h}^{\ell} \delta_m^{(s^*)}  \left[z_{i\ell}^{(s+1)} \right]^2, 1 \right),$ \\
        where $s^* = s + 1$ for $m < h$ and $s^* = s$ for $m>h$.
    \item Calculate $\omega^{(s+1)}_{\ell}$ by taking the cumulative product of $\delta_1^{(s+1)}$ to $\delta_{\ell}^{(s+1)}$.  
\end{enumerate}

% The geodesic distance used in the initialisation step is the shortest path between two nodes, measured as the number of edges that must be traversed to get from one node to the other. For a pair of nodes that are not connected, the geodesic distance is replaced with 1.5 times the largest geodesic distance in the network. The geodesic distance for every pair of nodes is calculated to form a $n \times n$ matrix. Using this pairwise geodesic distance, multidimensional scaling algorithm is used to construct a geometric representation of the nodes in a low-dimensional space, such that the pairwise distances between the nodes in the space approximate the original geodesic distance as closely as possible. The multidimensional scaling algorithm performs eigendecomposition to obtain the eigenvector which represents the positions for the nodes and the eigenvalue which represents the variance of the dimension.

Since the likelihood function considers the Euclidean distances between the latent positions, it is invariant to rotation, reflection, or translation of the latent positions. Thus, to ensure valid posterior inference, as in \cite{gormley_murphy_2010}, here a Procrustean transformation is performed which translates, reflects and rotates the configurations of the latent positions $\mathbf{Z}^{(1)}, \ldots, \mathbf{Z}^{(S)}$ to be as similar as possible to a reference configuration $\tilde{\mathbf{Z}}$. This reference configuration is the configuration with the highest log-likelihood in the burn-in period of the MCMC chain. While this choice is arbitrary, it is somewhat irrelevant as the configuration is used only to identify the model.

As inference from the MCMC algorithm is sensitive to its initial values, they are initialized using the following approach. When $s=0$:

% When $s=0$, calculate the geodesic distance of the nodes in the network. The geodesic distance \citep{Kolaczyk_Csárdi_2020} is the shortest path between two nodes, measured as the number of edges that must be traversed to get from one node to the other. Unconnected nodes are assigned a distance 1.5 times the largest computed geodesic distance. Then, apply classical multidimensional scaling (MDS) \citep{Cox_Cox_2001} to the geodesic distance of the network. To determine the initial truncation level $p_0$, K-means algorithm is used on the eigenvalues of the MDS and the number of data points in the smallest cluster is taken. The initial latent position $\mathbf{Z}^{(s)}$ is the resulting set of $n \times p_0$ positions from the MDS coordinates.

% Initialisation step. For $s=0$:
\begin{enumerate}
    % \item Set the truncation level $p$ where $p< \frac{n}{2}$. In practice, $p \approx 5$ has worked well.
    \item Calculate the geodesic distances between the nodes in the network. The geodesic distance \citep{Kolaczyk_Csárdi_2020} is the shortest path between two nodes, measured as the number of edges that must be traversed to move from one node to the other. 
    % Unconnected nodes are assigned a distance 1.5 times the largest computed geodesic distance.}
    \item Apply classical multidimensional scaling (MDS) \citep{Cox_Cox_2001} to the geodesic distances of the network.
    \item K-means clustering \citep{Wu_2012} is applied to the eigenvalues from the MDS and the number of eigenvalues in the smallest cluster is used as the initial truncation level, $p_0$. \label{infer:cluster_d}
    \item Set $\mathbf{Z}^{(s)}$ to be the set of $n \times p_0$ positions from the resulting MDS coordinates.
    \item Fit a standard regression model, with regression coefficients $\alpha$ and $\beta$, to the vectorised adjacency matrix, depending on the edge type, e.g.,
    \begin{enumerate}
        \item for binary edges, fit a logistic regression model where $\mbox{log odds} (q_{i,j}=1) 
                  = \alpha - \beta \Vert \mathbf{z}_i^{(s)}-\mathbf{z}_j^{(s)}\Vert^2 _2 $ to obtain estimates $\hat{\alpha}$ and $\hat{\beta} $;
        \item for count valued edges, fit a Poisson regression model where $\log (\lambda_{i,j}) = \alpha - \beta \Vert \mathbf{z}_i^{(s)}-\mathbf{z}_j^{(s)} \Vert^2_2,$ to obtain estimates $\hat{\alpha}$ and $\hat{\beta}.$
    \end{enumerate}
    \item Set $\alpha^{(s)} = \hat{\alpha}$. As the LSPM model (e.g., \eqref{eqn:binprob}) implicitly constrains $\beta = 1$, centre and rescale the latent positions by setting $\mathbf{Z}^{(s)} = \sqrt{|\hat{\beta}|} \tilde{\mathbf{Z}}^{(s)}$, where $\tilde{\mathbf{Z}}^{(s)}$ is the mean-centred matrix of the initial latent positions.
    
    \item Obtain $\omega_{\ell}^{(s)}$ for $\ell=1, \ldots, p_0$ by calculating the precision empirically from $\mathbf{Z}^{(s)}$.
    \item Set $\delta_1^{(s)} = \omega_{1}^{(s)}$ and calculate $\delta_h^{(s)} = \frac{\omega_{h}^{(s)}}{\omega_{h-1}^{(s)}}$ where $h = 2, \ldots, p_0$.
    \end{enumerate}

% Post-processing is used to deal with the identifiability issues that arise from the infinite configurations that can result from translation, reflection, and rotation operators on the latent positions.

\subsection{An informed proposal distribution for \texorpdfstring{$\alpha$}{} } \label{ssec:informA}

To improve mixing and the speed of convergence of the Markov chain, 
%rather than using a normal proposal distribution with fixed parameters, 
an informed proposal distribution is used for $\alpha$ in the Metropolis-Hastings algorithm. This proposal distribution has parameters that are updated as the chain progresses, ensuring that it shadows the target distribution well. This is achieved by approximating a non-linear term in the loglikelihood function using a quadratic Taylor expansion, similar to \cite{gormley_murphy_2010}. Details of the proposal distribution for the Poisson LSPM are provided below; the derivation of the informed proposal distribution for the logistic LSPM is given in the supplementary material in Appendix \ref{app:logit_inform_alpha}. 

For the Poisson LSPM, the loglikelihood is
% \begin{equation}
% \begin{split}
% \log L(\mathbf{Y}\mid\mathbf{Z},\alpha) &= \sum_{i \neq j} \hspace{2pt} [ (\alpha - \Vert \mathbf{z}_i-\mathbf{z}_j \Vert^2_2) y_{i,j} ] \\  & \qquad - \sum_{i \neq j} \hspace{2pt} [ \exp(\alpha - \Vert \mathbf{z}_i-\mathbf{z}_j \Vert^2_2) ]  - \sum_{i \neq j}\log[y_{i,j}!].
% \end{split}
% \label{eqn:Poisll}
% \end{equation}
%
\begin{equation}
\log L(\mathbf{Y}\mid\mathbf{Z},\alpha) = \sum_{i \neq j} \hspace{2pt} [ (\alpha - \Vert \mathbf{z}_i-\mathbf{z}_j \Vert^2_2) y_{i,j} ] - \sum_{i \neq j} \hspace{2pt} [ \exp(\alpha - \Vert \mathbf{z}_i-\mathbf{z}_j \Vert^2_2) ]  - \sum_{i \neq j}\log[y_{i,j}!] ,
\end{equation}
% \log  L(\mathbf{Y}_B|\mathbf{Z},\alpha)  &= \sum_{i \neq j} \hspace{2pt} [ \alpha y_{i,j} ] - \sum_{i \neq j} \hspace{2pt} [\Vert \mathbf{z}_i-\mathbf{z}_j \Vert^2_2y_{i,j} ] - \sum_{i \neq j} \hspace{2pt} [\log (1 + \exp(\alpha - \Vert \mathbf{z}_i-\mathbf{z}_j \Vert^2_2)) ] \\
the second term of which can be approximated by a quadratic Taylor expansion around $\alpha^{(s)}$ to give
\begin{equation}
\begin{aligned}  
%\notag
 - \sum_{i \neq j} \hspace{2pt} [ \exp(\alpha - \Vert \mathbf{z}_i-\mathbf{z}_j \Vert^2_2) ]  &\approx - \sum_{i \neq j} \hspace{2pt} [ \exp(\alpha^{(s)} - \Vert \mathbf{z}_i-\mathbf{z}_j \Vert^2_2) ] \\
 & \quad - (\alpha - \alpha^{(s)}) \sum_{i \neq j} \hspace{2pt} [ \exp(\alpha^{(s)} - \Vert \mathbf{z}_i-\mathbf{z}_j \Vert^2_2) ]  \\
 & \quad- 0.5(\alpha - \alpha^{(s)})^2 \sum_{i \neq j} \hspace{2pt} [ \exp(\alpha^{(s)} - \Vert \mathbf{z}_i-\mathbf{z}_j \Vert^2_2) ].
\end{aligned}
\label{eqn:Poisllapprox}
\end{equation}
% $N(\Bar{\mu}_{\alpha, B}, \Bar{\sigma}_{\alpha, B}^2)$, where the $B$ subscript indicates the binary setting or 
Substituting this expression %(\ref{eqn:Poisllapprox}) in (\ref{eqn:Poisll})
and 
%combiningSince the full conditional of $\alpha$ only needs to be proportional to $\alpha$, terms that are proportional constants to $\alpha$ are removed and a new proportional constant term $-\alpha^{(s)}y_{i,j}$ is introduced to the approximated loglikelihood. Then, by 
completing the square gives an approximated likelihood function that is quadratic in $\alpha$, which, when combined with the normal prior on $\alpha$, suggests the use of a normal proposal distribution $N(\Bar{\mu}_{\alpha, C}, \Bar{\sigma}_{\alpha, C}^2)$ with mean 
% \begin{equation}
% \begin{aligned}  
% \notag
% \Bar{\mu}_{\alpha, B} = \alpha^{(s)} + \Bar{\sigma}_{\alpha, B}^2\times \left[\sum_{i \neq j}  y_{i,j} - \sum_{i \neq j} \hspace{2pt} \frac{\exp(\alpha^{(s)} - \Vert \mathbf{z}_i^{(s)}-\mathbf{z}_j^{(s)} \Vert ^2_2 )}{1+\exp(\alpha^{(s)} - \Vert \mathbf{z}_i^{(s)}-\mathbf{z}_j^{(s)}\Vert ^2_2)}  + \frac{1}{\sigma^2_{\alpha}}(\mu_{\alpha}-\alpha^{(s)}) \right] ,
% \end{aligned}
% \end{equation}
\begin{equation}
\begin{aligned}  
\notag
\Bar{\mu}_{\alpha, C} = \alpha^{(s)} + \Bar{\sigma}^2_{\alpha, C}\left[ \sum_{i \neq j} y_{i,j} - \sum_{i \neq j} \hspace{2pt} [ \exp(\alpha^{(s)} - \Vert \mathbf{z}_i^{(s)}-\mathbf{z}_j^{(s)} \Vert ^2_2) ]  + \frac{1}{\sigma^2_{\alpha}}(\mu_{\alpha}-\alpha^{(s)}) \right],
\end{aligned}
\end{equation}
and variance
% \begin{equation}
% \begin{aligned}  
% \notag
% \Bar{\sigma}_{\alpha, B}^2 = \left[\sum_{i \neq j} \hspace{2pt} \frac{\exp(\alpha^{(s)} - \Vert \mathbf{z}_i^{(s)}-\mathbf{z}_j^{(s)} \Vert ^2_2)}{(1+\exp(\alpha^{(s)} - \Vert \mathbf{z}_i^{(s)} -\mathbf{z}_j^{(s)} \Vert ^2_2))^2} + \frac{1}{\sigma^2_{\alpha}}\right]^{-1} ,
% \end{aligned}
% \end{equation}
\begin{equation}
\begin{aligned}  
\notag
\Bar{\sigma}_{\alpha, C}^2 = \left[\sum_{i \neq j} \hspace{2pt} [ \exp(\alpha^{(s)} - \Vert \mathbf{z}_i^{(s)}-\mathbf{z}_j^{(s)} \Vert ^2_2) ] + \frac{1}{\sigma^2_{\alpha}}\right]^{-1}.
\end{aligned}
\end{equation}
The parameters of this informed proposal distribution depend on the current state of the chain and thus are automatically updated each iteration. This maintains the approximation of the target distribution across all iterations of the algorithm. A step size factor that multiplies the variance parameter of the proposal distribution is introduced to assist in achieving satisfactory acceptance and further improve mixing. In terms of computational gain when compared to using a random walk proposal distribution, in the case of a network with $n=50$ and 2 latent dimensions, for example, using the informed proposal distribution reduced the thinning required by up to 25\%.

\subsection{An adaptive MCMC sampler to infer the latent dimension} \label{ssec:adapt}
Inference through MCMC for the LSPM is implemented using an adaptive sampler. The sampler shrinks or augments the truncation level $p$, in order to have a finite number of active dimensions in each iteration of the MCMC chain. A similar sampler is used in \cite{bhattacharya_dunson_2011} and \cite{Murphy_Viroli_Gormley_2020}. At the $s$-th iteration, adaptation occurs with probability $\mathbb{P}(s) = \exp(-\kappa_0 - \kappa_1s)$, with $\kappa_0 \geq 0$ and $\kappa_1 > 0$ chosen so that adaptation occurs often at the beginning of the chain but decreases exponentially fast as the chain grows. Here, $3 \le \kappa_0 \le 5$ and $\kappa_1 = 3 \times 10^{-5}$ were found to be useful and are used across the studies that follow. Adaptation only occurs after the burn-in period, in order to ensure targeting of the posterior distribution.
% which adapts on average about every 25 steps at the start but drops to about every 500 steps after 100,000th iterations.

At an adaptation step, when $p>1$, a reduction in the truncation level is based on a criterion which considers the cumulative proportion of variance that the dimensions $\ell = 1, \ldots, p - 1$ contain. If the dimensions up to the $\ell$-th cumulatively contain at least a proportion $\epsilon_1$ of the total variance, then the dimensions from $\ell +1$ to $p$ add little information, and $p$ is reduced to $\ell$. In practice, $\epsilon_1 = 0.9$ has been found to work well.  
When the adaptation criterion for reducing $p$ is not met, an increase in $p$ is then considered by examining $\delta_p^{-1}$ and a threshold $\epsilon_2$.  If  $\delta_p^{-1} > \epsilon_2$, a new dimension is added, by sampling $\delta_{p+1}$ from the MTGP prior and the latent positions from a univariate zero mean Normal distribution with induced variance $\omega_{p+1}$. Under the MTGP, $\delta_p^{-1}< 1$ and when $\delta_p^{-1} \approx 1$ shrinkage in the $p$th dimension is weak and more dimensions may be required to fully describe the data. Therefore we consider $\epsilon_2 = 0.9$ which was found to work well in practice.

% Due to the truncated gamma prior, dimensions from 2 to $p$ have shrinkage strength greater than $1$, implying that  $\frac{1}{\delta_p}< 1$. If this ratio is close to $1$, there is no strong shrinkage in the $p$-th dimension, hence all dimensions up to $p$ may be required and may be not sufficient to fully describe the data. Therefore, we set $\epsilon_2 = 0.9$, which requires the shrinkage strength $\delta_p$ to be greater than 1.11. Reducing $\epsilon_2$ would make the sampler to add dimensions more often. This cutoff has been found to work well in practice and to provide a good exploration of the number of active dimensions when used in conjunction with the adaptive removal step. }

When $p=1$, only an increase in $p$ is possible. In this case, we consider the proportion of latent positions that have absolute deviation from the mean that exceeds the 95\% critical value of a standard Normal distribution. When this proportion is $\epsilon_3$ times greater than the expected 0.05 proportion, $p$ is increased to 2. Setting $\epsilon_3 = 5$ demonstrated good performance in simulation studies, with higher values of $\epsilon_3$ increasing the tendency to remain at 1 dimension.

The hyperparameters $\kappa_0$ and $a_2$ can influence mixing in the adaptive MCMC sampler. When $p$ is increased or decreased, it can take time for $\alpha$ to mix well as it compensates for the change in dimension. Small $\kappa_0$ increases the adaptation frequency, which can make achieving sufficient mixing of $\alpha$ challenging. Similarly, small $a_2$ encourages the addition of a dimension with variance similar to that of the $p$th dimension, which results in large distances between latent positions and $\alpha$ needs time to adjust. Conversely, an $a_2$ that is too large adds a dimension that has very little variance and encodes little information. The aforementioned $3 \le \kappa_0 \le 5$ and $a_2 = 3$ were found to be a good combination in practice. 

The adaptive sampler allows inference on the posterior distribution of the number of active dimensions $p$. The posterior mode $p_m$ is used as the estimate of the effective $p$, with credible intervals quantifying the associated uncertainty.

\subsection{Posterior predictive checking}

Posterior predictive checking is a useful way of assessing the fit of a model to the data. Any systematic differences between the networks simulated from the posterior predictive distribution and the observed network may indicate potential failings of the model \citep{Andrew_Gelman_John}. Sections \ref{sec:4} and \ref{sec:5} include posterior predictive checks to assess the fit of the LSPM to simulated and observed networks respectively. Samples drawn from the posterior predictive distribution are used to simulate replicate networks under the fitted model. These simulated networks are then compared to the observed network by checking similarity metrics, network properties, and distances between networks. The type of metrics used depends on the network type.

\subsubsection{Binary valued networks}
For binary networks, the similarity metrics considered here are the accuracy and the $F_1$-score between the observed network and the replicate networks drawn from the posterior predictive distribution. Accuracy is a measure of the correctly identified edges i.e., $ \text{accuracy} =  (\text{TP + TN})/[n(n-1)]$, where TP is the number of true positives and TN is the number of true negatives. The $F_1$-score is the harmonic mean of precision and recall i.e., $ F_1\text{-score} = 2 \times \frac{\text{precision} \times \text{recall}}{\text{precision}+\text{recall}}$ where $\text{precision} = \text{TP}/(\text{TP}+\text{FP})$ and $\text{recall} = \text{TP}/(\text{TP}+\text{FN})$, with FP the number of false positives, and FN the number of false negatives.

Network properties such as density and transitivity are also considered. A network's density is the ratio of the number of observed dyadic connections over the number of possible dyadic connections, while transitivity is three times the number of triangles divided by the total number of connected triples. These metrics assess if the LSPM captures network properties of the observed data. 

The Hamming distance is also considered, measuring the normalised difference between the observed network and a replicate network. The Hamming distance is given by $\frac{1}{n(n-1)} \sum_{1 \leq i \neq j \leq n} | y_{i,j} - y_{i,j}^{(r)}| $, where $y_{i,j}^{(r)}$ is the link between nodes $i$ and $j$ in the $r$th replicate network from the posterior predictive distribution.
% Frobenius norm is the Euclidean norm between the probabilities of forming a link between nodes in the observed network and the network simulated from the posterior predictive distribution. The Frobenius norm is only used in Section \hyperlink{section.4}{4} where the true probabilities are known.

\subsubsection{Count valued networks}
As binary network measures are not directly applicable to count valued networks, the (log) frequencies of counts in  replicate networks from the posterior predictive distribution are compared to those of the observed network. The mean absolute difference between replicate and observed counts is also considered.

% To assess model fit, a pseudo $R^2$ value \citep{Cameron_Windmeijer_1996, Sewell_Chen_2016} is used, which is a deviance based $R^2$ for count data. The pseudo $R^2$ is
% \begin{equation*}
%     R^2 = \frac{\sum_{i \neq j} y_{i,j} \log \left( \frac{\hat{\lambda}_{i,j}}{\Bar{y}} \right) - \hat{\lambda}_{i,j} + \Bar{y} }{\sum_{i' \neq j'} y_{i',j'} \log \left( \frac{y_{i',j'}}{\Bar{y}} \right)} 
% \end{equation*}
% where $\Bar{y} =  \frac{\sum_{i \neq j}y_{i,j}}{n(n-1)}$ and $\hat{\lambda}_{i,j}$ is the posterior mean of $\lambda_{i,j}$. Pseudo $R^2$ values close to 1 indicate good model fit.

% However, values close to 1 may not always be obtainable even when the true parameters are used as the nature of some data may not fit well with the assumption of the model distribution used. This is noticeable in the case of a overdispersed network where a Poisson distribution simply do not have an underlying parameter to account for overdispersion.

In the simulation studies outlined in Section \ref{sec:4}, the fit of the Poisson LSPM is also evaluated in terms of Euclidean distances for estimated and actual latent positions. Specifically, the ratios between the $n(n - 1)/2$ distances derived from the posterior mean latent positions and the distances derived from the true latent positions are computed. A distribution of these ratios tightly centered around 1 will indicate good model fit. 

\section{Simulation Studies}
\label{sec:4}
% The latent positions are sampled from a multivariate normal distribution where the mean vector is zero while the diagonal covariance matrix has variances that comes from the inverse cumulative product of the shrinkage strengths of the latent dimensions. This means that the latent positions will be independent, identically distributed, and uncorrelated between dimensions. 

The performance of the LSPM is assessed on simulated data scenarios. The simulated data are generated by Bernoulli trials for each node pair based on the probabilities derived from the distances between the nodes' latent positions. For count valued edges, a Poisson distribution is employed instead of the Bernoulli. 

The latent positions are simulated according to \eqref{eq:z_mvn} with the shrinkage strengths being manually set to explore their effect across different settings. Hyperparameters are set as $\mu_{\alpha} = 0, \sigma_{\alpha}=3, a_1=2, a_2=3, b_1 = b_2 = 1$. A total of 30 networks are simulated in each case. Different step sizes between 0.1 to 3 are used in the proposal distributions to ensure acceptance rates are within the $20\%-40\%$ range. The MCMC chains are run for 1,000,000 iterations for Section \ref{ssec:truncation} and \ref{ssec:density} with a burn-in period of 100,000 iterations, thinning every 1,500th. In Section \ref{ssec:size} and Section \ref{ssec:overdispersion}, 500,000 iterations are considered, the burn-in period is 50,000 with thinning every 1,500th iteration for the logistic LSPM and every 1,000th iteration for the Poisson LSPM.  

The simulation studies are structured as follows: Section \ref{ssec:truncation} examines the impact of different initial truncation levels, $p_0$. Section \ref{ssec:size} assesses LSPM capability under different network sizes, $n$. Section \ref{ssec:density} studies the performance of the logistic LSPM under different network densities; Section \ref{ssec:overdispersion} explores the performance of the Poisson LSPM  under different levels of overdispersion. Where relevant, violin plots are used to visualise posterior distributions for each of the 30  simulated networks. 
%NOT SURE WHAT THIS MEANS?here each violin plot corresponds to a posterior distribution that is conditioned on a specific number of active dimensions. These plots are overlaid with semi-transparency to display all possible numbers of active dimensions, as inferred by the LSPM results for the 30 simulated networks.
Red crosses or red lines within the violin plots represent the true values used to simulate the network. Throughout, the LSPM is compared with the LPM and posterior predictive checks are used to assess model fit. Additional results and posterior predictive checks for the simulation studies are available in the supplementary material, in Appendix \ref{app:simstudies}.

\subsection{Study 1: initial truncation level}
\label{ssec:truncation}
This section explores the effect of the initial number of
latent dimensions $p_0$ on the inference of the number of active dimensions. Networks with $n=100$ are generated with the true number of effective latent dimensions $p^*=4$ and shrinkage strengths of $\delta_1=0.5$, $\delta_2=1.1$, $\delta_3=1.05$, and $\delta_4=1.15$ which mean that the second dimension has importance similar to the first. Here, $\alpha=6$  meaning moderate network density (i.e. $\simeq 20\%$) in the case of binary networks. 
Four initial truncation levels are considered: $p_0$ initialised as described in step \ref{infer:cluster_d} of Section 3 (termed `auto') and $p_0 = \{2, 4, 10\}$ representing situations where the initial truncation has been underestimated, correctly specified, and overestimated, respectively. Across the simulated networks, $3 \le p_0 \le 6$ were found under the `auto' procedure, with $p_0 = \{4, 5\}$ in the majority of cases.

\begin{figure}[!ht]
\includegraphics[width=\linewidth]{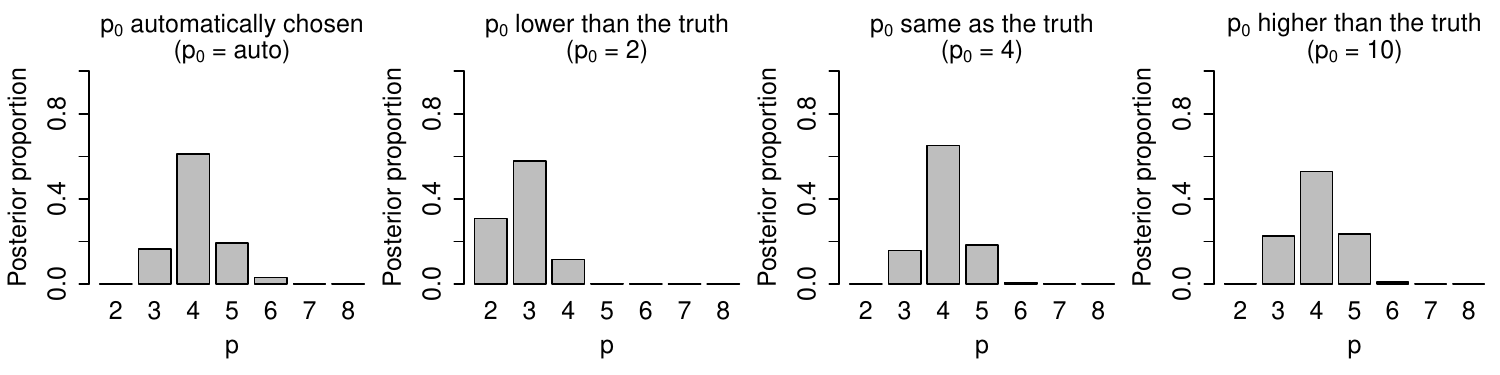}
\includegraphics[width=\linewidth]{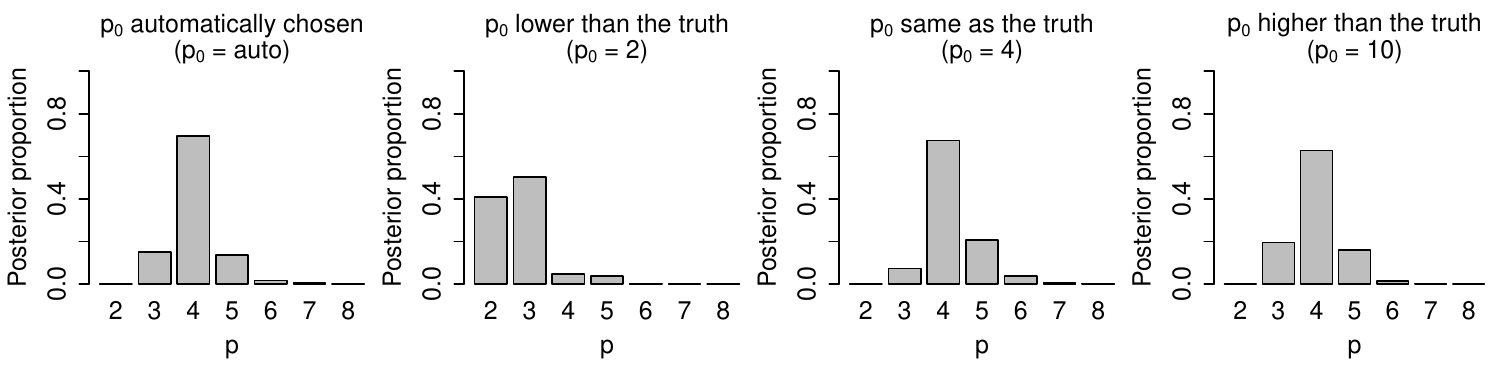}
        \caption{For different initial truncation levels, $p_0$, the posterior disrtibution of the number of active dimensions $p$ across 30 simulated networks under the logistic LSPM for binary networks (top) and under the Poisson LSPM for count networks (bottom).}
        \label{fig:dim_trunc}
\end{figure}

Figure \ref{fig:dim_trunc} shows the posterior of the number of active dimensions $p$ inferred from the 30 simulated networks. Under the `auto' initialisation procedure, and when $p_0 > p^*$, the posterior concentrates around the true number of dimensions. However, when $p_0 < p^*$, the posterior tends to concentrate on dimensions lower than $p^*$. Further results are summarised in Table \ref{tab:trunc_dim_protest}, which shows that for all initial values $p_0$, the 95\% credible intervals include the true dimension $p^*$. Across binary and count simulated networks, the proportion of times the posterior modal dimension $p_m = p^* = 4$ is greater than 0.63 under the `auto' initialisation, and is greater than 0.53 when $p_0 = 10$, with a poorer proportion of 0.13 or less when initialising $p_0 < p^*$.  Table \ref{tab:trunc_dim_protest} also reports the Procrustes correlations between the true latent positions
and the LSPM posterior mean positions, conditioned on $p_m$; there is good agreement across network types and initialisation strategies. These results suggest that the `auto' approach or starting with a reasonably large $p_0$ is advisable for accurate inference on the number of dimensions.

\begin{table}[b]
\caption{For different initial truncation levels, $p_0$, the posterior mode $p_m$ of the number of active dimensions, the proportion of the 30 simulated networks for which $p_m = p^* = 4$, and the Procrustes correlation of the 30 simulated networks' latent positions with the posterior mean positions. The 95\% credible intervals are given in the brackets.}
\label{tab:trunc_dim_protest}
\resizebox{\columnwidth}{!}{%
\begin{tabular}{|c||ccc||ccc|}
\hline
 & \multicolumn{3}{c||}{\textbf{Binary networks}} & \multicolumn{3}{c|}{\textbf{Count networks}} \\ \hline
$p_0$ & \multicolumn{1}{c|}{\begin{tabular}[c]{@{}c@{}}$p_m$ \\ \end{tabular}} & \multicolumn{1}{c|}{\begin{tabular}[c]{@{}c@{}}Proportion where \\ $p_{m} = p^*$\end{tabular}} & \begin{tabular}[c]{@{}c@{}}Procrustes correlation \\ \end{tabular} & \multicolumn{1}{c|}{\begin{tabular}[c]{@{}c@{}}$p_m$ \\ \end{tabular}} & \multicolumn{1}{c|}{\begin{tabular}[c]{@{}c@{}}Proportion where \\ $p_{m} = p^*$\end{tabular}} & \begin{tabular}[c]{@{}c@{}}Procrustes correlation \\ \end{tabular} \\ \hline
auto & \multicolumn{1}{c|}{4 (3, 6)} & \multicolumn{1}{c|}{0.63} & 0.96 (0.89, 1.00) & \multicolumn{1}{c|}{4 (3, 5)} & \multicolumn{1}{c|}{0.73} & 0.98 (0.84, 1.00) \\ \hline
2 & \multicolumn{1}{c|}{3 (2, 4)} & \multicolumn{1}{c|}{0.13} & 0.84 (0.69, 0.94) & \multicolumn{1}{c|}{3 (2, 5)} & \multicolumn{1}{c|}{0.03} & 0.81 (0.69, 0.92) \\ \hline
4 & \multicolumn{1}{c|}{4 (3, 5)} & \multicolumn{1}{c|}{0.67} & 0.97 (0.89, 0.99) & \multicolumn{1}{c|}{4 (3, 6)} & \multicolumn{1}{c|}{0.70} & 0.98 (0.80, 1.00) \\ \hline
10 & \multicolumn{1}{c|}{4 (3, 5)} & \multicolumn{1}{c|}{0.53} & 0.97 (0.89, 0.99) & \multicolumn{1}{c|}{4 (3, 5)} & \multicolumn{1}{c|}{0.67} & 0.98 (0.89, 1.00) \\ \hline
\end{tabular}%
}
\end{table}

\begin{figure}[htb]
\includegraphics[width=.95\linewidth]{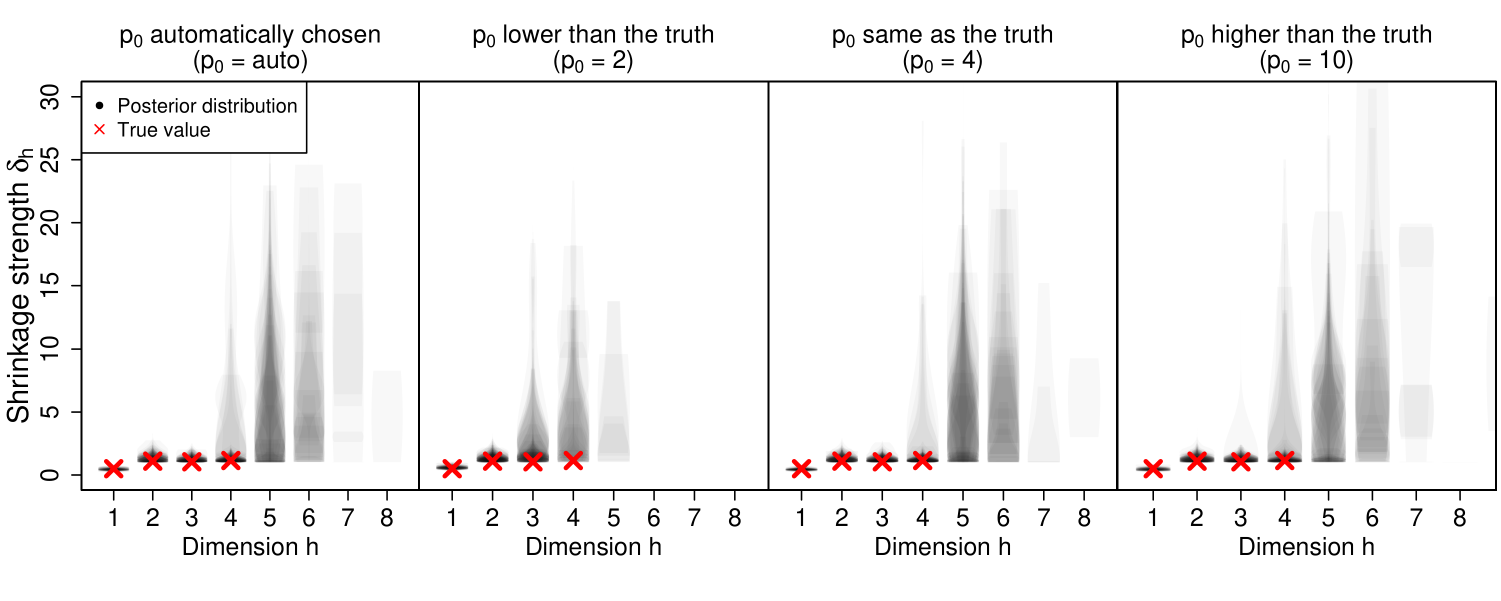} 
\includegraphics[width=.95\linewidth]{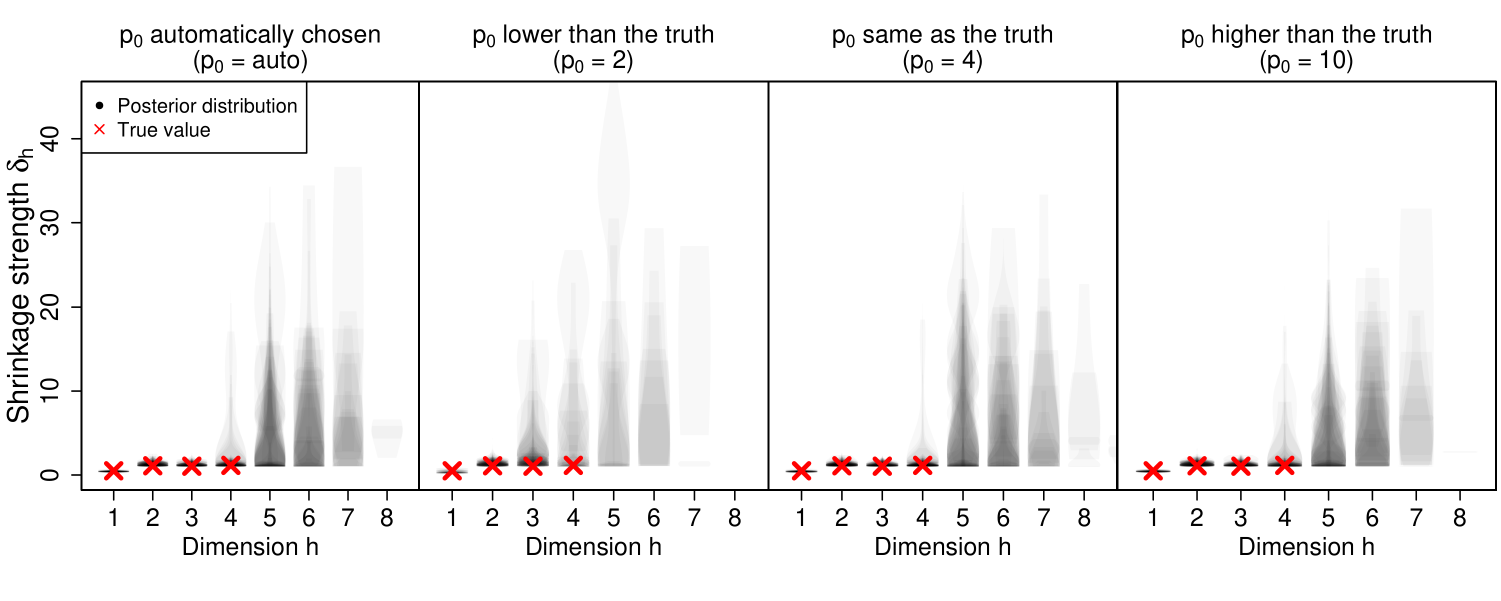} 
\caption{Posterior distributions of shrinkage strength across dimensions for different initial truncation levels $p_0$, across 30 simulated networks for the logistic LSPM for binary data (top) and the Poisson LSPM for count data (bottom).}
\label{fig:pmd_trunc}
\end{figure}

Figure \ref{fig:pmd_trunc} illustrates the posterior distributions of the shrinkage strength parameters, with the number of dimensions truncated at 8 for visualization purposes. 
%The range of shrinkage strength can be taken as an indication of the number of effective dimensions required. 
%An increase to a large value of $\delta_h$ suggests that $h - 1$ dimensions are sufficient to describe the data. 
When $p_0 = \{4, 10\}$ or is `auto' initialised, the shrinkage parameter $\delta_5$ tends to have a higher and more diffuse posterior than $\delta_4$, correctly indicating strong shrinkage after the 4th dimension. However, when $p_0 = 2 < p^*$, the shrinkage parameters $\delta_h$ for $h > p_0$ tend to be overestimated, causing excessive shrinkage below the true number of dimensions. When $p_0 < p^*$, the adaptive sampler can struggle to increase the dimension, with the shrinkage prior hyperparameter being very influential. These results support the suggestion that the `auto' approach or starting with a reasonably large $p_0$ is advisable for accurate inference on the number of dimensions.
%In fact, for large shrinkage strength, the addition of a new dimension is deemed unfavorable in the sampler. 
% Indeed, when $p_0 < p^*$, for dimensions $h > p_0$, overestimation of $\delta_h$ occurs; when a new dimension is added under the adaptive sampler, parameters are drawn from the shrinkage prior, which is strongly influenced byt the hyperparameter $a_2$, before data informs inference. This leads to a slow increase in dimension which is unfavourable when $p^* > p_0$.

\begin{figure}[htb]
     \centering
     \begin{subfigure}[b]{0.495\linewidth}
         \centering
         \includegraphics[width=\linewidth]{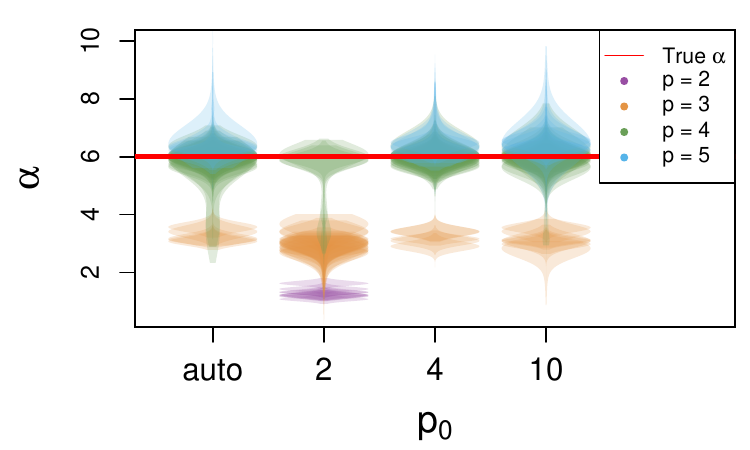}
         \caption{}
         \label{fig:trunc_alpha}
     \end{subfigure}
     \hfill
     \begin{subfigure}[b]{0.495\linewidth}
         \centering
         \includegraphics[width=\linewidth]{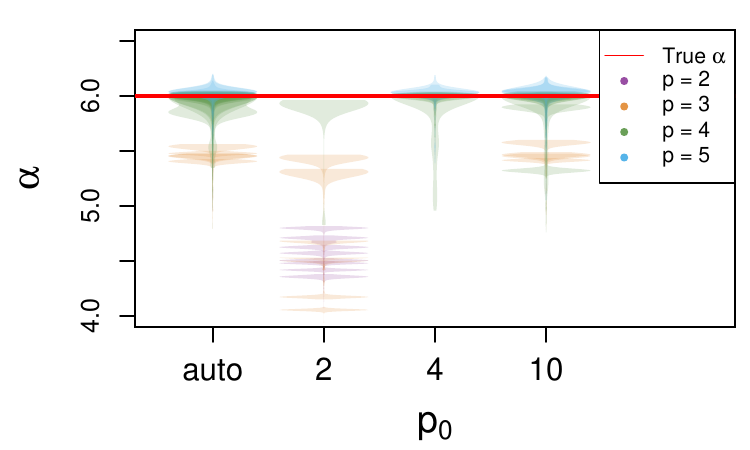}
         \caption{}
         \label{fig:trunc_alpha_c}
     \end{subfigure}
        \caption{Posterior distributions of $\alpha$ for different initial truncation levels $p_0$ across 30 simulated networks under (a) the logistic LSPM for binary networks and (b) the Poisson LSPM for count networks.}
        \label{fig:alpha_trunc}
\end{figure}

Posterior distributions for the intercept parameter $\alpha$, conditional on different numbers of active dimensions $p = \{2, 3, 4, 5\}$, are illustrated in Figure \ref{fig:alpha_trunc}.  When $p = p^* = 4$, for any $p_0$, $\alpha$ is accurately inferred, with more diffuse posteriors in the case of binary networks than for count networks. However, when $p \neq p^* $, the $\alpha$ parameter compensates for the additional or missing dimensions: when $p < p^*$ underestimation of $\alpha$ occurs, due to the missing contribution from the Euclidean distances in the omitted dimensions. 
%The lower $\alpha$ estimate compensates in the link probability for the missing dimensions in order to retain network density similar to the observed network. 
Conversely, when $p > p^*$, $\alpha$ is overestimated to account for the additional distance contributed by the extra dimensions, but with a smaller bias than when $p < p^*$.

\begin{figure}[htb]
\includegraphics[width=\linewidth]{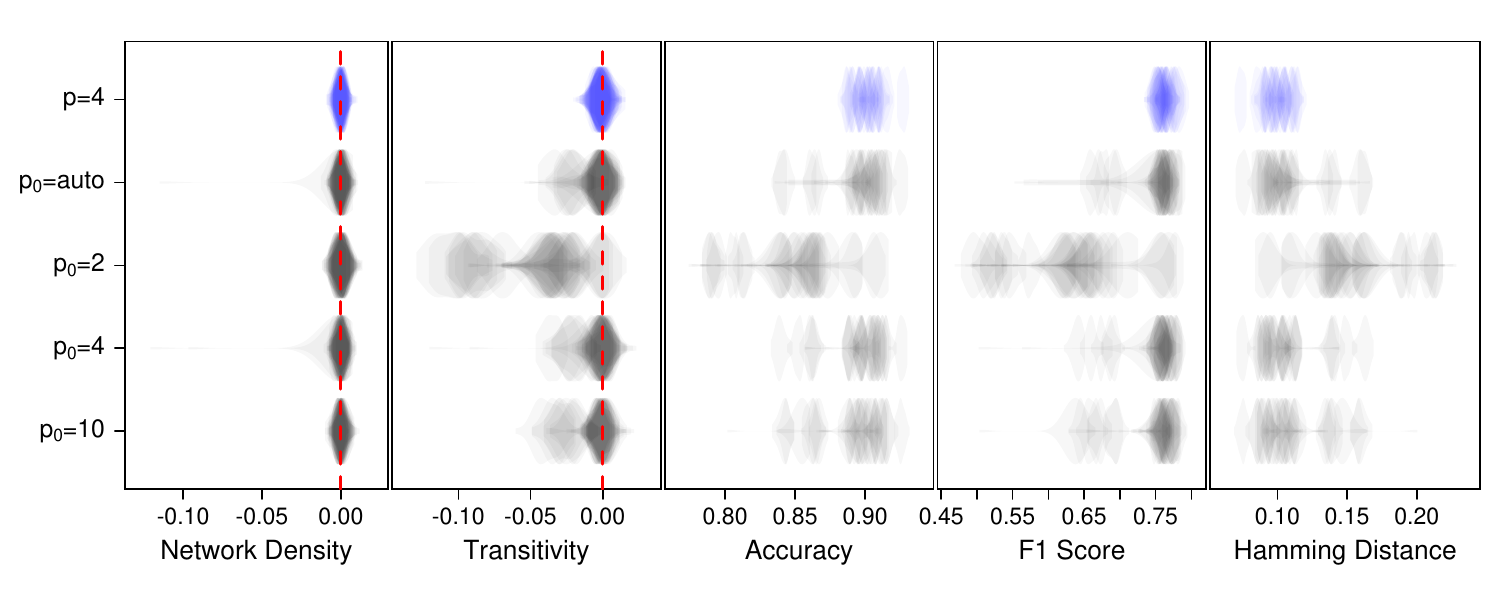} 
\caption{Posterior predictive checks, across 30 simulated binary networks, for the logistic LSPM (gray) under different initial truncation levels $p_0$ conditioned on $p_m$ and the logistic LPM (blue) with $p = 4$. The violin plots indicate the metric distributions from 30 replicate networks. The network density and transitivity plots illustrate the differences between the posterior predictive and observed networks.}
\label{fig:pred_trunc}
\end{figure}

\begin{figure}[htb]
%\nextfloat
\hfill
          \begin{subfigure}[b]{0.49\linewidth}
         \centering
         \includegraphics[width=\linewidth]{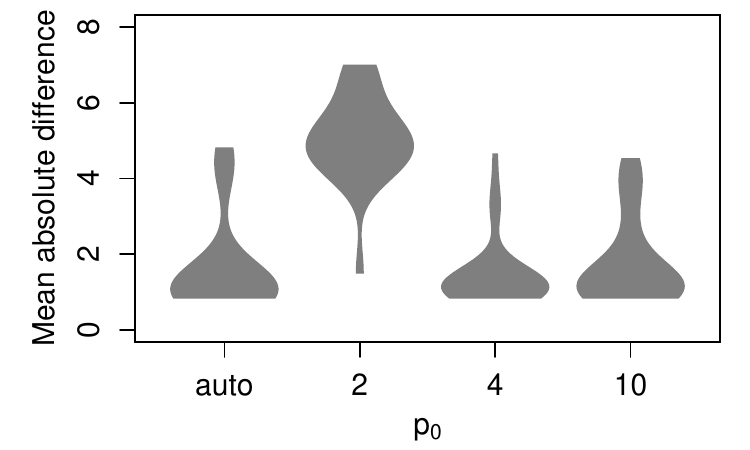}
         \caption{}
         \label{fig:trunc_abs}
     \end{subfigure}
          \begin{subfigure}[b]{0.49\linewidth}
         \centering
         \includegraphics[width=\linewidth]{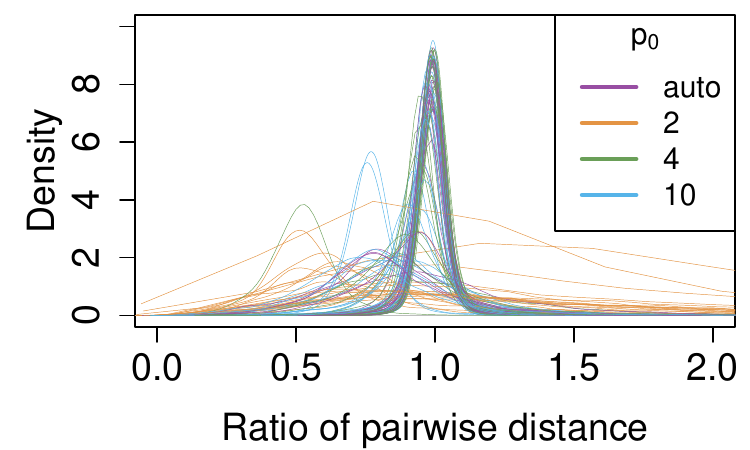}
         \caption{}
         \label{fig:trunc_ratio}
     \end{subfigure}

    \begin{subfigure}[b]{0.98\linewidth}
         \centering
         \includegraphics[width=\linewidth]{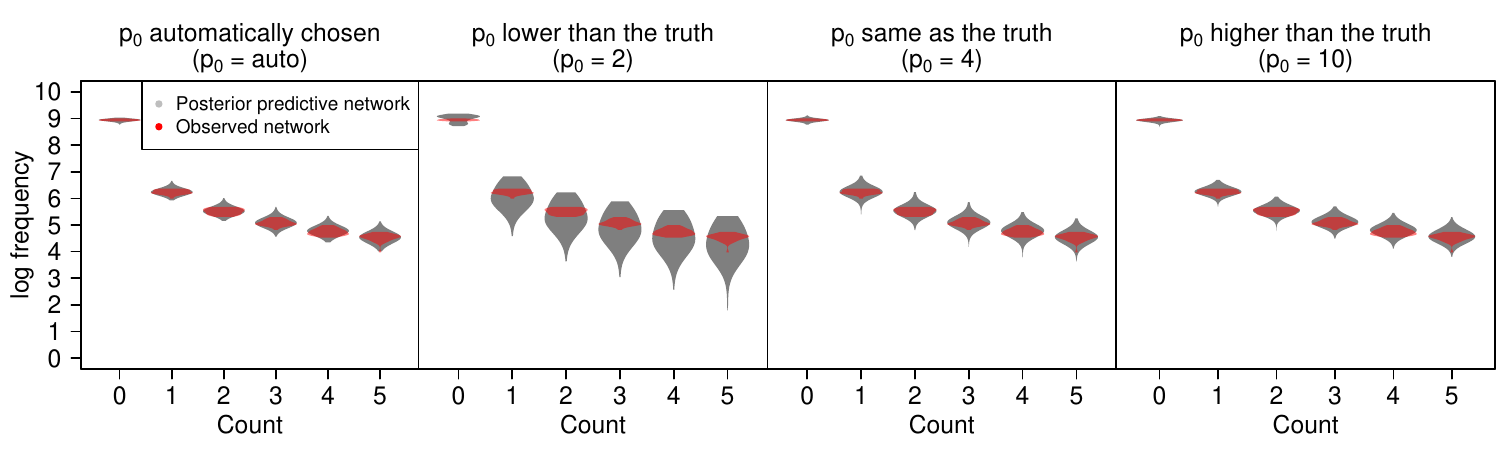}
         \caption{}
         \label{fig:trunc_logcount}
     \end{subfigure}
        \caption{Assessment of model fit for the Poisson LSPM via (a) the mean absolute difference between replicate and observed network counts, (b) ratios of the pairwise distances between inferred and true node locations, and (c) the log frequency of replicate and observed network counts. }
        \label{fig:pred_trunc_c}
\end{figure}

Posterior predictive checks under the logistic LSPM for binary simulated networks initialised using the `auto' approach and $p_0=\{2, 4, 10\}$ are shown in Figure \ref{fig:pred_trunc}. A number of LPM models were fitted using the \texttt{latentnet} R package \citep{latentnetpackage} with $p = \{1, \ldots, 8\}$; the BIC suggested $p = 4$ dimensions as optimal. 
%All posterior checks were conducted for each of the 30 simulated networks. 
The logistic LSPM fitted well and similarly across $p_0 \ge p^*$ while model fit was poorer across all metrics considered when $p_0 < p^*$. The LPM with $p=4$ also fitted well, but required multiple models to be fit and the use of a model selection criterion. Figure \ref{fig:pred_trunc_c} shows the posterior predictive checks for the Poisson LSPM fitted to simulated count networks. Similar behaviour to the binary case is observed with poor model fit when $p_0 < p^*$ and improved fit when $p_0 \ge p^*$: mean absolute count differences are small and ratios of pairwise distances between true and inferred latent locations are centred around 1. There is good agreement between the log frequency of counts in posterior predictive networks and the observed counts,  with $p_0=2$ having larger uncertainty.

In terms of computational cost, fitting the logistic LSPM on a computer with an i7-10510U CPU and 16GB RAM took on average 40 minutes for a simulated binary network with $n=100$. For comparison, fitting a single $p = 4$ LPM with the \texttt{latentnet} \citep{latentnetpackage} R package, with default settings but the same burn in period and thinning as LSPM, took on average 37 minutes. However, fitting a number of LPM models each with a different $p$ and choosing between them using the BIC is required in the LPM
setting, thus incurring additional computational cost.

\clearpage
\subsection{Study 2: network size}
\label{ssec:size}
Here the focus is on assessing the effect of different network sizes. Networks with $n=\{20, 50, 100, 200\}$ are generated with $p^*=2$, and shrinkage strengths of $\delta_1=0.5$ and $\delta_2=1.1$ giving $\omega_{1}^{-1} = 2$ and $\omega_{2}^{-1} = 1.82$ respectively. Here, $\alpha=3$ leading to a moderate network density (i.e. $\simeq 20\%$) in the binary network case. 

% \usepackage{graphicx}
\begin{table}[b]
\caption{For different $n$,  the posterior mode $p_m$ of the number of active dimensions,  the proportion of the 30 simulated networks for which $p_m = p^* = 2$, and the Procrustes correlation of the 30 simulated networks' positions with the LSPM posterior mean positions. The 95\% credible intervals are given in brackets.}
\label{tab:size_dim_protest}
\resizebox{\columnwidth}{!}{%
\begin{tabular}{|c||ccc||ccc|}
\hline
 & \multicolumn{3}{|c||}{\textbf{Binary networks}} & \multicolumn{3}{|c|}{\textbf{Count networks}} \\ \hline
$n$ & \multicolumn{1}{c|}{\begin{tabular}[c]{@{}c@{}} $p_m$\end{tabular}} & \multicolumn{1}{c|}{\begin{tabular}[c]{@{}c@{}}Proportion where \\ $p_{m} = p^*$\end{tabular}} & \begin{tabular}[c]{@{}c@{}}Procrustes correlation \\ \end{tabular} & \multicolumn{1}{c|}{$p_m$} & \multicolumn{1}{c|}{\begin{tabular}[c]{@{}c@{}}Proportion where \\ $p_{m} = p^*$\end{tabular}} & \begin{tabular}[c]{@{}c@{}}Procrustes correlation \\ \end{tabular} \\ \hline
20 & \multicolumn{1}{c|}{2 (2, 3)} & \multicolumn{1}{c|}{1.00} & 0.964 (0.882, 0.985) & \multicolumn{1}{c|}{2 (2, 3)} & \multicolumn{1}{c|}{0.93} & 0.992 (0.807, 0.998) \\ \hline
50 & \multicolumn{1}{c|}{2 (2, 4)} & \multicolumn{1}{c|}{0.80} & 0.992 (0.981, 0.995) & \multicolumn{1}{c|}{2 (2, 3)} & \multicolumn{1}{c|}{0.97} & 0.998 (0.995, 0.999) \\ \hline
100 & \multicolumn{1}{c|}{2 (2, 4)} & \multicolumn{1}{c|}{0.87} & 0.996 (0.992, 0.997) & \multicolumn{1}{c|}{2 (2, 4)} & \multicolumn{1}{c|}{0.93} & 0.999 (0.974, 1.000) \\ \hline
200 & \multicolumn{1}{c|}{2 (2, 5)} & \multicolumn{1}{c|}{0.87} & 0.998 (0.997, 0.998) & \multicolumn{1}{c|}{2 (2, 4)} & \multicolumn{1}{c|}{0.77} & 1.000 (0.999, 1.000) \\ \hline
\end{tabular}%
}
\end{table}

Inference on $p_m$, the posterior mode of $p$, and the Procrustes correlations between inferred and observed latent locations as $n$ varies are detailed in Table \ref{tab:size_dim_protest}. The posterior mode $p_m$ is accurate across different $n$, while the 95\% credible intervals' upper bounds tend to increase as $n$ increases. This affects the proportion of times the posterior mode is equal to the optimal $p^* = 2$. Networks with larger $n$ are characterized by larger uncertainty on the dimension of the latent space, as the sampler tends to explore higher dimensional solutions. In the majority of cases where $p_m$ is not 2, the posterior mode indicates 3 dimensions. 
The Procrustes correlation between true and posterior mean latent positions, conditioned on $p_m$, are increasingly accurate and precise as $n$ increases, with higher values for count than binary networks. As $n$ increases, posterior predictive checks under the LSPM indicate improved model fit (see supplementary material, Appendix \ref{app:simstudies}).
%Some aspects worth noting are that the variance tends to be overestimated in earlier effective dimensions, while being underestimated in later effective dimensions. 
%In general, Poisson LSPM performs slightly better than logistic LSPM may be attributed to the richer information from having count edges instead of binary edges between nodes. 

\subsection{Study 3: density in binary networks}
\label{ssec:density}
In the context of binary networks, the impact of network density on the performance of the LSPM is explored. Networks with $n=50$, $p^*=3$, $\delta_1=0.5$, $\delta_2=1.1$, and $\delta_3=1.05$ are generated. A range of $\alpha$ values between 0 and 30 are used to simulate networks with densities ranging from 2\% to 99\%. Table \ref{tab:dens_dim_protest} reports the posterior mode of $p$ and Procrustes correlations between inferred and true latent positions under different $\alpha$ values. Under the LSPM, the accuracy of $p_m$ increases as the network density increases from 2\% to 65\%, but accuracy deteriorates for denser networks with density $> 0.79$.
%Networks that are too sparse or too dense provide not sufficient information for the identification of the appropriate number of latent dimensions. 

\begin{figure}[t]
     \centering
     \begin{subfigure}[b]{0.495\linewidth}
         \centering
         \includegraphics[width=\linewidth]{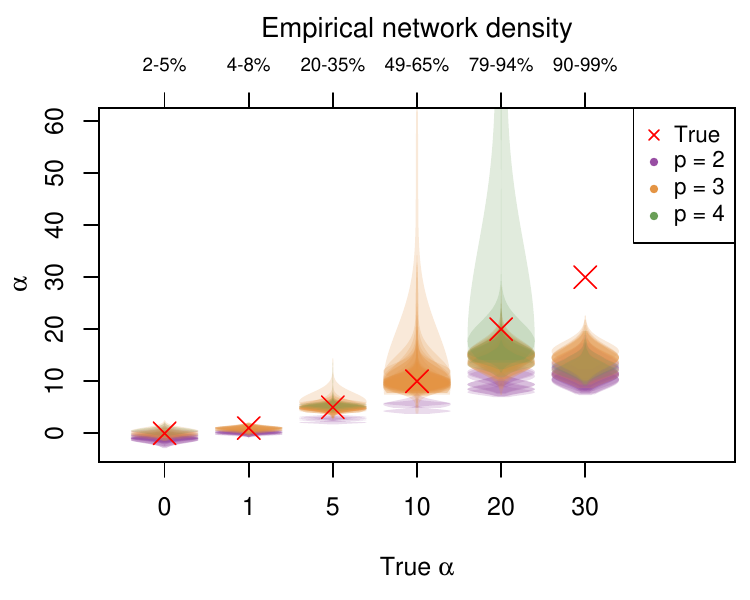}
         \caption{}
         \label{fig:dens_alpha}
     \end{subfigure}
     \hfill
     \begin{subfigure}[b]{0.495\linewidth}
         \centering
         \includegraphics[width=\linewidth]{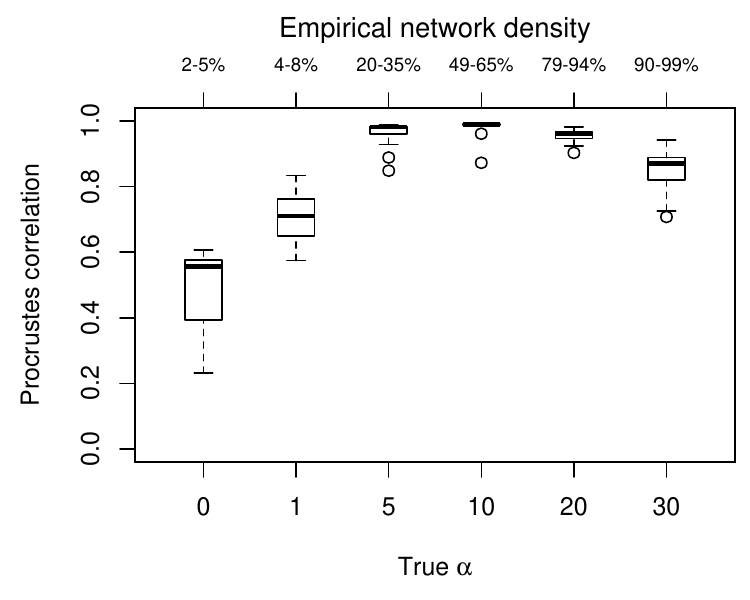}
         \caption{}
         \label{fig:dens_protest}
     \end{subfigure}
        \caption{For different network densities the (a) posterior distributions of $\alpha$ and (b) Procrustes correlations between true and LSPM posterior mean latent positions.}
        \label{fig:alpha_dens}
\end{figure}

Figure \ref{fig:dens_alpha} illustrates the effect of different network densities on the estimation of the $\alpha$ parameter. Values of $\alpha \leq 10$ are estimated accurately whereas higher values of $\alpha$ tend to be underestimated. Further, as it is evident from Figure \ref{fig:dens_protest} where the Procrustes correlations decrease for large values of $\alpha$, the true latent positions are difficult to recover since their influence on the link probabilities is small in comparison to $\alpha$ when $\alpha$ is large. Similarly, the Procrustes correlations decrease for lower values of $\alpha$ as there are many possible locations that will produce a mostly unconnected network. In summary, the logistic LSPM performs best for networks of moderate density. Additional posterior distributions of dimensions, variances and shrinkage strength parameters, and posterior predictive checks, are available in the supplementary material, in Appendix \ref{app:simstudies}.

\begin{table}[tb]
\caption{For different network densities, the posterior mode $p_m$ of the number active dimensions, the proportion of the 30 simulated networks for which $p_m = p^* = 3$, and the Procrustes correlation of the 30 simulated networks' positions against the LSPM posterior mean positions. The 95\% credible intervals are given in the brackets.}
\label{tab:dens_dim_protest}
\begin{tabular}{|c|c|c|c|c|}
\hline
True $\alpha$ & \begin{tabular}[c]{@{}c@{}}Empirical network\\ density (\%)\end{tabular} & \begin{tabular}[c]{@{}c@{}}$p_m$\end{tabular} & \begin{tabular}[c]{@{}c@{}}Proportion where \\ $p_{m} = p^*$\end{tabular} & \begin{tabular}[c]{@{}c@{}}Procrustes correlation \\ \end{tabular} \\ \hline
0 & 2 - 5 & 2 (2, 4) & 0.23 & 0.38 (0.22, 0.64) \\ \hline
1 & 4 - 8 & 2 (2, 3) & 0.48 & 0.50 (0.39, 0.75) \\ \hline
5 & 20 - 35 & 3 (2, 4) & 0.77 & 0.98 (0.87, 0.99) \\ \hline
10 & 49 - 65 & 3 (2, 4) & 0.90 & 0.99 (0.93, 0.99) \\ \hline
20 & 79 - 94 & 3 (2, 4) & 0.57 & 0.96 (0.91, 0.98) \\ \hline
30 & 90 - 99 & 2 (2, 4) & 0.33 & 0.70 (0.49, 0.90) \\ \hline
\end{tabular}
\end{table}

\subsection{Study 4: overdispersion in count networks}
\label{ssec:overdispersion}
Count network data are often characterised by overdispersion, whereby the variance of the counts is greater than their mean. Here, the impact on the performance of the Poisson LSPM of different levels of overdispersion in count network data is examined by varying the $\alpha$ and $\delta$ parameters. The first set of 30 networks are simulated with $\alpha = 0.5$ and $\delta_1 = \delta_2 = 1.5$ giving mean counts between 0.45 and 0.60 and variance between 0.65 and 0.85 (low overdispersion). Another 30 more overdispersed count networks are generated with $\alpha = 1.5$, $\delta_1=0.5$, and $\delta_2=1.5$ giving mean counts between 0.5 and 0.7 and variance between 1.4 and 2.0 (moderate overdispersion). The final set of 30 highly overdispersed count networks are generated with $\alpha = 5$, $\delta_1=0.1$, and $\delta_2=1.5$ giving mean counts between 3 and 6 and variance between 220 and 420 (high overdispersion).

\begin{figure}[tb]
\includegraphics[width=\linewidth]{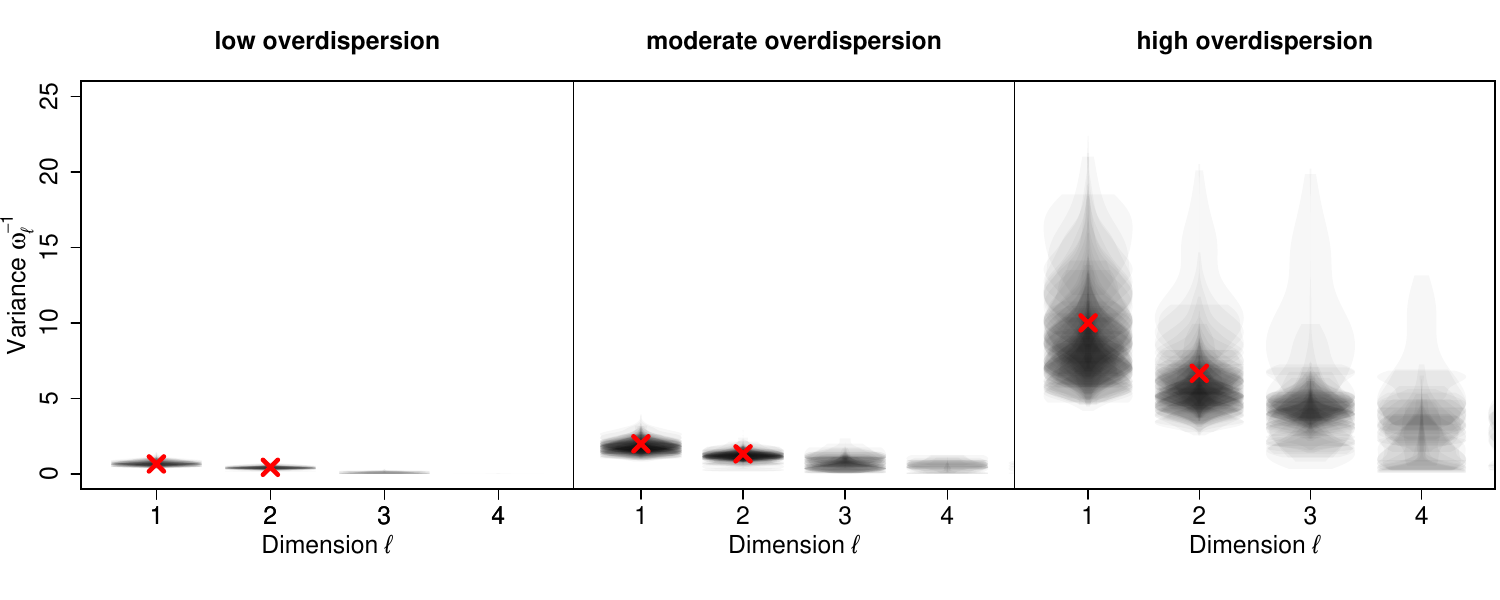} 
\caption{Posterior distributions of variance under the Poisson LSPM fitted to simulated networks with different levels of overdispersion.}
\label{fig:pmv_disp}
\end{figure}

Table \ref{tab:disp_dim_protest} reports on the posterior mode of $p$ and on the Procrustes correlation between true and inferred positions for different overdispersion levels. As overdispersion increases, the LSPM tends to overestimate the number of effective dimensions. Figure \ref{fig:pmv_disp} also illustrates that as overdispersion increases, inferential accuracy and precision for the variance parameters decreases.  Additional posterior
distributions of dimensions, variances and shrinkage strength parameters, and posterior
predictive checks, are available in the supplementary material, in Appendix \ref{app:simstudies}.

\begin{table}[tb]
\caption{For different levels of overdispersion, the  posterior mode $p_m$ of the number of active dimensions, the proportion of 30 simulated networks for which $p_m = p^* = 2$, and the Procrustes correlation of the 30 simulated networks' positions against the LSPM posterior mean positions. The 95\% credible intervals are given in the brackets.}
\label{tab:disp_dim_protest}
\begin{tabular}{|l|c|c|c|}
\hline
\begin{tabular}[l]{@{}c@{}}Level of \\overdispersion\end{tabular} & \begin{tabular}[c]{@{}c@{}}  $p_m$\end{tabular} & \begin{tabular}[c]{@{}c@{}}Proportion where \\ $p_{m} = p^*$\end{tabular} & \begin{tabular}[c]{@{}c@{}}Procrustes correlation \\ \end{tabular} \\ \hline
Low & 2 (2, 3) & 0.93 & 0.991 (0.950, 0.993) \\ \hline
Moderate & 2 (2, 4) & 0.80 & 0.997 (0.885, 0.998) \\ \hline
High & 3 (2, 4) & 0.43 & 0.855 (0.732, 0.914) \\ \hline
\end{tabular}
\end{table}

\section{Illustrative applications}
\label{sec:5}
The logistic and Poisson LSPMs are fitted to a range of real network datasets with different characteristics: the well-known Zachary karate club binary network which has a small number of nodes, a cat brain connectivity binary network with moderate number of nodes and density, a relatively large but sparse worm nervous system binary network, and an overdispersed phone calls count network. In all cases hyperparameters, initial values, and step sizes are set as in Section \ref{sec:4}.

\subsection{Zachary karate club binary network data}
The well-known Zachary's karate club network \citep{Zachary_1977} is a social network with $n=34$ members and describes the relationships in a university karate club. The karate teacher, Mr. Hi, and the club president, John, are the two central figures and the club was divided into two new clubs after an argument between them. The network is binary, undirected and has a density of 13.9\%. For inference, 10 MCMC chains are run, each for 100,000 iterations with a burn-in period of 1,000 and every 400th sample thinned.

\begin{figure}[htb]
     \centering
     \begin{subfigure}[b]{0.25\linewidth}
         \centering
         \includegraphics[width=\linewidth]{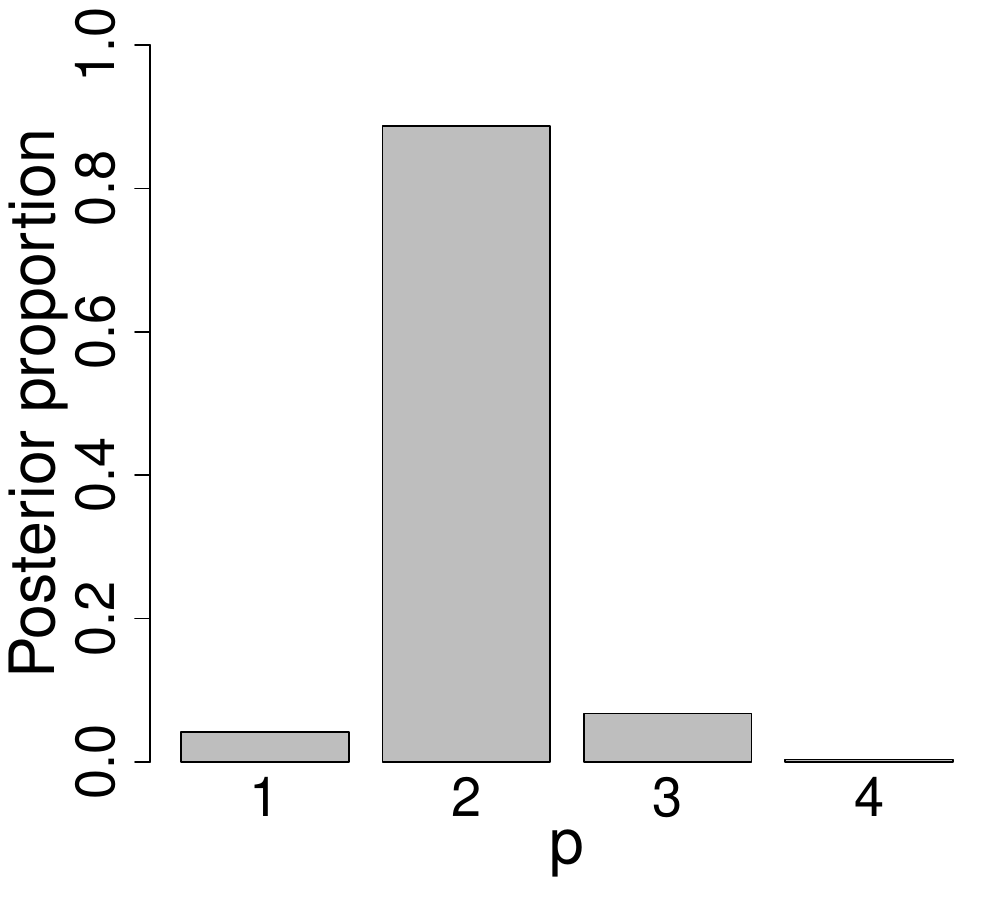}
         \caption{}
         \label{fig:zach_dim_bar}
     \end{subfigure}
     \hfill
     \begin{subfigure}[b]{0.74\linewidth}
         \centering
         \includegraphics[width=\linewidth]{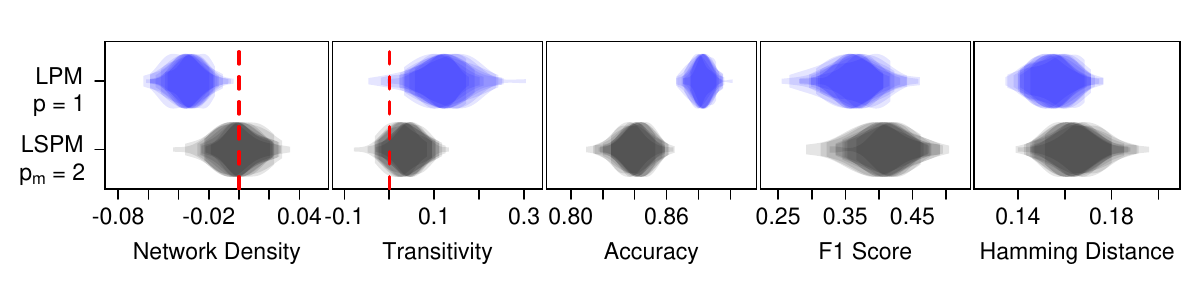}
         \caption{}
         \label{fig:zach_pred}
     \end{subfigure}
        \caption{Zachary karate club network: (a) the posterior distribution of $p$ and 
        (b) posterior predictive checks for LPM with $p = 1$ and the logistic LSPM for $p_m = 2$.}
        \label{fig:zachary}
\end{figure}

Figure \ref{fig:zach_dim_bar} indicates that the posterior modal dimension is 2, with low associated uncertainty. Upon fitting multiple LPMs with $p = \{1, \ldots, 5\}$, the BIC suggests 1 dimension is optimal.  Posterior predictive checks for the LPM with $p = 1$ and the logistic LSPM with $p_m=2$ (Figure \ref{fig:zach_pred}) indicate better fit for network density, transitivity and $F_1$ score under the LSPM while the LPM has better accuracy and Hamming distance. The Procrustes correlation between the posterior mean positions from the 1 dimensional LPM and the first dimension of the LSPM has a median value of 0.92 (standard deviation of 0.06) across 10 logistic LSPM and LPM chains. Additional results regarding the posterior distributions of variance and shrinkage strength parameters can be found in the supplementary material, Appendix \ref{app:application}.

\subsection{The cat brain connectivity binary network data}

The logistic LSPM is used to analyse a cat brain connectivity network which includes $n=65$ non-overlapping regions in the cortex treated as nodes and 1139 interregional macroscopic axonal projections treated as edges \citep{Scannell_Blakemore_Young_1995, de_Reus_van_den_Heuvel_2013}. The cortex regions were classified into four categories namely: visual regions (18 areas), auditory regions (10 areas), somatomotor regions (18 areas), and frontolimbic regions (19 areas). These classifications were based on neurophysiological information about the functional role of each brain area. 
%Details of how these network data were gathered and constructed can be found in the appendix of the original study by \cite{Scannell_Blakemore_Young_1995}. 
Here, the network is viewed as a binary directed network, with $y_{i,j} = 1$ indicating there is a connection between regions $i$ and $j$, while $y_{i,j} = 0$ indicating there is not. The network density is $27.37\%$.

%Hyperparameters, initial values, and step sizes are set as in Section \ref{sec:4} and the fitted truncation level is $p=5$. 
To assess convergence, the logistic LSPM is fitted 10 times using different initial latent positions by introducing noise to the initial configuration obtained as outlined in Section \ref{sec:3}. The MCMC chains are run for 500,000 iterations with a burn-in period of 50,000 and thinning every 2,000th sample. Model fitting took on average 10 minutes on a computer with an i7-10510U CPU and 16GB RAM. For comparison, fitting an LPM with $p=5$ with the \texttt{latentnet} \citep{latentnetpackage} R package with default settings but 500,000 iterations took 7 minutes. However, the LPM setting requires fitting multiple models, thus incurring additional computational cost.
%via a multivariate normal distribution of 1/100th of the empirical latent positions variance. 
The Gelman-Rubin convergence criterion \citep{Andrew_Gelman_John} was satisfied with $1.0 \leq \hat{R} < 1.1$ for $\alpha$ and the shrinkage parameters. Visual inspection of trace plots also suggests convergence; an example is provided in the supplementary material, Appendix \ref{app:application}.

\begin{figure}[htb]
     \centering
     \begin{subfigure}[b]{0.28\linewidth}
         \centering
         \includegraphics[width=\linewidth]{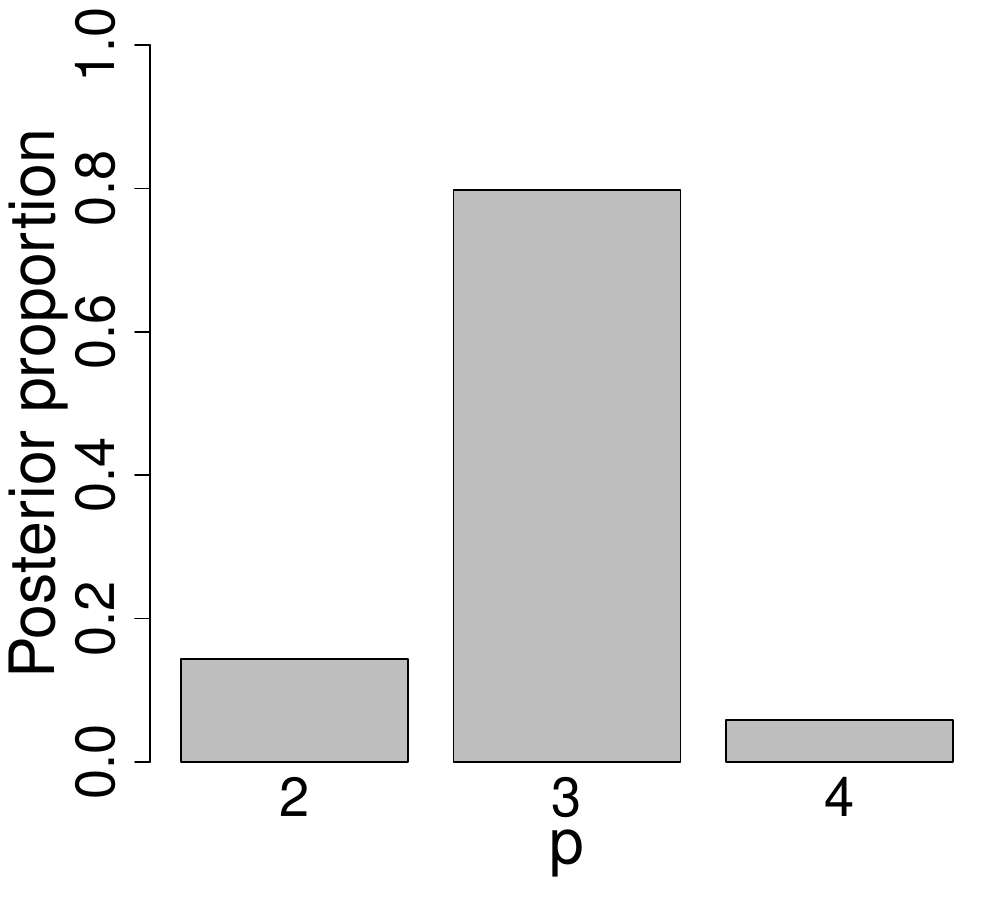}
         \caption{}
         \label{fig:cat_dim_bar}
     \end{subfigure}
     \hfill
     \begin{subfigure}[b]{0.71\linewidth}
         \centering
         \includegraphics[width=\linewidth]{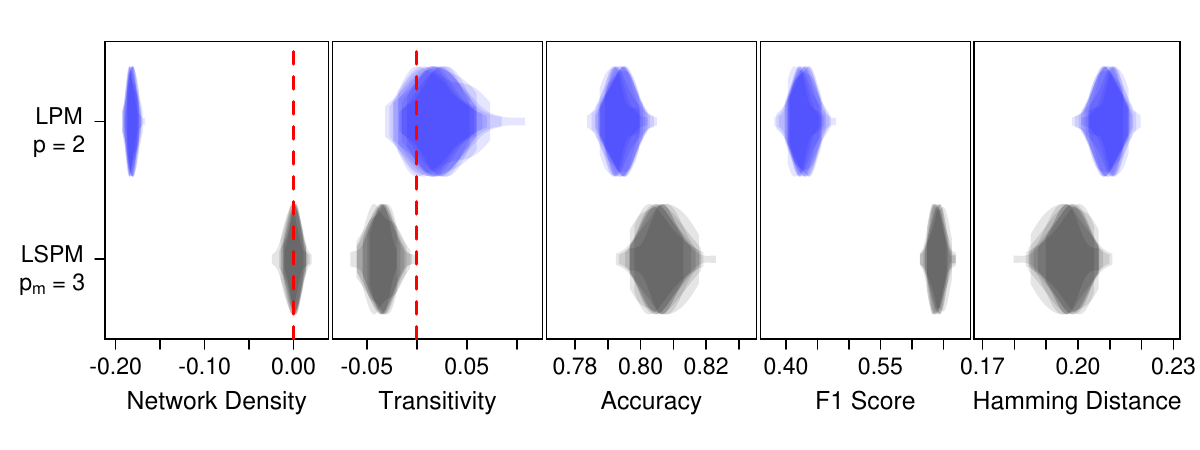}
         \caption{}
         \label{fig:cat_pred}
     \end{subfigure}
        \caption{Cat brain connectivity data: (a) the posterior distribution of $p$ and (b) posterior predictive checks for the LPM with $p=2$ and the logistic LSPM with $p_m = 3$.}
        \label{fig:cat}
\end{figure}

Figure \ref{fig:cat_dim_bar} suggests that $p_m = 3$ effective dimensions are required for these data, with low associated uncertainty. Upon fitting multiple LPMs with $p = \{1, \ldots, 5\}$, the BIC suggests 2 dimensions are optimal, with 3 dimensions being next best. Posterior predictive check metrics from 30 posterior predictive replicate networks for each of the 10 logistic LSPM and LPM MCMC chains are shown in Figure \ref{fig:cat_pred}. While the logistic LSPM tends to underestimate network transitivity, it performs better than the LPM across the majority of metrics, with the estimated density close to the network data density, higher accuracy and $F_1$ score, and lower Hamming distance.

\begin{figure}[htb]
\includegraphics[width=\linewidth]{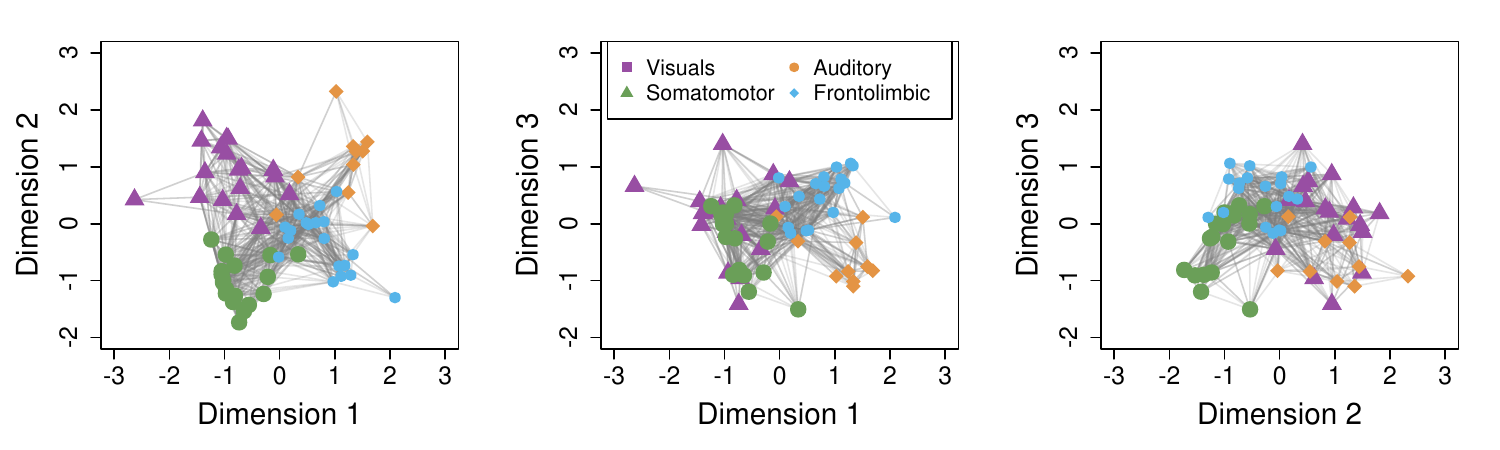} 
\caption{Posterior mean latent positions under the logistic LSPM with $p_m = 3$ for the cat brain connectivity network.}
\label{fig:cat_lspm}
\end{figure}

Figure \ref{fig:cat_lspm} shows posterior mean latent positions under the logistic LSPM with $p_m = 3$. 
%the latent positions from LSPM have lower variance across dimensions than LPM, but 
The inferred latent positions under both the LPM and the LSPM are very similar with a mean Procrustes correlation of 0.97 (standard deviation 0.01) across the 10 chains. Nodes from the same cortical region also lie close to each other in the latent space under both the LPM and the logistic LSPM.

\subsection{Worm nervous system binary network data}
This binary directed network contains $n=272$ nodes of neurons from the nervous system of the \emph{Caenorhabditis elegans} adult male worm, with each of the 4451 edges representing the presence of either a chemical or electrical interaction between nodes. The data were reconstructed from serial electron micrograph sections by \cite{Jarrell_Wang_Bloniarz_Brittin_Xu_Thomson_Albertson_Hall_Emmons_2012}. Three unconnected nodes are removed prior to analysis. The network is sparse with a density of 6.09\%.
%and is it expected to be harder to infer the effective dimension as indicated from Subsection \ref{ssec:density}. 
To fit the logistic LSPM to these data, each of 10 MCMC chains were run for 1,000,000 iterations with a burn-in period of 100,000 iterations and every 3,000th thinned.

\begin{figure}[htb]
     \centering
          \begin{subfigure}[b]{0.33\linewidth}
         \centering
         \includegraphics[width=\linewidth]{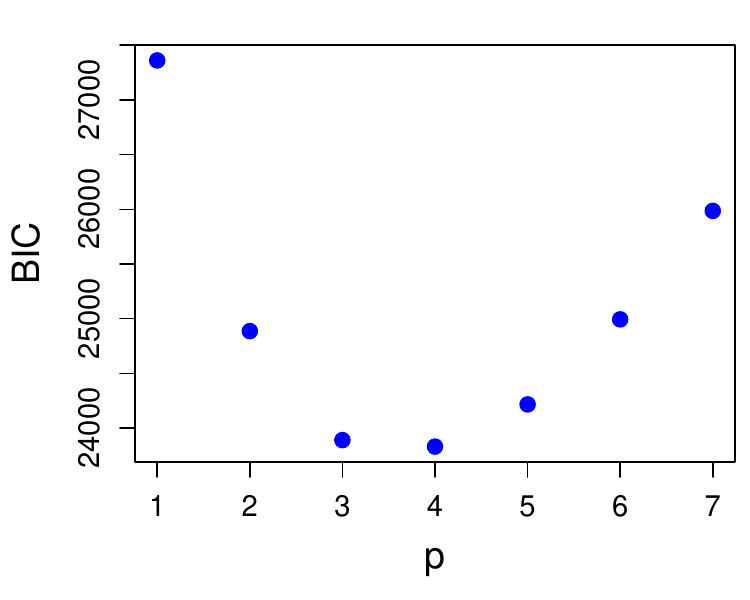}
         \caption{}
         \label{fig:worm_bic}
     \end{subfigure}
     \hfill
     \begin{subfigure}[b]{0.33\linewidth}
         \centering
         \includegraphics[width=\linewidth]{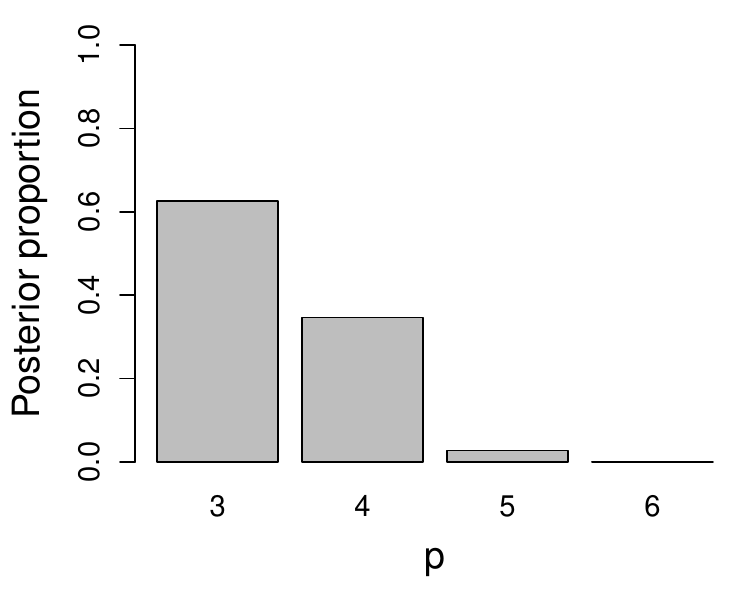}
         \caption{}
         \label{fig:worm_dim_bar}
     \end{subfigure}
               \begin{subfigure}[b]{0.31\linewidth}
         \centering
         \includegraphics[width=\linewidth]{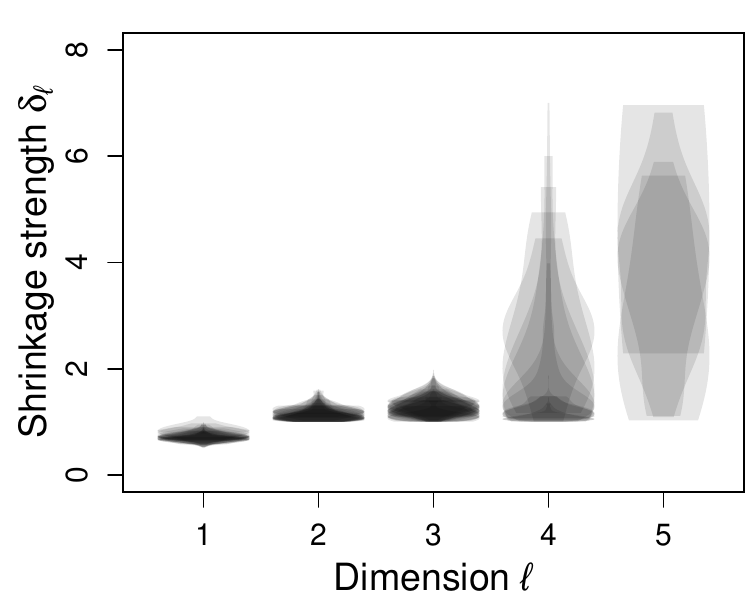}
         \caption{}
         \label{fig:worm_pmd}
     \end{subfigure}
        \caption{For the worm nervous system network: (a) the BIC for the LPM for $p = \{1, \ldots, 7\}$ and under the logistic LSPM (b) the posterior distribution of $p$ and (c) the posterior distribution of shrinkage strength parameters across the 10 MCMC chains.}
        \label{fig:worm}
\end{figure}

From a range of LPMs with $p= \{1, \ldots, 7\}$, the BIC suggests $p = 4$ as optimal, closely followed by $p = 3$, as shown in Figure \ref{fig:worm_bic}. Under the logistic LSPM, Figure \ref{fig:worm_dim_bar} indicates that the posterior modal dimension is $p_m = 3$, with 4 dimensions the next most probable, \emph{a posteriori}; in addition Figure \ref{fig:worm_pmd} illustrates that the shrinkage strength parameter $\delta_4$ has large uncertainty. The LSPM approach indicates that while 3 dimensions are required to adequately describe the network, there is high, quantifiable posterior uncertainty as to whether or not an additional dimension is needed.

\begin{figure}[htb]
\includegraphics[width=\linewidth]{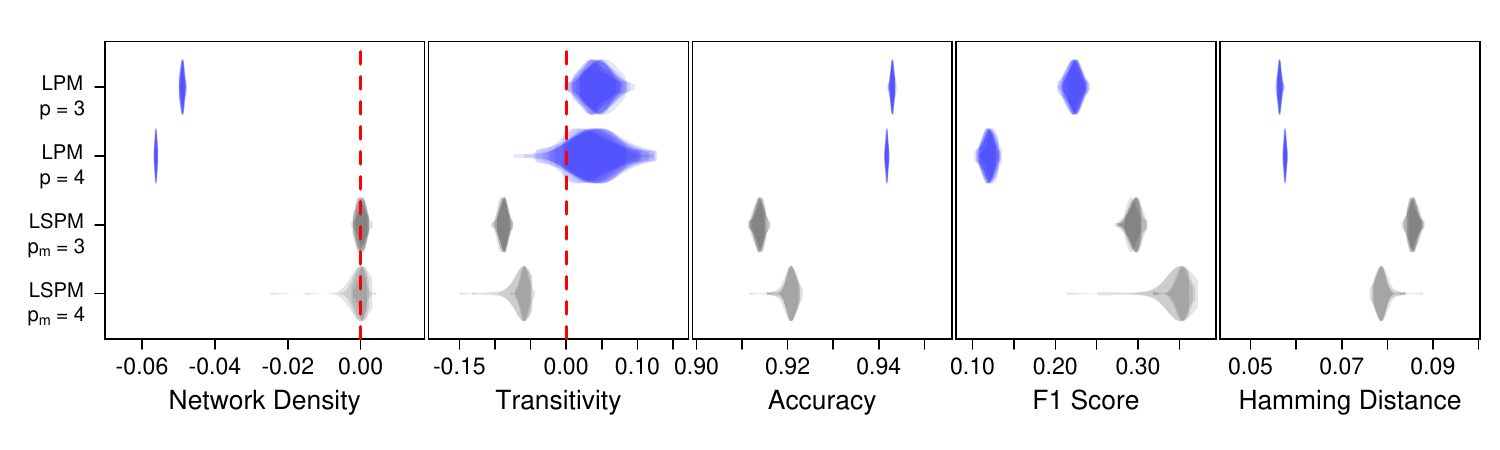} 
\caption{Posterior predictive checks on the worm nervous system network.}
\label{fig:worm_pred}
\end{figure}

In terms of posterior predictive checks, Figure \ref{fig:worm_pred} shows that the LSPM with both 3 and 4 active dimensions fit better than the LPM with 3 and 4 dimensions in terms of network density and $F_1$ score, but worse in terms of transitivity, accuracy, and Hamming distance. The LSPM with $p_m = 3$ fits only slightly worse than its 4-dimensional counterpart, and so may be preferred as a lower dimensional representation. The Procrustes correlation between locations under the LSPM with $p_m=\{3, 4\}$ and the LPM with $p=\{3, 4\}$ have similar median values of more than 0.88 (standard deviation 0.05).

\subsection{Phone calls count network data}
Mobile phone call logs are available from the ``Friends and Family" data collected by the MIT Human Dynamics Lab \citep{Aharony_Pan_Ip_Khayal_Pentland_2011}. Here, the set of call logs from the November 2010 period is considered; the associated network contains the number of (directed) calls within a young community of $n=120$ people. Modelling the number of calls rather than reducing the data to a binary network representing whether or not a call was made or received allows for deeper insight on the exchange of phone calls within the network. The Poisson LSPM is therefore fitted to this count network data. Fourteen disconnected nodes are removed prior to analysis. The data are overdispersed, with a mean count of 0.51 and a variance of 33.29. 

\begin{figure}[htb]
     \centering
     \begin{subfigure}[b]{0.495\linewidth}
         \centering
         \includegraphics[width=\linewidth]{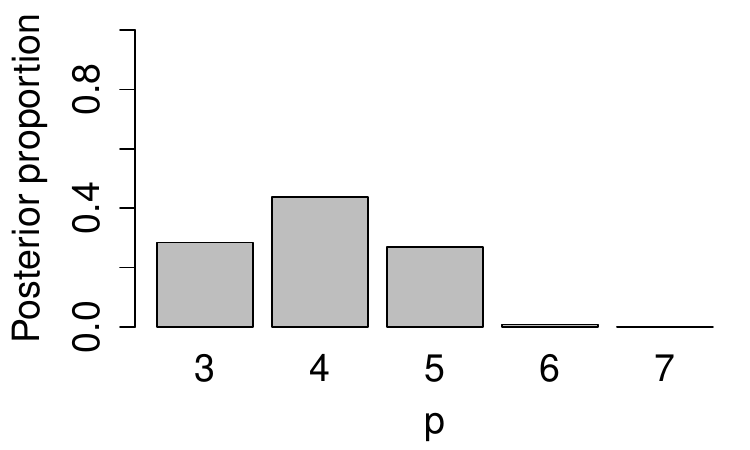}
         \caption{}
         \label{fig:phones_dim_bar}
     \end{subfigure}
     \hfill
     \begin{subfigure}[b]{0.495\linewidth}
         \centering
\includegraphics[width=\linewidth]{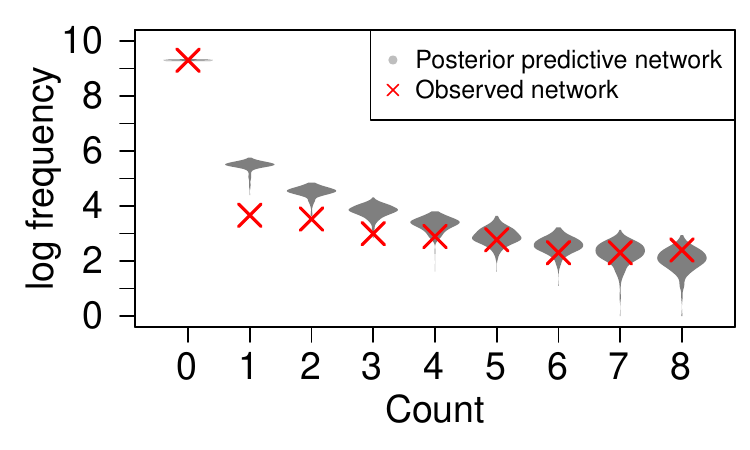}
         \caption{}
         \label{fig:phones_logcount}
     \end{subfigure}
        \caption{For the phone calls count network data: (a) the posterior distribution of $p$ and (b) the log frequency of the observed and posterior predictive call counts for the first 8 counts in the network.}
        \label{fig:phonenov}
\end{figure}

A total of 10 MCMC chains were run, each from different starting configurations and with 500,000 iterations, a burn-in of 100,000 iterations and  every 2,000th thinned. Figure \ref{fig:phones_dim_bar} shows that under a Poisson LSPM the posterior explored models with 3 to 5 active dimensions, with $p_m = 4$. Fitting the Poisson LPM with $p = 1, \ldots, 5$ suggested 1 dimension as optimal via the BIC. Comparing the posterior mean positions under the $p=1$ LPM and the first dimension of the $p_m=4$ Poisson LSPM gave a Procrustes correlation of 0.7, indicating some discrepancy. In terms of posterior predictive checks, Figure \ref{fig:phones_logcount} shows that the Poisson LSPM posterior predictive networks tend to be more overdispersed than the observed network; considering a model that accounts for such overdispersion is likely to give improved model fit.

\section{Discussion}
\label{sec:6}
The proposed latent shrinkage position model is a nonparametric Bayesian approach to model network data that focuses on the issue of inferring the dimension of the latent space. The LSPM extends the latent position model by employing a multiplicative truncated gamma process prior to allow an infinite number of dimensions in the latent space. In practice, the LSPM intrinsically infers the number of effective dimensions required to describe the network. The LSPM therefore circumvents the more computationally intensive approach where model selection criteria are used to choose the dimension after fitting a range of models with different dimensions $p$. Importantly, the LSPM retains the ease of interpretability inherent to the original LPM. The logistic and Poisson LSPM are developed for the analysis of binary and count network data respectively; extensions to alternative models for different link types are similarly feasible.

While fitting the LSPM is computationally feasible on the networks considered here, fitting the LSPM to large networks is relatively slow. There are many improvements that could be made to shorten run time, for example, by embedding the case-control concepts of \cite{Raftery_Niu_Hoff_Yeung_2012} and by implementing the variational Bayesian inference of \cite{Salter-Townshend_Murphy_2013}.

Since their introduction, both the concepts of the LPM for network data and the Bayesian nonparametric multiplicative gamma shrinkage prior have received much attention in a range of fields; as such the LSPM framework developed here should be of interest to a wide array of researchers and practitioners. The provision of the associated \href{https://gitlab.com/gwee95/lspm}{\texttt{lspm}} R package will assist in usage of the Bayesian nonparametric LSPM models.

% On a wider perspective, future research can be done to extend the LSPM to automatically infer the number of clusters within a network and the number of dimension required of the model. The interaction between the allocation of information between more dimensions or more clusters will be interesting to explore. Although automatic estimation of number of clusters has been approached before by a few authors such as \cite{dangelo2020modelbased} and \cite{Ryan_Wyse_Friel_2017}, they do not simultaneously infer both the number of clusters and the number of dimensions. For network data, such simultaneous and automatic approach has only been shown recently in the stochastic blockmodel setting by \cite{yang_priebe_park_marchette_2020} in a frequentist framework and \cite{passino_heard_2020} in a Bayesian framework. Adaptation can be made to supplement the LSPM so that such approach can work in the latent space setting.

% \begin{supplement}
% %\stitle{Supplementary material}
% The supplementary material contains the derivations of the properties of the LSPM and of the full conditional distributions required for the MCMC algorithm. Additional plots relating to the simulation studies and data applications are also included\end{supplement}

\section*{Acknowledgments}
This publication has emanated from research conducted with the financial support of Science Foundation Ireland under Grant number 18/CRT/6049. For the purpose of Open Access, the author has applied a CC BY public copyright licence to any Author Accepted Manuscript version arising from this submission. The authors are grateful to the anonymous reviewers for their suggestions which greatly contributed to improving this work.

\clearpage
\appendix
% \appendix
\section*{Supplementary Material and Appendix}
\section{Derivation of properties of the latent shrinkage position model}
\label{app:properties}

\subsection{Notation}
$n$: Number of nodes. \\
$p$: Truncation level i.e. the number of dimensions in the fitted LSPM.\\
$p^*$: The true effective dimension of the latent space.\\
$p_0$: The initial truncation level.\\
$p_m$: The posterior modal dimension. \\
$\mathbf{Y}$: $n \times n$ network adjacency matrix. \\
$y_{i,j}$: Edge between node $i$ and node $j$. \\
$\mathbf{Z}$: $n \times p$ matrix of latent positions.  \\
$z_{i\ell}$: Latent position of node $i$ in dimension $\ell$. \\
$q_{i,j}$: Probability of forming an edge between node $i$ and node $j$.  \\
$\alpha$: Global parameter that captures the overall connectivity level in the network.  \\
$\mathbf{\Omega}$: $p \times p$ precision matrix of latent positions .   \\
$\omega_{\ell}$: Precision/global shrinkage parameter for dimension $\ell$. \\
$\delta_{h}$: Shrinkage strength from dimension $h$.  \\
$\eta_{i,j}$: The linear predictor $\alpha - \Vert \mathbf{z}_i-\mathbf{z}_j \Vert^2_2 $. \\
$a_h$: Shape parameter of gamma distribution for $\delta_h$ in dimension $h$. \\
$b_h$: Rate parameter of the gamma distribution for $\delta_h$ in dimension $h$. \\
$c_h$: Left truncation point of the gamma distribution for $\delta_h$. \\
$P(\mathbf{Z} | -)  $: Probability of $\mathbf{Z}$ conditional on all other parameters. \\
$C$: Generic constant term used for the derivations of the full conditional distributions.
$\kappa_0, \kappa_1$: The dimension adaptation probability parameters. \\
$\epsilon_1$: The threshold for decreasing dimension(s) in the dimension adaptation step. \\
$\epsilon_2$: The threshold for increasing a dimension in the dimension adaptation step. \\
$\epsilon_3$: The threshold for increasing a dimension in the dimension adaptation step when the active dimension is 1.

%$s$: Current step in the MCMC chain 

%Using a truncated gamma distribution for the multiplicative gamma process prior resulted in different model properties.

\subsection{Expectation of \texorpdfstring{$\frac{1}{\delta_h}$}{Lg}}
The prior for $\delta_h$ is assumed to be
\begin{equation}
\begin{aligned}
\notag
\delta_h &\sim \text{Gam}^{\text{T}}(a_2, b_2 = 1, c_2 = 1) \\
\text{giving } f(\delta_h) &= \frac{\delta_h^{a_2-1} \exp(-b_2\delta_h)}{\int_{c_2}^{\infty} \delta_h^{a_2-1} \exp(-\delta_h) d\delta_h} = \frac{\delta_h^{a_2-1} \exp(-\delta_h)}{\Gamma(a_2, 1)} \\
\text{Thus: } \mathbb{E}\left(\frac{1}{\delta_h}\right) &= \int_{-\infty}^{\infty} \frac{1}{\delta_h} \frac{\delta_h^{a_2-1} \exp(-\delta_h)}{\Gamma(a_2, 1)} d\delta_h\\
&= \frac{1}{\Gamma(a_2, 1)} \int_{1}^{\infty} \delta_h^{(a_2-1)-1} \exp(-\delta_h) d\delta_h \\
&= \frac{\Gamma(a_2-1, 1)}{\Gamma(a_2, 1)}. 
\end{aligned}
\end{equation}

\subsection{Expectation of the pairwise squared distance in the \texorpdfstring{$\ell$}{Lg}-th dimension}
Since latent space dimensions are assumed to be independent, i.e. 
$$
\mathbf{\Omega}^{-1} =  \omega_{\ell}^{-1} \mathbf{I} =\text{diag}(\omega_1^{-1},\ldots,\omega_p^{-1}) ,
$$
then $\mathbf{Z}$ is such that $z_{i\ell} | \omega_{\ell} \sim N(0, \omega_{\ell}^{-1})$  for every dimension $\ell$ and node $i$. \\ Since $z_{i\ell} | \omega_{\ell} \sim N(0, \omega_{\ell}^{-1})$ and $ z_{j\ell} | \omega_{\ell} \sim N(0, \omega_{\ell}^{-1}) $,
\\ then $ (z_{i\ell} - z_{j\ell}) | \omega_{\ell} \sim N(0, 2\omega_{\ell}^{-1}) $. 
Hence,
\begin{equation}
\begin{aligned}
\notag
\frac{(z_{i\ell} - z_{j\ell})}{\sqrt{2\omega_{\ell}^{-1}}} | \omega_{\ell} &\sim N(0, 1)\\
\frac{(z_{i\ell} - z_{j\ell})^2}{2\omega_{\ell}^{-1}} | \omega_{\ell} &\sim \left[N(0, 1)\right]^2 = \chi^2(1) \\
(z_{i\ell} - z_{j\ell})^2 | \omega_{\ell} &\sim  2\omega_{\ell}^{-1} \chi^2_{(1)} \\
\mathbb{E}[(z_{i\ell} - z_{j\ell})^2 | \omega_{\ell}] &= \mathbb{E}[2\omega_{\ell}^{-1} \chi^2_{(1)}] \\
&= 2 \mathbb{E}(\omega_{\ell}^{-1}) (1). \\
\end{aligned}
\end{equation}
Since $\omega_{\ell} = \prod_{h=1}^{\ell} \delta_h $, then $ \mathbb{E}[(z_{i\ell} - z_{j\ell})^2 | \omega_{\ell}] = 2 \mathbb{E}\left[\frac{1}{\prod_{h=1}^{\ell} \delta_h}\right] = 2 \prod_{h=1}^{\ell} \mathbb{E}\left[\frac{1}{\delta_h}\right] $. \\
Since $ \delta_1 \sim \text{Gam}(a_1, b_1)$, then $\frac{1}{\delta_1} \sim \text{InvGam}(a_1, b_1)$, giving $ \mathbb{E}\left[\frac{1}{\delta_1}\right] = \frac{b_1}{a_1-1}$. \\
Hence, $$\mathbb{E}[(z_{i\ell} - z_{j\ell})^2 | \omega_{\ell}]  = 2 \left(\frac{b_1}{a_1-1}\right)\prod_{h=2}^{\ell} \mathbb{E}\left[\frac{1}{\delta_h}\right].$$
For $h=2, \ldots,\ell$, it is assumed that $\delta_h \sim \text{Gam}^{\text{T}}(a_2, b_2 = 1, c_2 = 1)$, 
giving $\mathbb{E}\left[\frac{1}{\delta_h}\right] = \frac{ \Gamma(a_2-1, 1)}{ \Gamma(a_2, 1)} $. \\
Thus, $$
\mathbb{E}[(z_{i\ell} - z_{j\ell})^2 | \omega_{\ell}]  = 2 \left(\frac{b_1}{a_1-1} \right) \left[\frac{ \Gamma(a_2-1, 1)}{ \Gamma(a_2, 1)} \right]^{\ell -1}.$$

\subsection{Expected sum of the pairwise squared distance across \texorpdfstring{$p$}{Lg} dimensions}
The distance contribution from $p$ dimensions is a geometric series with starting term of $2 \left( \frac{b_1}{a_1-1}\right)$ and a common ratio of $\frac{ \Gamma(a_2-1, 1)}{ \Gamma(a_2, 1)} $.
Thus, the geometric sum is:
$$
\mathbb{E}[(z_{i\ell} - z_{j\ell})^2 | \omega_{1}, \ldots, \omega_{p}]  = 2 \left(\frac{b_1}{a_1-1} \right) \left\{ \frac{1-\left[\frac{\Gamma(a_2-1, 1)}{ \Gamma(a_2, 1)} \right]^{p}} {1-\frac{\Gamma(a_2-1, 1)}{ \Gamma(a_2, 1)}  } \right\}.
$$

\newpage
\section{Derivation of full conditional distributions}
\label{app:fullconditionals}

\subsection{The logistic latent shrinkage position model}
Links between nodes $i$ and $j$ are assumed to form probabilistically from a binomial distribution:
\begin{equation}
\begin{aligned}
\notag
    y_{i,j} &\sim \text{Bin}(q_{i,j}) \\
    &= q_{i,j}^{y_{i,j}} (1-q_{i,j})^{1-y_{i,j}}. \\
\end{aligned}
\end{equation}
For the binary network setting, the probability $q_{i,j}$ follows a logistic model i.e.
$$\log \frac{q_{i,j}}{1-q_{i,j}} = \alpha - \Vert \mathbf{z}_i-\mathbf{z}_j \Vert^2_2.$$
Denoting $ \eta_{i,j} = \alpha - \Vert \mathbf{z}_i-\mathbf{z}_j \Vert^2_2 $, then $ q_{i,j} = \frac{\exp(\eta_{i,j})}{1+\exp(\eta_{i,j})}. \\$
The likelihood function of the LSPM is
\begin{equation}
\begin{aligned}
\notag
    L(\mathbf{Y}|\mathbf{Z},\alpha) &= \prod_{i \neq j} P(y_{i,j} | \mathbf{z}_i, \mathbf{z}_j, \alpha)  \\
                &= \prod_{i \neq j} 
                \left[ \frac{\exp(\eta_{i,j})}{1+\exp(\eta_{i,j})} \right]^{y_{i,j}}
                \left[1 - \frac{\exp(\eta_{i,j})}{1+\exp(\eta_{i,j})} \right]^{1 - y_{i,j}} \\
                &= \prod_{i \neq j} 
                 \frac{\exp(\eta_{i,j} y_{i,j})}{1+\exp(\eta_{i,j})}, \\
\end{aligned}
\end{equation}
and the log-likelihood function is
$$
\log  L(\mathbf{Y}|\mathbf{Z},\alpha)  = \sum_{i \neq j} \hspace{2pt} \{ \eta_{i,j} y_{i,j} - 
                    \log [ 1 + \exp(\eta_{i,j})] \}.
$$

The joint posterior distribution is
\begin{equation}
\begin{aligned}  
\notag
P(\alpha, \mathbf{Z}, \bm{\delta} | \mathbf{Y}) 
    &= \frac{P(\mathbf{Y}|\alpha, \mathbf{Z}, \bm{\delta}) P(\alpha, \mathbf{Z}, \bm{\delta})}{P(\mathbf{Y})} 
        &&   \\
    &\propto P(\mathbf{Y}|\alpha, \mathbf{Z}, \bm{\delta}) P(\alpha, \mathbf{Z}, \bm{\delta})     \\
    &\propto P(\mathbf{Y}|\alpha, \mathbf{Z}) P(\alpha, \mathbf{Z}, \bm{\delta})
        && \text{since $\mathbf{Y}$ only depends on $\alpha$ and $\mathbf{Z}$}, \\
    &\propto P(\mathbf{Y}|\alpha, \mathbf{Z}) P(\alpha | \mathbf{Z}, \bm{\delta}) P(\mathbf{Z}, \bm{\delta}) 
        && \\
    &\propto P(\mathbf{Y}|\alpha, \mathbf{Z}) P(\alpha) P(\mathbf{Z}, \bm{\delta}) 
        && \\
    &\propto P(\mathbf{Y}|\alpha, \mathbf{Z}) P(\alpha) P(\mathbf{Z} | \bm{\delta}) P(\bm{\delta}).
        &&  \\
\end{aligned}
\end{equation}

Assumptions and prior specifications are as follows:
\begin{equation}
\begin{aligned}  
\notag
Y_{i,j}|\alpha, \mathbf{z}_i, \mathbf{z}_j \sim \text{Ber}(q_{i,j}) \quad 
\alpha \sim N(\mu_{\alpha},\sigma_{\alpha}^2)  \quad 
\mathbf{z}_{i} \sim \text{MVN}(\mathbf{0}, \mathbf{\Omega}^{-1}) \\
\omega_{\ell} = \prod_{h=1}^{\ell} \delta_{h}   \quad 
\delta_1 \sim \text{Gam}(a_1, b_1=1) \quad
\delta_{h} \sim \text{Gam}^{\text{T}}(a_2, b_2=1, c_2 = 1)
\end{aligned}
\end{equation}
\begin{equation}
\begin{aligned}  
\notag
P(\alpha, \mathbf{Z}, \bm{\delta} | \mathbf{Y}) 
    &\propto 
        \left\{\prod_{i \neq j} \left[ \frac{\exp(\eta_{i,j} y_{i,j})}{1+\exp(\eta_{i,j})} \right] \right\}
    \times \left\{\frac{1}{\sqrt{2\pi\sigma_{\alpha}^2}}
        \exp\left[-\frac{1}{2\sigma_{\alpha}^2}(\alpha-\mu_{\alpha})^2\right] \right\} \\
        &\qquad  
    \times \left\{\prod_{i=1}^n \left(\frac{1}{2\pi}\right)^{\frac{p}{2}} 
        (\text{det} \mathbf{\Omega})^{\frac{1}{2}} 
        \exp\left[-\frac{1}{2} (\mathbf{z}_i - \mathbf{0})^T 
        \mathbf{\Omega} (\mathbf{z}_i - \mathbf{0})\right] \right\} \span \span \\
    &\qquad  
    \times \left\{\frac{b_1^{a_1}}{\Gamma(a_1)} (\delta_1)^{a_1-1} \exp\left[-b_1(\delta_1)\right] \right\} \\
    &\qquad  
    \times \left\{\prod_{h=2}^p 
        \frac{b_2^{a_2}}{\Gamma(a_2)} (\delta_{h})^{a_2-1} \exp\left[-b_2(\delta_{h})\right] \right\}. \span \span \\
\end{aligned}
\end{equation}
Since latent space dimensions are assumed independent, then 
$$ \mathbf{\Omega}^{-1} =  \omega_{\ell}^{-1} \mathbf{I} =\text{diag}(\omega_1^{-1},\ldots,\omega_p^{-1}) , $$
then, $\mathbf{Z}$ can be treated as $z_{i\ell} | \omega_{\ell} \sim N(0, \omega_{\ell}^{-1})$  for every dimension $\ell$. Therefore 
\begin{equation}
\begin{aligned}  
\notag
P(\alpha, \mathbf{Z}, \bm{\delta} | \mathbf{Y}) 
    &\propto 
        \left\{\prod_{i \neq j} \left[ \frac{\exp(\eta_{i,j} y_{i,j})}{1+\exp(\eta_{i,j})} \right] \right\}
    \times \left\{\frac{1}{\sqrt{2\pi\sigma_{\alpha}^2}}
        \exp\left[-\frac{1}{2\sigma_{\alpha}^2}(\alpha-\mu_{\alpha})^2\right] \right\} \\
        &\qquad 
    \times \left\{\prod_{i=1}^n \prod_{\ell=1}^p  {\sqrt{\frac{\omega_{\ell}}{2\pi}}} 
        \exp\left[-\frac{\omega_{\ell}}{2} (z_{i\ell} - 0)^2\right] \right\} \span \span \\
    &\qquad  
    \times \left\{\frac{b_1^{a_1}}{\Gamma(a_1)} (\delta_1)^{a_1-1} \exp\left[-b_1(\delta_1)\right] \right\} \\
    &\qquad  
    \times \left\{\prod_{h=2}^p 
        \frac{b_2^{a_2}}{\Gamma(a_2)} (\delta_{h})^{a_2-1} \exp\left[-b_2(\delta_{h})\right] \right\}. \span \span
\end{aligned}
\end{equation}
The full conditional distribution for $\mathbf{Z}$ is:
\begin{equation}
\begin{aligned}  
\notag
P(\mathbf{Z} | -)        
    &\propto \left\{\prod_{i \neq j} 
                 \left[ \frac{\exp(\eta_{i,j} y_{i,j})}{1+\exp(\eta_{i,j})} \right] \right\}
                 \times \prod_{i=1}^n \prod_{\ell=1}^p  {\sqrt{\frac{\omega_{\ell}}{2\pi}}} 
        \exp\left[-\frac{\omega_{\ell}}{2} (z_{i\ell} - 0)^2\right],  \\
\log P(\mathbf{Z} | -)     &= \sum_{i \neq j} \hspace{2pt} \left\{ \eta_{i,j} y_{i,j} - 
                    \log \left[ 1 + \exp(\eta_{i,j}) \right] \right\} +
                    \sum_{\ell=1}^p \left[ -\frac{n}{2} \log(2\pi\sigma_{\ell}^2) \right]
                    \\ 
                    &\qquad - \sum_{i=1}^n \sum_{\ell=1}^p  
                    \frac{\omega_{\ell}}{2} z_{i\ell}^2  + C\\
                &= \sum_{i \neq j} \hspace{2pt} \left\{ \eta_{i,j} y_{i,j} - 
                    \log \left[ 1 + \exp(\eta_{i,j}) \right] \right\} 
                    - \sum_{i=1}^n \sum_{\ell=1}^p  
                    \frac{\omega_{\ell}}{2}  z_{i\ell}^2 + C\\
                &= \sum_{i \neq j} \hspace{2pt} \left\{ \Vert \mathbf{z}_i-\mathbf{z}_j \Vert^2_2 y_{i,j} 
                - \log \left[ 1 + \exp(\alpha - \Vert \mathbf{z}_i-\mathbf{z}_j \Vert^2_2\right)] \right\} 
                - \sum_{i=1}^n \sum_{\ell=1}^p  \frac{\omega_{\ell}}{2}  z_{i\ell}^2 + C.
\end{aligned}
\end{equation}
As this is not a recognisable distribution, the Metropolis-Hasting algorithm is employed.

The full conditional distribution for $\delta_1$ is: 
\begin{equation}
\begin{aligned}  
\notag
P(\delta_1 |-)  &\propto \prod_{i=1}^n \prod_{\ell=1}^p
{\sqrt{\frac{\omega_{\ell}}{2\pi}}}  \exp\left(-\frac{\omega_{\ell}}{2}  z_{i\ell}^2\right)
                    \times \left[\frac{b_1^{a_1}}{\Gamma(a_1)} (\delta_1)^{a_1-1} \exp(-b_1\delta_1) \right] \\
                &\propto \prod_{i=1}^n \prod_{\ell=1}^p
                    {\sqrt{\frac{ \prod_{m=1}^{\ell} \delta_{m}}{2\pi}}} 
                    \exp\left(-\frac{\prod_{m=1}^{\ell} \delta_{m}}{2} z_{i\ell}^2\right)
                \times \left[\frac{b_1^{a_1}}{\Gamma(a_1)} (\delta_1)^{a_1-1} \exp(-b_1\delta_1) \right] \\
                &\propto \prod_{i=1}^n \prod_{\ell=1}^p \delta_{1}^{\frac{1}{2}}
                    \exp\left(-\frac{\prod_{m=1}^{\ell} \delta_{m}}{2} z_{i\ell}^2\right)
                \times \left[(\delta_1)^{a_1-1} 
                    \exp(-b_1\delta_1) \right] \\     
                &\propto \delta_{1}^{\frac{np}{2}}
                    \exp\left(- \frac{1}{2}\sum_{i=1}^{n}\sum_{\ell=1}^{p} \prod_{m=1}^{\ell} \delta_{m} z_{i\ell}^2 \right)
                \times \left[(\delta_1)^{a_1-1} 
                    \exp(-b_1\delta_1) \right] \\    
                &\propto \delta_{1}^{\frac{np}{2}+a_1-1}
                    \exp\left[- \left(b_1 + \frac{1}{2}\sum_{i=1}^{n}\sum_{\ell=1}^{p} \prod_{m=2}^{\ell} \delta_{m} z_{i\ell}^2\right) \delta_1 \right],  \\
                    \text{hence,}\\
                &\delta_1 \sim \text{Gam}\left(\frac{np}{2}+a_1, 
                    \quad b_1 + \frac{1}{2}\sum_{i=1}^{n}\sum_{\ell=1}^{p} \prod_{m=2}^{\ell} \delta_{m} z_{i\ell}^2  \right), \\
                & \text{let} \quad \omega_{\ell}^{(h)} = \prod_{m=1, m \neq h}^{\ell} \delta_m,  \\
                &\delta_1 \sim \text{Gam}\left(\frac{np}{2}+a_1, 
                    \quad b_1 + \frac{1}{2}\sum_{i=1}^{n}\sum_{\ell=1}^{p} \omega_{\ell}^{(1)} z_{i\ell}^2 \right). \\
\end{aligned}
\end{equation}
The full conditional distribution for $\delta_h$, where $h \geq 2$ is:
\begin{equation}
\begin{aligned}  
\notag
P(\delta_h |-)  &\propto \prod_{i=1}^n \prod_{\ell=1}^p
                        {\sqrt{\frac{\omega_{\ell}}{2\pi}}} 
                        \exp\left(-\frac{\omega_{\ell}}{2}  z_{i\ell}^2\right)
                    \times \left[\frac{b_2^{a_2}}{\Gamma(a_2)} (\delta_h)^{a_2-1} \exp(-b_2\delta_h) \right] \\
                &\propto \prod_{i=1}^n \prod_{\ell=1}^p
                    {\sqrt{\frac{\omega_{\ell}}{2\pi}}} 
                    \exp\left(-\frac{\omega_{\ell}}{2} z_{i\ell}^2\right)
                \times \left[\frac{b_2^{a_2}}{\Gamma(a_2)} (\delta_h)^{a_2-1} \exp(-b_2\delta_h) \right] \\
                &\propto \prod_{i=1}^n \prod_{\ell=1}^p
                    {\sqrt{\frac{\prod_{m=1}^{\ell} \delta_{m}}{2\pi}}} 
                    \exp\left(-\frac{\prod_{m=1}^{\ell} \delta_{m}}{2} z_{i\ell}^2\right)
                \times \left[\frac{b_2^{a_2}}{\Gamma(a_2)} (\delta_h)^{a_2-1} \exp(-b_2\delta_h) \right] \\
                &\propto \prod_{i=1}^n \prod_{\ell=1}^p 
                    \left[ \prod_{m=1}^{\ell} \delta_{m}^{\frac{1}{2}} \right]
                    \exp\left(-\frac{\prod_{m=1}^{\ell} \delta_{m}}{2} z_{i\ell}^2\right)
                \times \left[(\delta_h)^{a_2-1} 
                    \exp(-b_2\delta_h) \right] \\    
                &\propto \delta_{h}^{\frac{n(p-h+1)}{2}}
                    \exp\left(- \frac{1}{2}\sum_{i=1}^{n}\sum_{\ell=1}^{p} \prod_{m=1}^{\ell} \delta_{m} z_{i\ell}^2\right)
                \times \left[(\delta_h)^{a_2-1} 
                    \exp(-b_2\delta_h) \right] \\    
                &\propto \delta_{h}^{\frac{n(p-h+1)}{2}} (\delta_h)^{a_2-1} 
                    \exp\left[- \frac{1}{2}\left(\sum_{i=1}^{n}\sum_{\ell=1}^{p} \prod_{m=1}^{\ell} \delta_{m} z_{i\ell}^2 \right) - b_2\delta_h \right],  \\  
\end{aligned}
\end{equation}

\begin{equation}
\begin{aligned}  
\notag
                \text{let} \quad \omega_{\ell}^{(h)} = \prod_{m=1, m \neq h}^{\ell} \delta_m,  \\
P(\delta_h |-)  \propto \delta_{h}^{\frac{n(p-h+1)}{2}+a_2-1}
                &    \exp\left[- \left(b_2+\frac{1}{2}\sum_{i=1}^{n}\sum_{\ell=h}^{p} \omega_{\ell}^{(h)} z_{i\ell}^2 \right)\delta_h\right], \\ 
\text{since $\delta_h$ is bounded between $[1, \infty)$,} 
\end{aligned}
\end{equation}

\begin{equation}
\begin{aligned}  
\notag
                \delta_h \sim \text{Gam}^{\text{T}}\left(\frac{n(p-h+1)}{2}+a_2, 
                    \quad b_2 + \frac{1}{2}\sum_{i=1}^{n}\sum_{\ell=h}^{p} \omega_{\ell}^{(h)} z_{i\ell}^2,
                    \quad 1 \right).
\end{aligned}
\end{equation}

The full conditional distribution for $\alpha$ is:
\begin{equation}
\begin{aligned}  
\notag
P(\alpha | -)   &\propto \left\{\prod_{i \neq j} 
                 \left[ \frac{\exp(\eta_{i,j} y_{i,j})}{1+\exp(\eta_{i,j})} \right] \right\} 
                 \times \frac{1}{\sqrt{2\pi\sigma_{\alpha}^2}} \exp\left[-\frac{1}{2}\frac{(\alpha-\mu_{\alpha})^2}{\sigma_{\alpha}^2}\right] \\
                 &\propto \left\{\prod_{i \neq j} 
                 \left[ \frac{\exp(\eta_{i,j} y_{i,j})}{1+\exp(\eta_{i,j})} \right] \right\}
                 \times \exp\left[-\frac{1}{2}\frac{(\alpha-\mu_{\alpha})^2}{\sigma_{\alpha}^2}\right], \\
\log P(\alpha | -)&= \sum_{i \neq j} \hspace{2pt} \left\{ \eta_{i,j} y_{i,j} - 
                    \log \left[ 1 + \exp(\eta_{i,j})\right] \right\}
                    - \frac{1}{2}\frac{(\alpha-\mu_{\alpha})^2}{\sigma_{\alpha}^2} + C\\
                &= \sum_{i \neq j} \hspace{2pt} \left\{ \eta_{i,j} y_{i,j} - 
                    \log \left[ 1 + \exp(\eta_{i,j})\right] \right\} - (\alpha-\mu_{\alpha})^2 + C\\
                &= \sum_{i \neq j} \hspace{2pt} \left\{ (\alpha - \Vert \mathbf{z}_i-\mathbf{z}_j \Vert^2_2) y_{i,j} - \log \left[  1 + \exp(\alpha - \Vert \mathbf{z}_i-\mathbf{z}_j \Vert^2_2) \right] \right\} \\
                &\qquad - (\alpha-\mu_{\alpha})^2  + C,\\
\log P(\alpha | -)&= \sum_{i \neq j} \hspace{2pt} \left\{ \alpha y_{i,j} - 
                    \log \left[ 1 + \exp(\alpha - \Vert \mathbf{z}_i-\mathbf{z}_j \Vert^2_2)\right] \right\} - (\alpha-\mu_{\alpha})^2  + C.
\end{aligned}
\end{equation}
\noindent A Metropolis-Hasting algorithm is used to sample from this full conditional distribution since it is not in a recognisable form.

\subsection{Deriving an informed proposal distribution for \texorpdfstring{$\alpha$}{Lg} for the logistic latent shrinkage position model}
\label{app:logit_inform_alpha}
The log-likelihood term is given by
\begin{equation}
\begin{aligned}  \label{eq:logbinary}
\log  L(\mathbf{Y}|\mathbf{Z},\alpha)   &= \sum_{i \neq j} \hspace{2pt} \left\{ \eta_{i,j} y_{i,j} - 
                    \log [ 1 + \exp(\eta_{i,j})] \right\} \\
                &= \sum_{i \neq j} \hspace{2pt} [ \alpha y_{i,j} ] - \sum_{i \neq j} \hspace{2pt} [\Vert \mathbf{z}_i-\mathbf{z}_j \Vert^2_2y_{i,j} ] - \sum_{i \neq j} \hspace{2pt} \log [ 1 + \exp(\eta_{i,j})].
\end{aligned}
\end{equation}
The 3rd term in equation \eqref{eq:logbinary} can be approximated using a quadratic Taylor expansion, i.e. let $ g(\alpha) = - \sum_{i \neq j} \hspace{2pt}  \log ( 1 + \exp(\eta_{i,j}))$, then the quadratic Taylor expansion around $\alpha$ is
\begin{equation}
\begin{aligned}  
\notag
g(\alpha) &\approx g(\alpha^{(s)}) + (\alpha - \alpha^{(s)}) g'(\alpha^{(s)}) + 0.5(\alpha - \alpha^{(s)})^2 g''(\alpha^{(s)}), \\
\text{with } g'(\alpha^{(s)})  &= - \sum_{i \neq j} \hspace{2pt} \frac{\exp(\alpha^{(s)} - \Vert \mathbf{z}_i-\mathbf{z}_j \Vert^2_2)}{1+\exp(\alpha^{(s)} - \Vert \mathbf{z}_i-\mathbf{z}_j \Vert^2_2)}, \\
\text{and } g''(\alpha^{(s)})  &= - \sum_{i \neq j} \hspace{2pt}  \frac{\exp(\alpha^{(s)} - \Vert \mathbf{z}_i-\mathbf{z}_j \Vert^2_2)}{\left[1+\exp(\alpha^{(s)} - \Vert \mathbf{z}_i-\mathbf{z}_j \Vert^2_2)\right]^2}. \\
\text{Thus, } g(\alpha) &\approx - \sum_{i \neq j} \hspace{2pt} [ \log ( 1 + \exp(\eta_{i,j})) ] - (\alpha - \alpha^{(s)}) \sum_{i \neq j} \hspace{2pt} \frac{\exp(\alpha^{(s)} - \Vert \mathbf{z}_i-\mathbf{z}_j \Vert^2_2)}{1+\exp(\alpha^{(s)} - \Vert \mathbf{z}_i-\mathbf{z}_j \Vert^2_2)}  \\
& \qquad - 0.5(\alpha - \alpha^{(s)})^2 \sum_{i \neq j} \hspace{2pt}  \frac{\exp(\alpha^{(s)} - \Vert \mathbf{z}_i-\mathbf{z}_j \Vert^2_2)}{\left[1+\exp(\alpha^{(s)} - \Vert \mathbf{z}_i-\mathbf{z}_j \Vert^2_2)\right]^2}.
\end{aligned}
\end{equation}
Here, $\alpha^{(s)}$ is the current value of $\alpha$ in the MCMC chain. Removing terms that are constant with respect to $\alpha$,
\begin{equation}
\begin{aligned}  
\notag
\log  L(\mathbf{Y}|\mathbf{Z},\alpha) &\approx \sum_{i \neq j} \hspace{2pt} \alpha y_{i,j} - (\alpha - \alpha^{(s)}) \sum_{i \neq j} \hspace{2pt} \frac{\exp(\alpha^{(s)} - \Vert \mathbf{z}_i-\mathbf{z}_j \Vert^2_2)}{1+\exp(\alpha^{(s)} - \Vert \mathbf{z}_i-\mathbf{z}_j \Vert^2_2)}  \\
& - 0.5(\alpha - \alpha^{(s)})^2 \sum_{i \neq j} \hspace{2pt}  \frac{\exp(\alpha^{(s)} - \Vert \mathbf{z}_i-\mathbf{z}_j \Vert^2_2)}{\left[1+\exp(\alpha^{(s)} - \Vert \mathbf{z}_i-\mathbf{z}_j \Vert^2_2)\right]^2}. \\
\end{aligned}
\end{equation}
Introducing the constant (with respect to $\alpha$) term,  $-\alpha^{(s)}y_{i,j}$:
\begin{equation}
\begin{aligned}  
\notag
\log  L(\mathbf{Y}|\mathbf{Z},\alpha)    &\approx \sum_{i \neq j} \hspace{2pt} (\alpha - \alpha^{(s)}) y_{i,j} - (\alpha - \alpha^{(s)}) \sum_{i \neq j} \hspace{2pt} \frac{\exp(\alpha^{(s)} - \Vert \mathbf{z}_i-\mathbf{z}_j \Vert^2_2)}{1+\exp(\alpha^{(s)} - \Vert \mathbf{z}_i-\mathbf{z}_j \Vert^2_2)}  \\
    &\hspace{10pt} -0.5(\alpha - \alpha^{(s)})^2 \sum_{i \neq j} \hspace{2pt}  \frac{\exp(\alpha^{(s)} - \Vert \mathbf{z}_i-\mathbf{z}_j \Vert^2_2)}{\left[1+\exp(\alpha^{(s)} - \Vert \mathbf{z}_i-\mathbf{z}_j \Vert^2_2)\right]^2}, \\
\log  L(\mathbf{Y}|\mathbf{Z},\alpha) &\approx  -\frac{1}{2}\left[ \sum_{i \neq j} \hspace{2pt} \frac{\exp(\alpha^{(s)} - \Vert \mathbf{z}_i-\mathbf{z}_j \Vert^2_2)}{\left[1+\exp(\alpha^{(s)} - \Vert \mathbf{z}_i-\mathbf{z}_j \Vert^2_2)\right]^2} \right] \left\{ (\alpha - \alpha^{(s)})^2 \right. \\
&\quad \left. - 2 \frac{(\alpha - \alpha^{(s)})}{\sum_{i \neq j} \hspace{2pt} \frac{\exp(\alpha^{(s)} - \Vert \mathbf{z}_i-\mathbf{z}_j \Vert^2_2)}{\left[1+\exp(\alpha^{(s)} - \Vert \mathbf{z}_i-\mathbf{z}_j \Vert^2_2)\right]^2}} \left(\sum_{i \neq j} y_{i,j} \right. \right. \\
&\quad \left. \left. - \sum_{i \neq j} \hspace{2pt} \frac{\exp(\alpha^{(s)} - \Vert \mathbf{z}_i-\mathbf{z}_j \Vert^2_2)}{1+\exp(\alpha^{(s)} - \Vert \mathbf{z}_i-\mathbf{z}_j \Vert^2_2)}\right)   \right\}. \\
\end{aligned}
\end{equation}
Completing the square gives: 

\begin{equation}
\begin{aligned}  
\notag
\log  L(\mathbf{Y}|\mathbf{Z},\alpha)&\approx -\frac{1}{2} \left[\sum_{i \neq j} \hspace{2pt} \frac{\exp(\alpha^{(s)} - \Vert \mathbf{z}_i-\mathbf{z}_j \Vert^2_2)}{\left[1+\exp(\alpha^{(s)} - \Vert \mathbf{z}_i-\mathbf{z}_j \Vert^2_2)\right]^2} \right] \left\{ \left[(\alpha - \alpha^{(s)}) - \right. \right. \\
&\quad \left. \left. \frac{(\alpha - \alpha^{(s)})}{\sum_{i \neq j} \hspace{2pt} \frac{\exp(\alpha^{(s)} - \Vert \mathbf{z}_i-\mathbf{z}_j \Vert^2_2)}{\left[1+\exp(\alpha^{(s)} - \Vert \mathbf{z}_i-\mathbf{z}_j \Vert^2_2)\right]^2}}\left(\sum_{i \neq j}y_{i,j} \right. \right. \right. \\
&\quad \left. \left. \left. - \sum_{i \neq j} \hspace{2pt} \frac{\exp(\alpha^{(s)} - \Vert \mathbf{z}_i-\mathbf{z}_j \Vert^2_2)}{1+\exp(\alpha^{(s)} - \Vert \mathbf{z}_i-\mathbf{z}_j \Vert^2_2)}\right) \right]^2 \right\}, \\
\end{aligned}
\end{equation}
which is in the form of a Gaussian distribution with parameters
\begin{equation}
\begin{aligned}  
\notag
\mbox{variance} &= \left\{\sum_{i \neq j} \hspace{2pt} \frac{\exp(\alpha^{(s)} - \Vert \mathbf{z}_i-\mathbf{z}_j \Vert^2_2)}{\left[1+\exp(\alpha^{(s)} - \Vert \mathbf{z}_i-\mathbf{z}_j \Vert^2_2)\right]^2}\right\}^{-1}, \\
\mbox{mean} &= \mbox{variance} \times \left[\sum_{i \neq j} y_{i,j} - \sum_{i \neq j} \hspace{2pt} \frac{\exp(\alpha^{(s)} - \Vert \mathbf{z}_i-\mathbf{z}_j \Vert^2_2)}{1+\exp(\alpha^{(s)} - \Vert \mathbf{z}_i-\mathbf{z}_j \Vert^2_2)} \right].
\end{aligned}
\end{equation}
The prior on $\alpha$ is $N(\mu_{\alpha}, \sigma^2_{\alpha})$, which makes $(\alpha - \alpha^{(s)}) \sim N(\mu_{\alpha}- \alpha^{(s)},\sigma^2_{\alpha})$, (since $\alpha^{(s)}$ is a constant, it only affects the mean). Hence, putting the approximated log likelihood function and the log of the prior distribution on $(\alpha - \alpha^{(s)}) $, this results in a sum of (the log of) two normal distributions, which gives the distribution of $(\alpha - \alpha^{(s)}) $ to be $N(\mu_{\alpha, B}, \Bar{\sigma}^2_{\alpha, B})$, where

\begin{equation}
\begin{aligned}  
\notag
\Bar{\sigma}^2_{\alpha, B} = \left\{\sum_{i \neq j} \hspace{2pt} \frac{\exp(\alpha^{(s)} - \Vert \mathbf{z}_i-\mathbf{z}_j \Vert^2_2)}{\left[1+\exp(\alpha^{(s)} - \Vert \mathbf{z}_i-\mathbf{z}_j \Vert^2_2)\right]^2} + \frac{1}{\sigma^2_{\alpha}}\right\}^{-1},
\end{aligned}
\end{equation}
and mean
\begin{equation}
\begin{aligned}  
\notag
\mu_{\alpha, B} = \Bar{\sigma}^2_{\alpha, B}\times \left[\sum_{i \neq j} y_{i,j} - \sum_{i \neq j} \hspace{2pt} \frac{\exp(\alpha^{(s)} - \Vert \mathbf{z}_i-\mathbf{z}_j \Vert^2_2)}{1+\exp(\alpha^{(s)} - \Vert \mathbf{z}_i-\mathbf{z}_j \Vert^2_2)}  + \frac{1}{\sigma^2_{\alpha}}(\mu_{\alpha}-\alpha^{(s)}) \right].
\end{aligned}
\end{equation}
Thus, here the full conditional distribution of $\alpha$ is approximated by a normal distribution with variance $\Bar{\sigma^2}_{\alpha, B}$ and mean

\begin{equation}
\begin{aligned}  
\notag
\Bar{\mu}_{\alpha, B} = \alpha^{(s)} + \Bar{\sigma}^2_{\alpha, B}\times \left[ \sum_{i \neq j}y_{i,j} - \sum_{i \neq j} \hspace{2pt} \frac{\exp(\alpha^{(s)} - \Vert \mathbf{z}_i-\mathbf{z}_j \Vert^2_2)}{1+\exp(\alpha^{(s)} - \Vert \mathbf{z}_i-\mathbf{z}_j \Vert^2_2)}  + \frac{1}{\sigma^2_{\alpha}}(\mu_{\alpha}-\alpha^{(s)}) \right].
\end{aligned}
\end{equation}
This distribution is used as the proposal distribution in the Metropolis-Hastings algorithm.

\subsection{The Poisson latent shrinkage position model}
A link between nodes $i$ and $j$ is formed probabilistically from a Poisson distribution:
\begin{equation}
\begin{aligned}
\notag
    y_{i,j} &\sim \text{Pois}(\lambda_{i,j}) \\
    &= \frac{\lambda_{i,j}^{y_{i,j}} {\exp(-\lambda_{i,j})}}{y_{i,j}!}. \\
\end{aligned}
\end{equation}
A natural logarithm is used as the link function: 
$$\log (\lambda_{i,j}) = \alpha - \Vert \mathbf{z}_i-\mathbf{z}_j \Vert^2_2.$$

Denoting $ \eta_{i,j} = \alpha - \Vert \mathbf{z}_i-\mathbf{z}_j \Vert^2_2 $, then $ \lambda_{i,j} = \exp(\alpha - \Vert \mathbf{z}_i-\mathbf{z}_j \Vert^2_2) = \exp(\eta_{i,j}) $. \\
The likelihood function of the LSPM is

\begin{equation}
\begin{aligned}
\notag
    L(\mathbf{Y}|\mathbf{Z},\alpha) &= \prod_{i \neq j} P(y_{i,j} | \mathbf{z}_i, \mathbf{z}_j, \alpha)  \\
                &= \prod_{i \neq j} 
                \frac{\exp(\eta_{i,j} y_{i,j})\exp\left[-\exp(\eta_{i,j})\right]}{{y_{i,j}!}},  \\
\end{aligned}
\end{equation}
giving log-likelihood
$$
\log  L(\mathbf{Y}|\mathbf{Z},\alpha)  = \sum_{i \neq j} \hspace{2pt} [ \eta_{i,j} y_{i,j} - 
                     \exp(\eta_{i,j}) - \log(y_{i,j}!) ].
                     $$

The joint posterior distribution is the same as the binary case with the exception that $Y_{i,j}|\alpha, \mathbf{Z}_i \sim Pois(\lambda_{i,j})$ which gives:

\begin{equation}
\begin{aligned}  
\notag
P(\alpha, \mathbf{Z}, \bm{\delta} | \mathbf{Y}) 
    &\propto 
        \left\{\prod_{i \neq j} 
                \frac{\exp(\eta_{i,j} y_{i,j})\exp\left[-\exp(\eta_{i,j})\right]}{{y_{i,j}!}} \right\}
   \\
    &\qquad   \times \left\{\frac{1}{\sqrt{2\pi\sigma_{\alpha}^2}}
        \exp\left[-\frac{1}{2\sigma_{\alpha}^2}(\alpha-\mu_{\alpha})^2\right] \right\} \\
        &\qquad \times \left\{\prod_{i=1}^n \prod_{\ell=1}^p  {\sqrt{\frac{\omega_{\ell}}{2\pi}}} 
        \exp\left[-\frac{\omega_{\ell}}{2}  (z_{i\ell} - 0)^2\right] \right\} \span \span \\
        &\qquad  
    \times \left[\frac{b_1^{a_1}}{\Gamma(a_1)} (\delta_1)^{a_1-1} \exp(-b_1\delta_1) \right]
    \\
    &\qquad  \times \left\{\prod_{h=2}^p 
        \frac{b_2^{a_2}}{\Gamma(a_2)} (\delta_{h}-1)^{a_2-1} \exp\left[-b_2(\delta_{h}-1)\right] \right\}. \span \span \\
\end{aligned}
\end{equation}

The full conditional distribution for $\mathbf{Z}$ is:
\begin{equation}
\begin{aligned}  
\notag
P(Z | -)        
                &\propto \left\{\prod_{i \neq j} 
                \frac{\exp(\eta_{i,j} y_{i,j})\exp\left[-\exp(\eta_{i,j})\right]}{{y_{i,j}!}} \right\}
                  \\
                 &\quad \times \prod_{i=1}^n \left(\frac{1}{2\pi}\right)^{\frac{p}{2}}
                 (\text{det} \mathbf{\Sigma})^{-\frac{1}{2}} 
                 \exp\left[-\frac{1}{2} (\mathbf{Z}_i - \mathbf{0})^T \mathbf{\Sigma}^{-1} (\mathbf{Z}_i - \mathbf{0})\right] \\
    &\propto \left\{\prod_{i \neq j} 
                \frac{\exp(\eta_{i,j} y_{i,j})\exp[-\exp(\eta_{i,j})]}{{y_{i,j}!}} \right\} \\
                & \quad \times \prod_{i=1}^n 
                 \prod_{\ell=1}^p  {\sqrt{\frac{\omega_{\ell}}{2\pi}}}  \exp\left[-\frac{\omega_{\ell}}{2}  (z_{i\ell}-0)^2\right], \\
\log P(Z | -)     &= \sum_{i \neq j} \hspace{2pt} [ \eta_{i,j} y_{i,j} - 
                    \exp(\eta_{i,j}) -\log(y_{i,j}!) ] + C\\
                    &\quad +
                    \sum_{\ell=1}^p \left[ -\frac{n}{2} \log(2\pi\sigma_{\ell}^2) \right]
                    - \sum_{i=1}^n \sum_{\ell=1}^p  
                    \frac{\omega_{\ell}}{2}  z_{i\ell}^2  + C\\
                &= \sum_{i \neq j} \hspace{2pt} [ \eta_{i,j} y_{i,j} - 
                    \exp(\eta_{i,j}) -\log(y_{i,j}!)] 
                    - \sum_{i=1}^n \sum_{\ell=1}^p  
                    \frac{\omega_{\ell}}{2}  z_{i\ell}^2 + C \\
                &= \sum_{i \neq j} \hspace{2pt} \left\{ (\alpha - \Vert \mathbf{z}_i-\mathbf{z}_j \Vert^2_2) y_{i,j} 
                - \log [ 1 + \exp(\alpha - \Vert \mathbf{z}_i-\mathbf{z}_j \Vert^2_2)] \right\} \\
                &\qquad
                - \sum_{i=1}^n \sum_{\ell=1}^p  \frac{\omega_{\ell}}{2}  z_{i\ell}^2 + C\\
                &= \sum_{i \neq j} \hspace{2pt} \left\{\Vert \mathbf{z}_i-\mathbf{z}_j \Vert^2_2 y_{i,j} 
                - \log [ 1 + \exp(\alpha - \Vert \mathbf{z}_i-\mathbf{z}_j \Vert^2_2)] \right\} 
                - \sum_{i=1}^n \sum_{\ell=1}^p  \frac{\omega_{\ell}}{2}  z_{i\ell}^2 + C.
\end{aligned}
\end{equation}
As this is not a recognisable distribution, the Metropolis-Hasting algorithm is employed. \\
The full conditional distributions for $\delta_1$ and $\delta_h$, where $h \geq 2$ for count networks is the same as the binary networks case since the likelihood function does not affect the full conditionals.
\\
The full conditional distribution for $\alpha$ is:
\begin{equation}
\begin{aligned}  
\notag
P(\alpha | -)   &\propto 
        \left\{\prod_{i \neq j} 
                \frac{\exp(\eta_{i,j} y_{i,j})\exp\left[-\exp(\eta_{i,j})\right]}{{y_{i,j}!}} \right\}
    \times \left\{\frac{1}{\sqrt{2\pi\sigma_{\alpha}^2}}
        \exp\left[-\frac{1}{2\sigma_{\alpha}^2}(\alpha-\mu_{\alpha})^2\right] \right\} \\
                 &\propto \left\{\prod_{i \neq j} 
                \frac{\exp(\eta_{i,j} y_{i,j})\exp\left[-\exp(\eta_{i,j})\right]}{{y_{i,j}!}} \right\}
                 \times \exp\left[-\frac{1}{2}\frac{(\alpha-\mu_{\alpha})^2}{\sigma_{\alpha}^2}\right], \\
                 \end{aligned}
\end{equation}
\begin{equation}
\begin{aligned}  
\notag
\log P(\alpha | -)&= \sum_{i \neq j} \hspace{2pt} [ \eta_{i,j} y_{i,j} - 
                     \exp(\eta_{i,j}) - \log(y_{i,j}!) ]
                    - \frac{1}{2}\frac{(\alpha-\mu_{\alpha})^2}{\sigma_{\alpha}^2} + C\\
                &= \sum_{i \neq j} \hspace{2pt} [ \eta_{i,j} y_{i,j} - 
                   \exp(\eta_{i,j}) -\log(y_{i,j}!)] - (\alpha-\mu_{\alpha})^2 + C \\
                &= \sum_{i \neq j} \hspace{2pt} [ (\alpha -  \Vert \mathbf{z}_i-\mathbf{z}_j \Vert^2_2) y_{i,j} -  \exp(\alpha - \Vert \mathbf{z}_i-\mathbf{z}_j \Vert^2_2) -\log(y_{i,j}!)] \\
                &\qquad- (\alpha-\mu_{\alpha})^2 + C\\
                &= \sum_{i \neq j} \hspace{2pt} [ \alpha y_{i,j} - 
                    \exp(\alpha - \Vert \mathbf{z}_i-\mathbf{z}_j \Vert^2_2) -\log(y_{i,j}!) ] - (\alpha-\mu_{\alpha})^2 + C.
\end{aligned}
\end{equation}

\noindent A Metropolis-Hasting algorithm is used to sample from this full conditional distribution since it is not in a recognisable form.

\subsection{Deriving an informed proposal distribution for \texorpdfstring{$\alpha$}{Lg} for the Poisson latent shrinkage position model}
For count network, the log-likelihood function is 
\begin{equation}
\begin{aligned}  
\log  L(\mathbf{Y}|\mathbf{Z},\alpha)   &= \sum_{i \neq j} \hspace{2pt} [ \eta_{i,j} y_{i,j} - 
                    \exp(\eta_{i,j}) -\log(y_{i,j}!)] \\
                &= \sum_{i \neq j} \hspace{2pt} [ \alpha y_{i,j} ] - \sum_{i \neq j} \hspace{2pt} [\Vert \mathbf{z}_i-\mathbf{z}_j \Vert^2_2y_{i,j} ] - \sum_{i \neq j} \hspace{2pt} [ \exp(\eta_{i,j}) ] - \sum_{i \neq j}\log[y_{i,j}!].
                \label{eq:logpoisson}
\end{aligned}
\end{equation}
The 3rd term in equation \ref{eq:logpoisson} can be approximated using a quadratic Taylor expansion i.e. let $g(\alpha)  = - \sum_{i \neq j} \hspace{2pt} [ \exp(\eta_{i,j}) ] $, then the quadratic Taylor expansion around $g(\alpha)$ is

\begin{equation}
\begin{aligned}  
\notag
g(\alpha) &\approx g(\alpha^{(s)}) + (\alpha - \alpha^{(s)}) g'(\alpha^{(s)}) + 0.5(\alpha - \alpha^{(s)})^2 g''(\alpha^{(s)}), \\
\text{with } g'(\alpha^{(s)})  &= g''(\alpha^{(s)})  = - \sum_{i \neq j} \hspace{2pt} [ \exp(\alpha^{(s)} - \Vert \mathbf{z}_i-\mathbf{z}_j \Vert^2_2) ]. \\
\text{Thus, } g(\alpha) &\approx - \sum_{i \neq j} \hspace{2pt} [ \exp(\alpha^{(s)} - \Vert \mathbf{z}_i-\mathbf{z}_j \Vert^2_2) ] - (\alpha - \alpha^{(s)}) \sum_{i \neq j} \hspace{2pt} [ \exp(\alpha^{(s)} - \Vert \mathbf{z}_i-\mathbf{z}_j \Vert^2_2) ]  \\
&\quad - 0.5(\alpha - \alpha^{(s)})^2 \sum_{i \neq j} \hspace{2pt} [ \exp(\alpha^{(s)} - \Vert \mathbf{z}_i-\mathbf{z}_j \Vert^2_2) ].
\end{aligned}
\end{equation}
Here, $\alpha^{(s)}$ is the current value of $\alpha$ in the MCMC chain. Removing terms that are constant with respect to $\alpha$,
\begin{equation}
\begin{aligned}  
\notag
\log  L(\mathbf{Y}|\mathbf{Z},\alpha) &\approx\sum_{i \neq j} \hspace{2pt} \alpha y_{i,j} - (\alpha - \alpha^{(s)}) \sum_{i \neq j} \hspace{2pt} [ \exp(\alpha^{(s)} - \Vert \mathbf{z}_i-\mathbf{z}_j \Vert^2_2) ]  \\
& - 0.5(\alpha - \alpha^{(s)})^2 \sum_{i \neq j} \hspace{2pt} [ \exp(\alpha^{(s)} - \Vert \mathbf{z}_i-\mathbf{z}_j \Vert^2_2) ].\\
\end{aligned}
\end{equation}
Introducing the constant (with respect to $\alpha$) term,  $-\alpha^{(s)}y_{i,j}$:
\begin{equation}
\begin{aligned}  
\notag
    \log  L(\mathbf{Y}|\mathbf{Z},\alpha) &\approx \sum_{i \neq j} \hspace{2pt} (\alpha - \alpha^{(s)}) y_{i,j} - (\alpha - \alpha^{(s)}) \sum_{i \neq j} \hspace{2pt} [ \exp(\alpha^{(s)} - \Vert \mathbf{z}_i-\mathbf{z}_j \Vert^2_2) ]  \\
    &\hspace{10pt} -0.5(\alpha - \alpha^{(s)})^2 \sum_{i \neq j} \hspace{2pt} [ \exp(\alpha^{(s)} - \Vert \mathbf{z}_i-\mathbf{z}_j \Vert^2_2) ] \\
&\approx  -\frac{1}{2}\left\{ \sum_{i \neq j} \hspace{2pt} [ \exp(\alpha^{(s)} - \Vert \mathbf{z}_i-\mathbf{z}_j \Vert^2_2) ] \right\} \left\{ (\alpha - \alpha^{(s)})^2 \right.\\
&\quad \left. - 2 \frac{(\alpha - \alpha^{(s)})}{\sum_{i \neq j} \hspace{2pt} [ \exp(\alpha^{(s)} - \Vert \mathbf{z}_i-\mathbf{z}_j \Vert^2_2) ]} \left[\sum_{i \neq j}y_{i,j} - 
\right. \right.\\
&\quad \left. \left. \left.  \sum_{i \neq j} \hspace{2pt}  \exp(\alpha^{(s)} - \right. 
\Vert \mathbf{z}_i-\mathbf{z}_j \Vert^2_2) \right]   \right\} \\
\end{aligned}
\end{equation}

Completing the square gives
\begin{equation}
\begin{aligned}  
\notag
\log  L(\mathbf{Y}|\mathbf{Z},\alpha) &\approx -\frac{1}{2} \left\{\sum_{i \neq j} \hspace{2pt} [ \exp(\alpha^{(s)} - \Vert \mathbf{z}_i-\mathbf{z}_j \Vert^2_2) ] \right\} \left\{ \left[(\alpha - \alpha^{(s)}) - \right. \right. \\
&\quad \left. \left. \frac{1}{\sum_{i \neq j} \hspace{2pt} [ \exp(\alpha^{(s)} - \Vert \mathbf{z}_i-\mathbf{z}_j \Vert^2_2) ]}\left(\sum_{i \neq j}y_{i,j} - \right. \right. \right. \\
&\quad \left. \left. \left. \sum_{i \neq j} \hspace{2pt}  \exp(\alpha^{(s)} - \Vert \mathbf{z}_i-\mathbf{z}_j \Vert^2_2) \right) \right]^2 \right\}, \\
\end{aligned}
\end{equation}
which is in the form of a Gaussian distribution with parameters
\begin{equation}
\begin{aligned}  
\notag
\mbox{variance} &= \left\{\sum_{i \neq j} \hspace{2pt} [ \exp(\alpha^{(s)} - \Vert \mathbf{z}_i-\mathbf{z}_j \Vert^2_2) ]\right\}^{-1}, \\
\mbox{mean} &= \mbox{variance} \times \left\{\sum_{i \neq j} y_{i,j} - \sum_{i \neq j} \hspace{2pt} [ \exp(\alpha^{(s)} - \Vert \mathbf{z}_i-\mathbf{z}_j \Vert^2_2) ] \right\}.
\end{aligned}
\end{equation}
The prior on $\alpha$ is $N(\mu_{\alpha}, \sigma^2_{\alpha})$, which makes $(\alpha - \alpha^{(s)}) \sim N(\mu_{\alpha}- \alpha^{(s)},\sigma^2_{\alpha})$, (since $\alpha^{(s)}$ is a constant, it only affects the mean). Hence, putting the approximated log likelihood function and the log of the prior distribution on $(\alpha - \alpha^{(s)}) $, this results in a sum of (the log of) two normal distributions, which gives the distribution of $(\alpha - \alpha^{(s)}) $ to be $N(\mu_{\alpha, C}, \Bar{\sigma^2}_{\alpha, C})$, where

\begin{equation}
\begin{aligned}  
\notag
\Bar{\sigma}^2_{\alpha, C} = \left[\sum_{i \neq j} \hspace{2pt} \exp(\alpha^{(s)} - \Vert \mathbf{z}_i-\mathbf{z}_j \Vert^2_2) + \frac{1}{\sigma^2_{\alpha}}\right]^{-1},
\end{aligned}
\end{equation}

and

\begin{equation}
\begin{aligned}  
\notag
\mu_{\alpha, C} = \Bar{\sigma}^2_{\alpha, C} \times \left[ \sum_{i \neq j} y_{i,j} - \sum_{i \neq j} \hspace{2pt}  \exp(\alpha^{(s)} - \Vert \mathbf{z}_i-\mathbf{z}_j \Vert^2_2)   + \frac{1}{\sigma^2_{\alpha}}(\mu_{\alpha}-\alpha^{(s)}) \right].
\end{aligned}
\end{equation}
Thus, here the full conditional distribution of $\alpha$ is approximated by a normal distribution with variance $\Bar{\sigma}^2_{\alpha, C}$ and mean

\begin{equation}
\begin{aligned}  
\notag
\Bar{\mu}_{\alpha, C} = \alpha^{(s)} + \Bar{\sigma}^2_{\alpha, C} \times \left[ \sum_{i \neq j} y_{i,j} - \sum_{i \neq j} \hspace{2pt}  \exp(\alpha^{(s)} - \Vert \mathbf{z}_i-\mathbf{z}_j \Vert^2_2)   + \frac{1}{\sigma^2_{\alpha}}(\mu_{\alpha}-\alpha^{(s)}) \right].
\end{aligned}
\end{equation}
This distribution is used as the proposal distribution in the Metropolis-Hastings algorithm.

\newpage

\section{Comparison of LSPM performance under the MGP and MTGP priors} \label{app:MGPvsMTGP}

As discussed in Section 2.2 of the paper, it is of interest to explore the substantive impact of using the MTGP rather than the MGP prior on the performance of the LSPM. In this appendix we consider the analyses from the simulation studies and the applications examined in the paper, but where the LSPM employs the MGP rather than the MTGP prior.

\begin{figure}[htb]
\includegraphics[width=\linewidth]{figures/trunc_dim_bar.pdf}
\includegraphics[width=\linewidth]{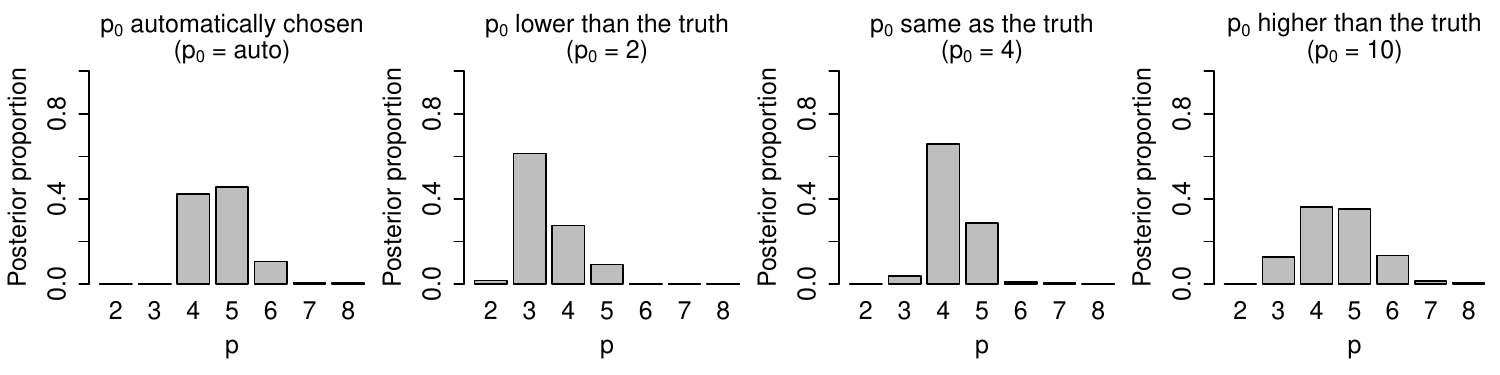}
\caption{Posterior proportion of dimensions for different initial truncation levels, $p_0$. The top row shows logistic LSPM results for the MTGP prior fitted to the simulated binary networks while the bottom row shows results with the MGP prior.}
\label{fig:dim_bar_trunc_nonT}
\end{figure}

Figure \ref{fig:dim_bar_trunc_nonT} illustrates the posterior distributions of $p$ for the logistic LSPM with the MTGP prior (top row) and the MGP prior (bottom row) for the simulated binary networks from Section 4.1 of the main paper. While the posterior modal dimension when $p_0 = \{2, 4, 10\}$ is broadly similar under both priors, the posterior under the LSPM with the MTGP prior has higher precision. The LSPM with the MGP prior tends to explore higher dimensions more often. The performance of the LSPM with the MGP prior is further summarised in Table \ref{tab:trunc_nonT_dim_protest} where $p_m$ is only accurately inferred when $p_0 = \{4, 10\}$. The proportion of simulated networks for which $p_m = p^*$ under the MGP prior is lower than when MTGP prior is used (see Table 1 in the main paper) for $p_0 = 10$ or `auto'.

\begin{figure}[htb]
\includegraphics[width=\linewidth]{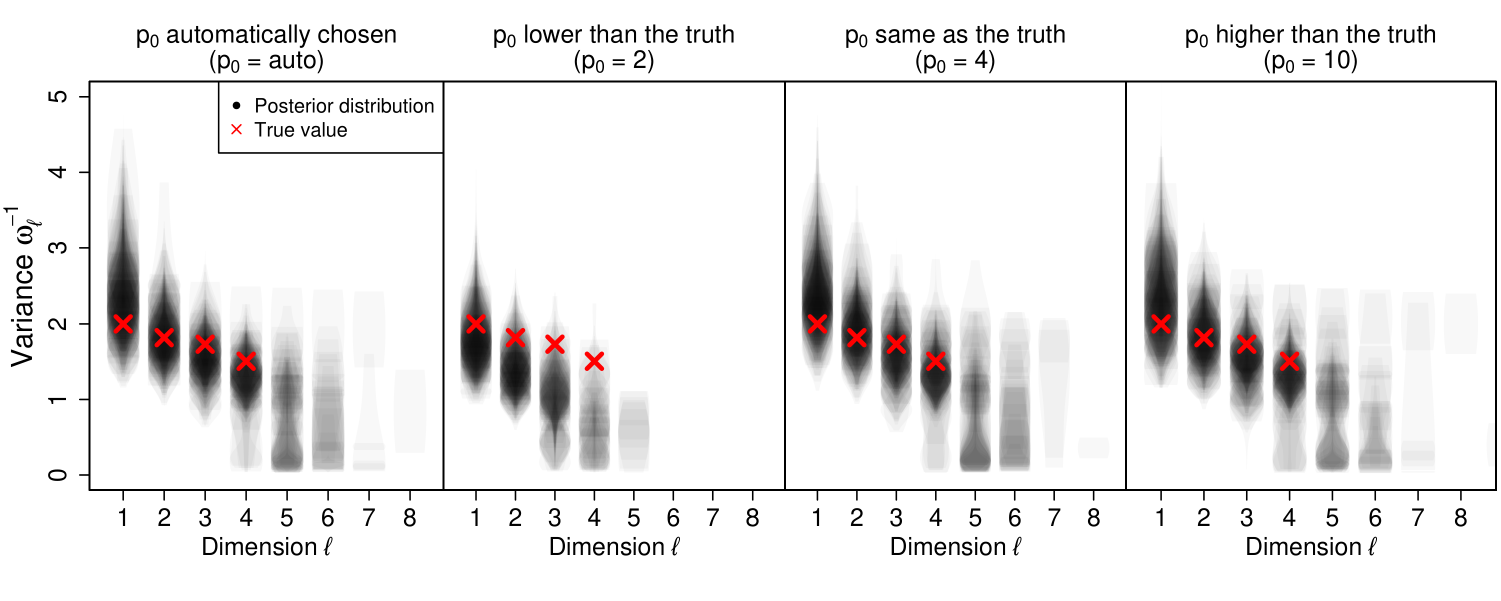}
\includegraphics[width=\linewidth]{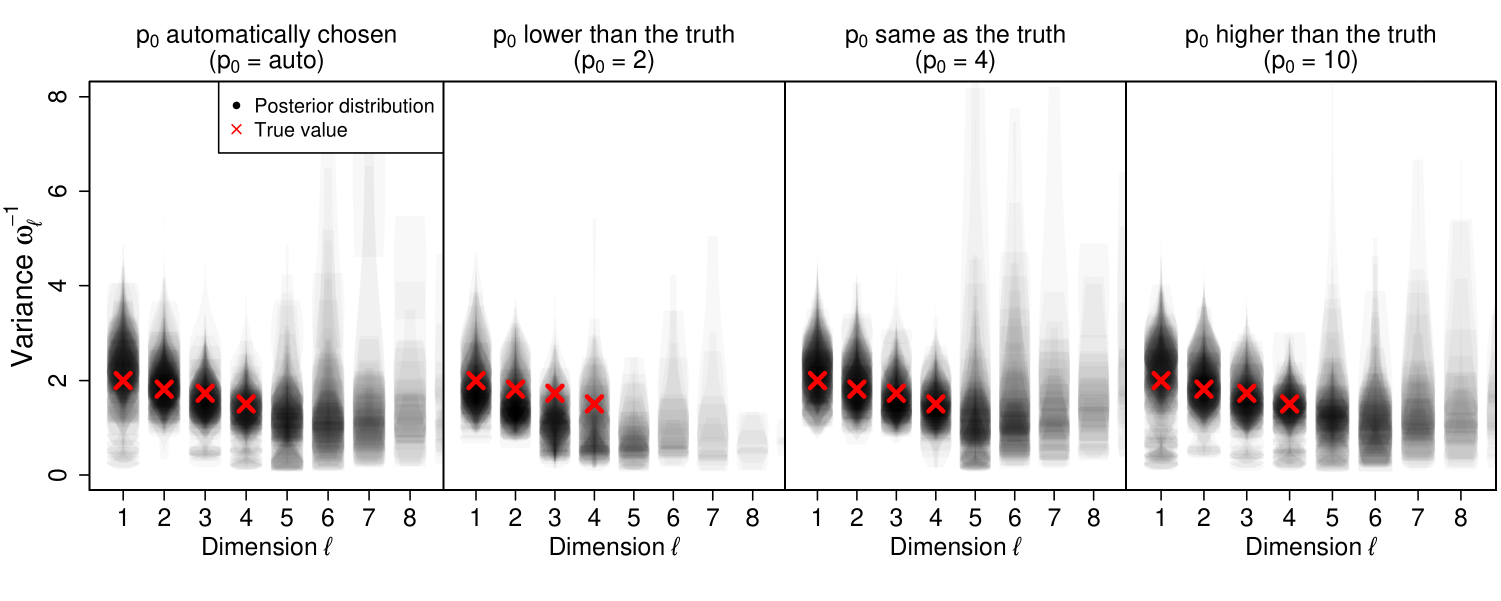}
\caption{Posterior distribution of variances under the logistic LSPM for simulated binary networks across dimensions for different initial truncation levels, $p_0$ under the MTGP prior (top) and the MGP prior (bottom).}
\label{fig:pmv_trunc_nonT}
\end{figure}

\begin{figure}[htb]
     \begin{subfigure}[b]{0.47\linewidth}
         \centering
         \includegraphics[width=\linewidth]{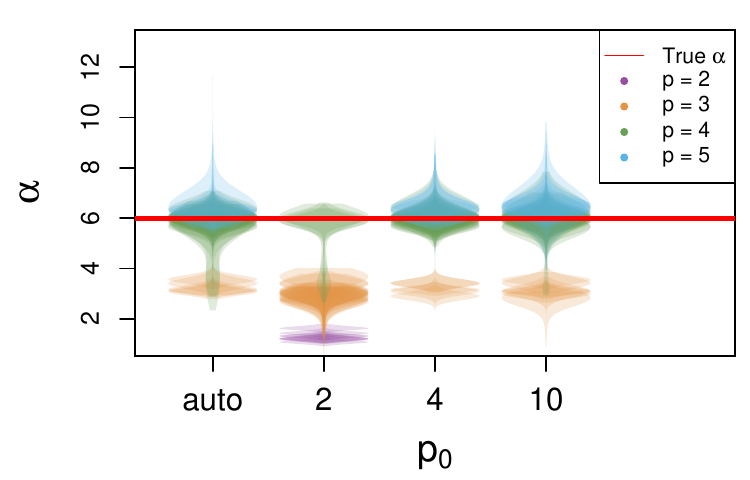}
         \caption{MTGP prior}
         \label{fig:mtgp_trunc_alpha}
     \end{subfigure}
     \hfill
    \begin{subfigure}[b]{0.47\linewidth}
         \centering
         \includegraphics[width=\linewidth]{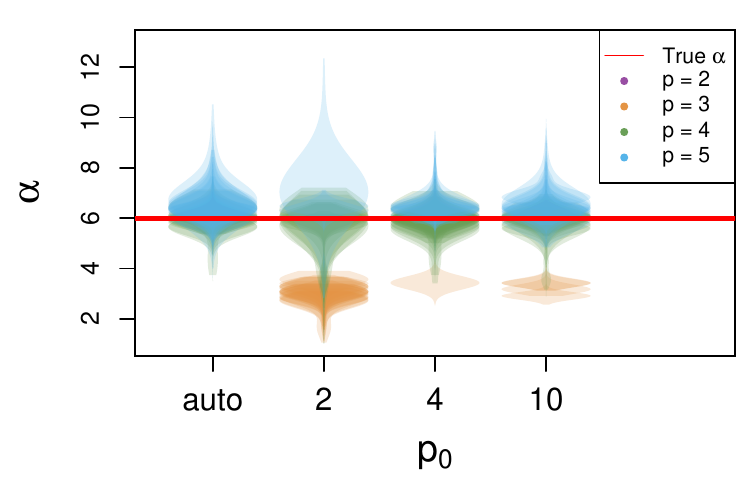}
         \caption{MGP prior}
         \label{fig:mgp_trunc_alpha}
     \end{subfigure}
\caption{Posterior distributions of $\alpha$ across dimensions for different initial truncation levels, $p_0$, for the logistic LSPM fitted to simulated binary networks under the MTGP prior and the MGP prior.}
\label{fig:alpha_trunc_nonT}
\end{figure}

\begin{figure}[htb]
\includegraphics[width=\linewidth]{figures/trunc_pred.pdf}
\includegraphics[width=\linewidth]{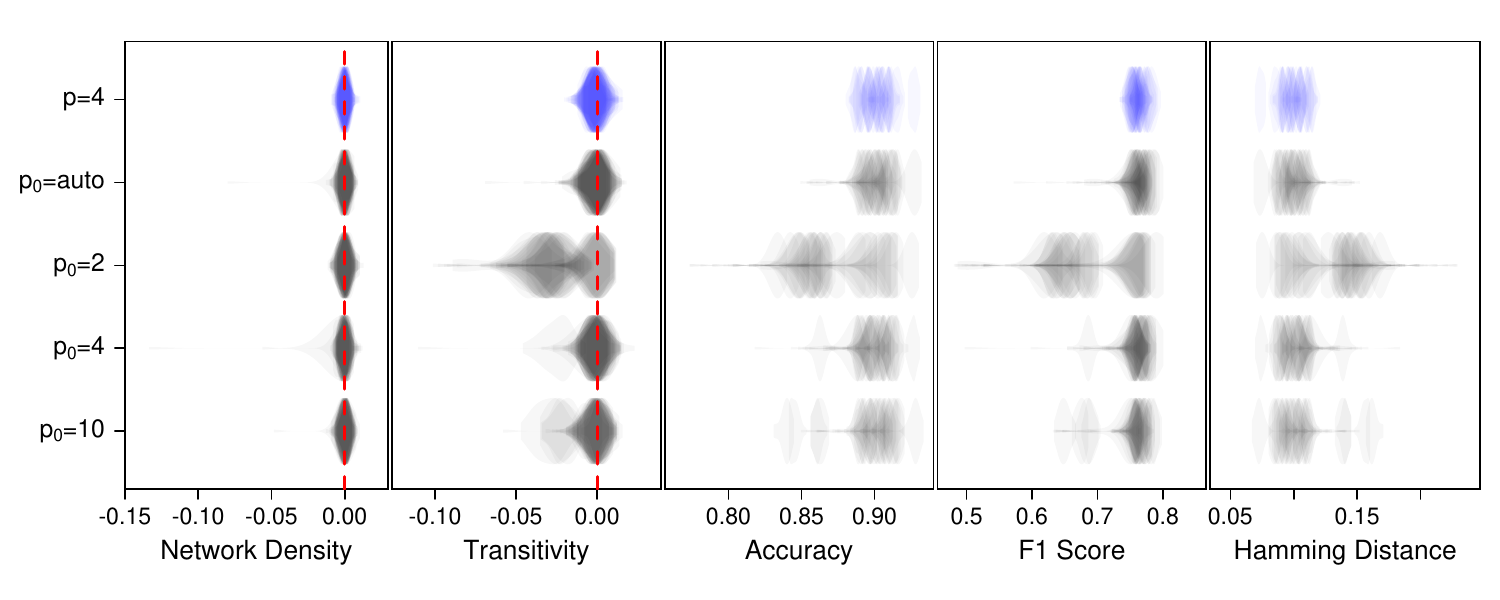}
\caption{Posterior predictive check results conditioned on $p_m$ across different initial truncation levels $p_0$ for simulated binary networks under the logistic LSPM with the MTGP prior (top) and the MGP (bottom). The violin plots indicate the metrics' distributions from the predicted networks for the LPM (blue) and the LSPM (gray). The network density and transitivity plots illustrate the differences between the posterior predictive networks and the observed networks.}
\label{fig:pred_trunc_nonT}
\end{figure}

Figure \ref{fig:pmv_trunc_nonT} illustrates posterior distributions of the variance parameters under a logistic LSPM with the  MTGP and the MGP priors. Again the exploration of higher dimensions is evident under the MGP prior, giving more diffuse posteriors. Figure \ref{fig:alpha_trunc_nonT} shows similar behaviour for the intercept parameter for both MTGP and MGP priors when $p$ varies, with the posteriors under the MTGP prior being more precise.
% The posterior mean variance does not necessarily shrink as dimension increases. While admittedly these occurrences do not happen frequently in practice, the possibility that they can contradicts the model's inherent aim of having increasingly small variance across higher dimensions. Besides, using the MTGP prior also results in a smaller range of posterior distribution of variance.
Figure \ref{fig:pred_trunc_nonT} provides posterior predictive checks for the logistic LSPM with the MTGP prior and the MGP prior for the simulated binary networks. These checks show very similar model fit behaviour, conditional on accurate inference on the number of dimensions.

\begin{figure}[htb]
     \begin{subfigure}[b]{0.32\linewidth}
         \centering
         \includegraphics[width=\linewidth]{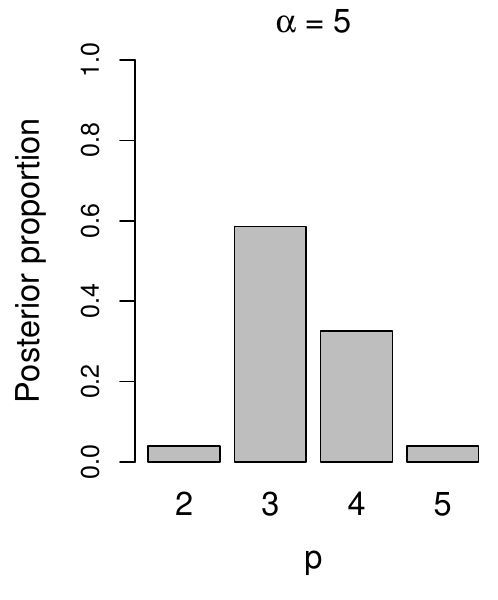}
         \caption{Posterior distribution of $p$}
         \label{fig:dens_nonT_dim_bar}
     \end{subfigure}
     \hfill
    \begin{subfigure}[b]{0.32\linewidth}
         \centering
         \includegraphics[width=\linewidth]{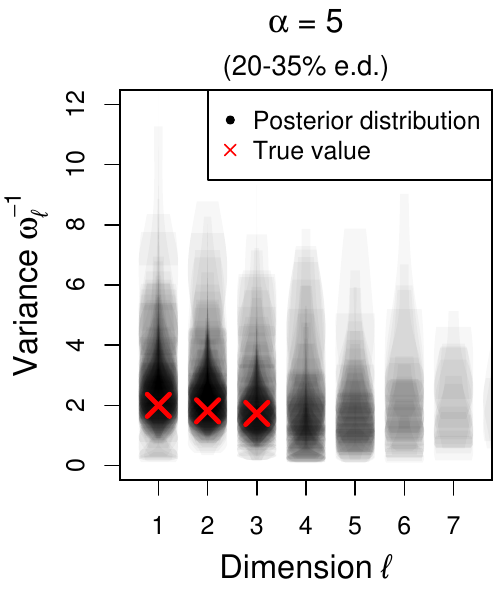}
         \caption{Posterior distributions of variance parameters}
         \label{fig:dens_nonT_pmv}
     \end{subfigure}
          \hfill
    \begin{subfigure}[b]{0.32\linewidth}
         \centering
         \includegraphics[width=\linewidth]{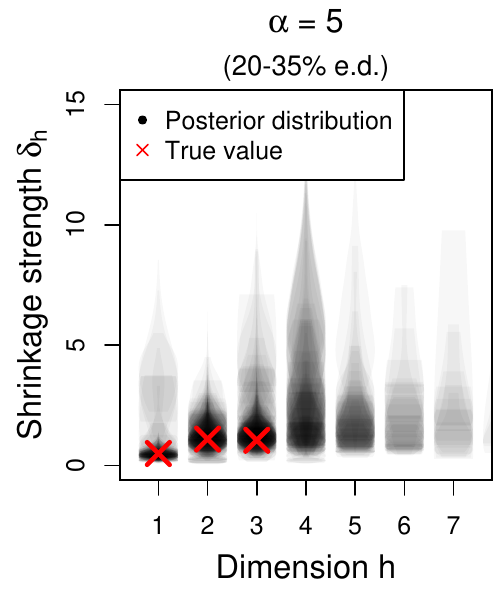}
         \caption{Posterior distributions of shrinkage strength parameters}
         \label{fig:dens_nonT_pmd}
     \end{subfigure}
\caption{Inference under the logistic LSPM with the MGP prior based on networks from Section 4.3 with $\alpha=5$ (e.d. = emprirical density.}
\label{fig:dens_nonT}
\end{figure}

\begin{figure}[htb]
     \begin{subfigure}[b]{0.195\linewidth}
         \centering
         \includegraphics[width=\linewidth]{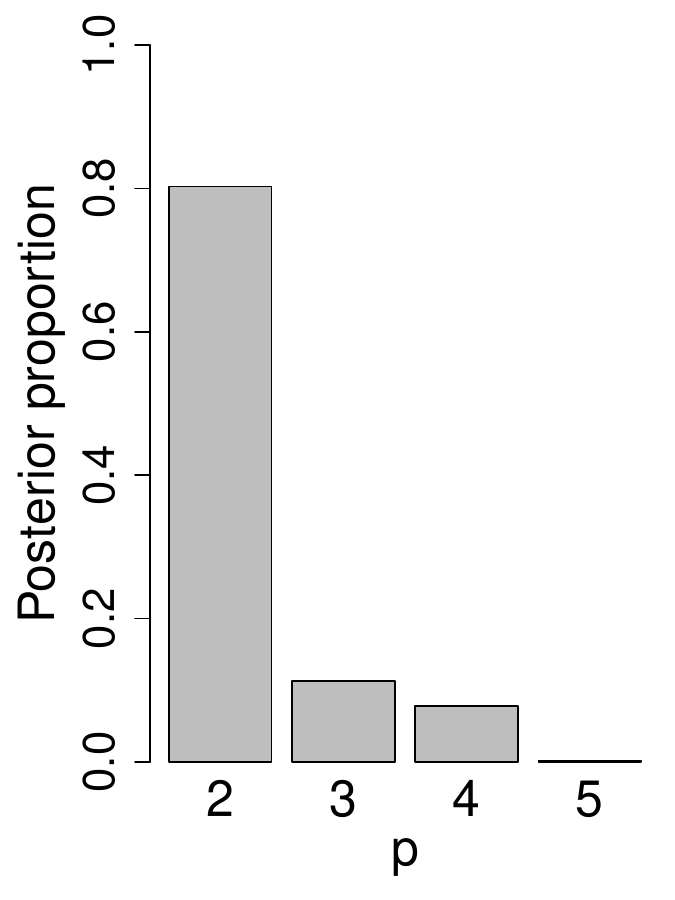}
         \caption{Posterior distribution of $p$}
         \label{fig:zach_nonT_dim_bar}
     \end{subfigure}
     \hfill
    \begin{subfigure}[b]{0.795\linewidth}
         \centering
         \includegraphics[width=\linewidth]{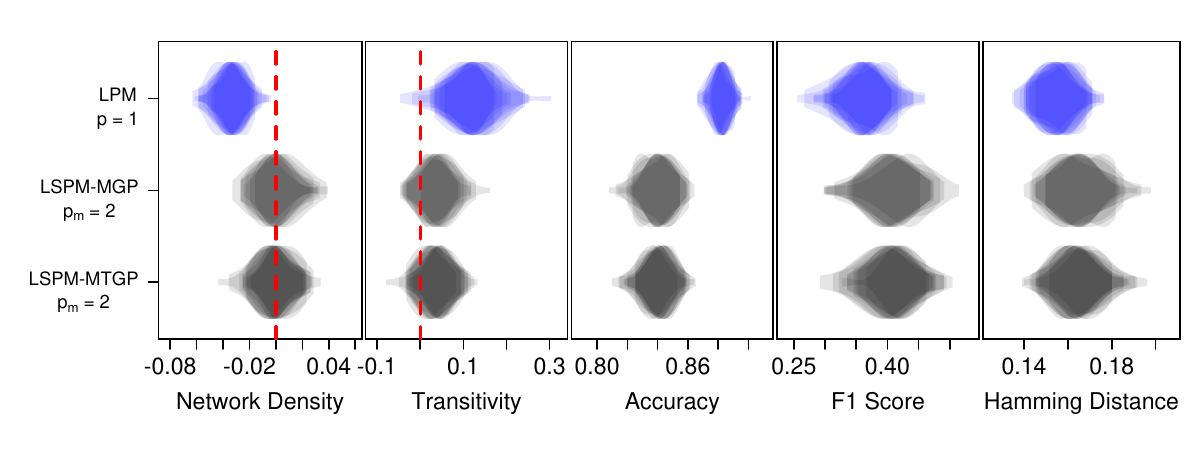}
         \caption{Posterior predictive checks}
         \label{fig:zach_nonT_pred}
     \end{subfigure}
\caption{Analysis of the Zachary karate network under the logistic LSPM with the MGP prior.}
\label{fig:zach_nonT}
\end{figure}

\begin{figure}[htb]
     \begin{subfigure}[b]{0.195\linewidth}
         \centering
         \includegraphics[width=\linewidth]{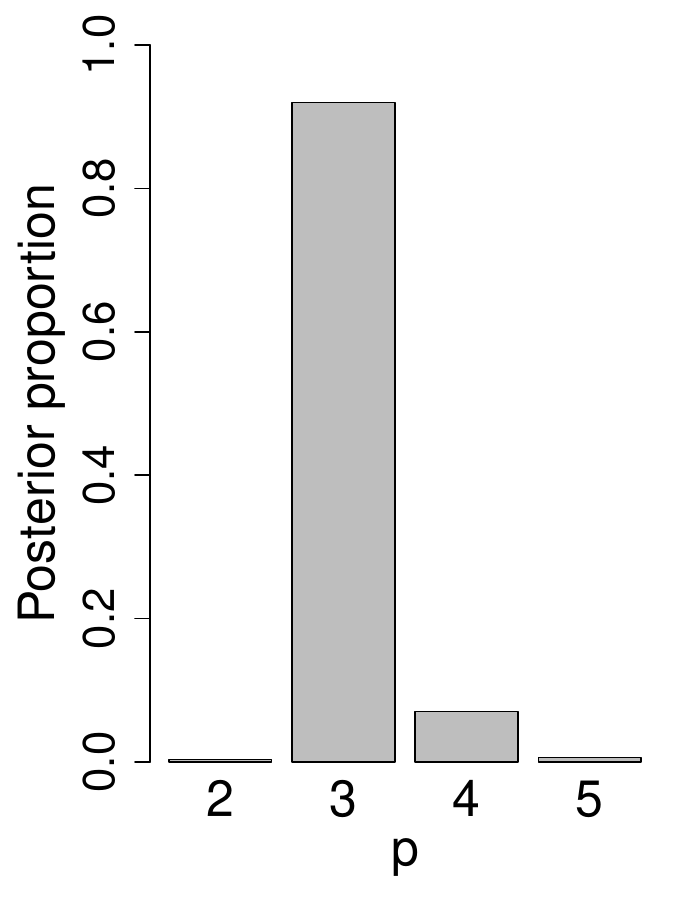}
         \caption{Posterior distribution of $p$}
         \label{fig:cat_nonT_dim_bar}
     \end{subfigure}
     \hfill
    \begin{subfigure}[b]{0.795\linewidth}
         \centering
         \includegraphics[width=\linewidth]{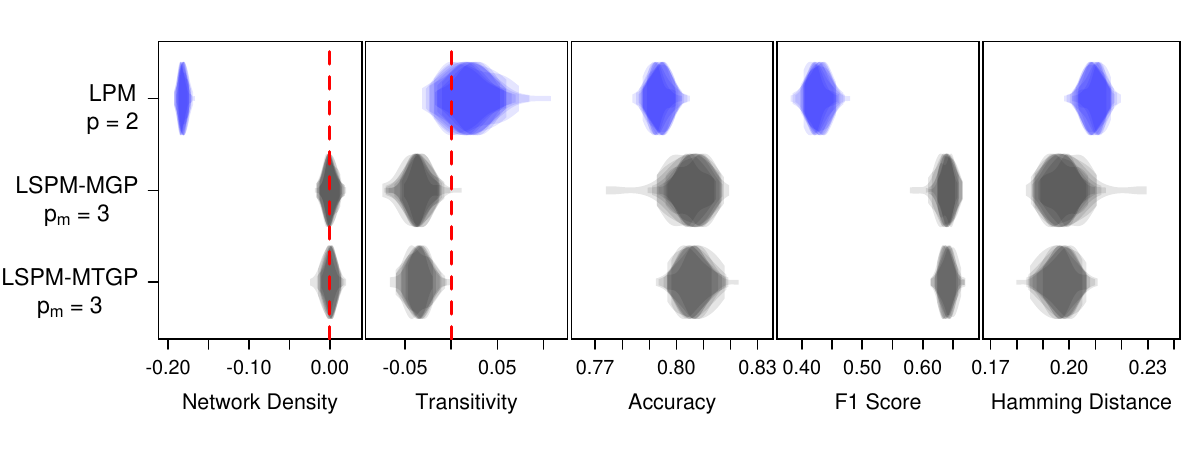}
         \caption{Posterior predictive checks}
         \label{fig:cat_nonT_pred}
     \end{subfigure}
\caption{Analysis of the cat connectome network under the logistic LSPM with the MGP prior.}
\label{fig:cat_nonT}
\end{figure}

\begin{figure}[htb]
     \begin{subfigure}[b]{0.195\linewidth}
         \centering
         \includegraphics[width=\linewidth]{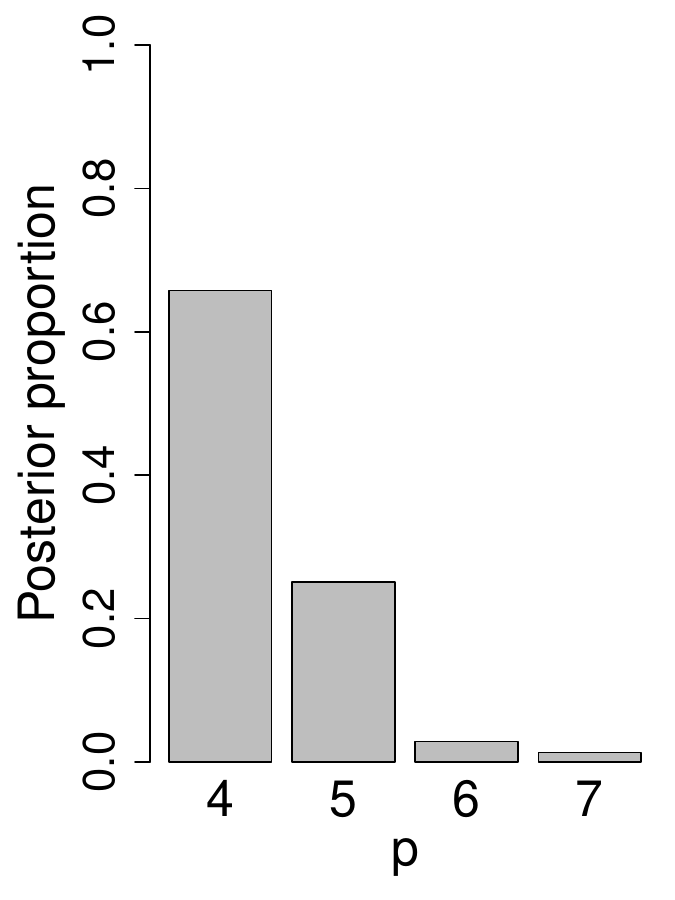}
         \caption{Posterior distribution of $p$}
         \label{fig:worm_nonT_dim_bar}
     \end{subfigure}
     \hfill
    \begin{subfigure}[b]{0.795\linewidth}
         \centering
         \includegraphics[width=\linewidth]{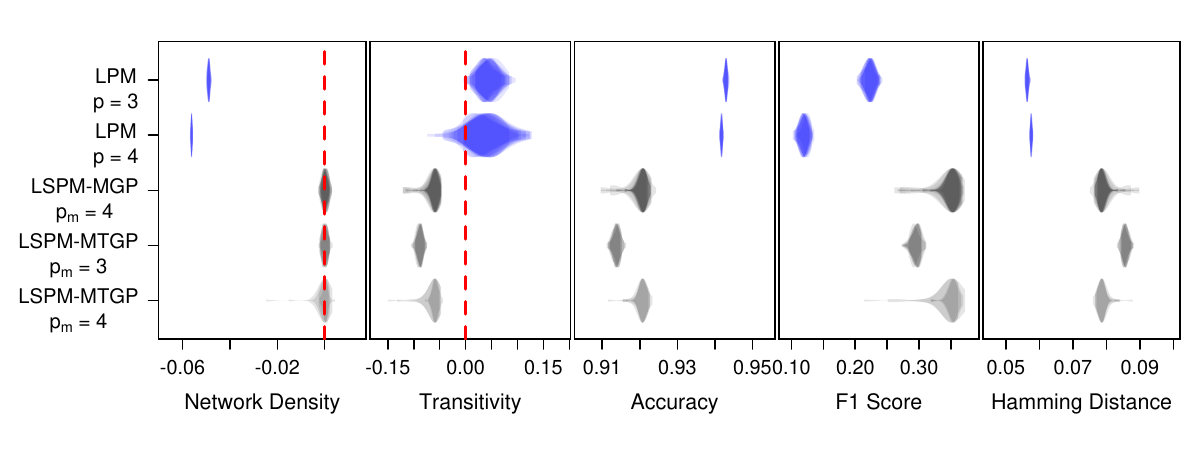}
         \caption{Posterior predictive checks}
         \label{fig:worm_nonT_pred}
     \end{subfigure}
\caption{Analysis of the worm network under the logistic LSPM with the MGP prior.}
\label{fig:worm_nonT}
\end{figure}

Figures \ref{fig:dens_nonT}, \ref{fig:zach_nonT}, \ref{fig:cat_nonT} and \ref{fig:worm_nonT} illustrate inference after applying the logistic LSPM with the MGP prior on simulated networks from Section 4.3, the Zachary network, the cat connectome network, and the worm neuron network respectively. All again show similar behaviour in that posterior distributions are more diffuse than when the MTGP is employed and model fit is similar, condition on the correct dimension being inferred.

\begin{figure}
\centering
     \begin{subfigure}[b]{0.243\linewidth}
         \centering
         \includegraphics[width=\linewidth]{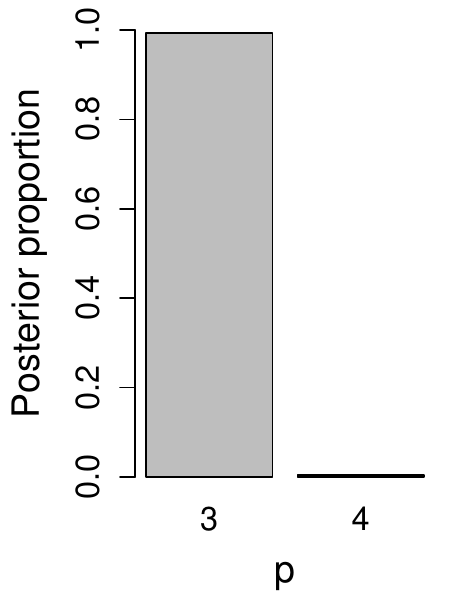}
         \caption{Simulated network}
         \label{fig:mgp_prior_sim_d}
     \end{subfigure}
     \hfill
     \begin{subfigure}[b]{0.243\linewidth}
         \centering
         \includegraphics[width=\linewidth]{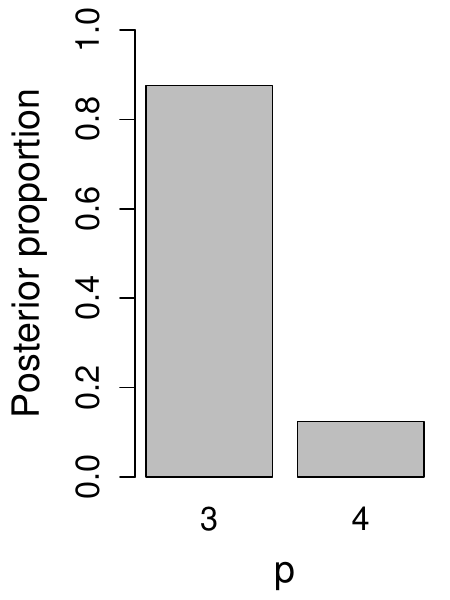}
         \caption{Cat brain network}
         \label{fig:mgp_prior_cat_d}
     \end{subfigure}
          \begin{subfigure}[b]{0.243\linewidth}
         \centering
         \includegraphics[width=\linewidth]{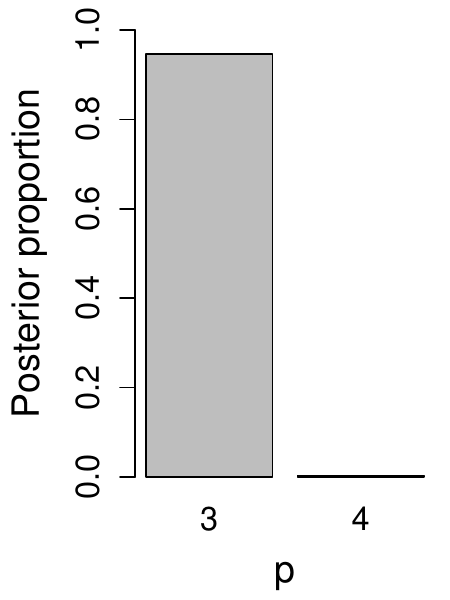}
         \caption{Simulated network}
         \label{fig:mtgp_prior_sim_d}
     \end{subfigure}
     \hfill
     \begin{subfigure}[b]{0.243\linewidth}
         \centering
         \includegraphics[width=\linewidth]{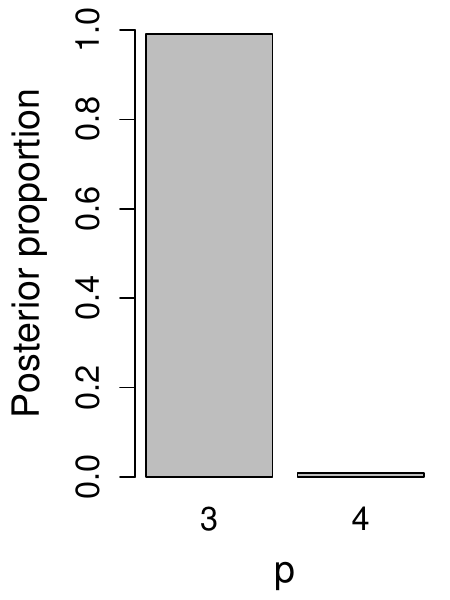}
         \caption{Cat brain network}
         \label{fig:mtgp_prior_cat_d}
     \end{subfigure}
        \caption{Posterior distributions of $p$ under the logistic LSPM with the MGP prior in (a) and (b), and with the MTGP prior in (c) and (d) fitted on a simulated network from Section 4.3 where $\alpha=5$, and on the cat brain network from Section 5.2.}
\label{fig:mgp_prior_d}
\end{figure}

\begin{figure}
\centering
     \begin{subfigure}[b]{0.243\linewidth}
         \centering
         \includegraphics[width=\linewidth]{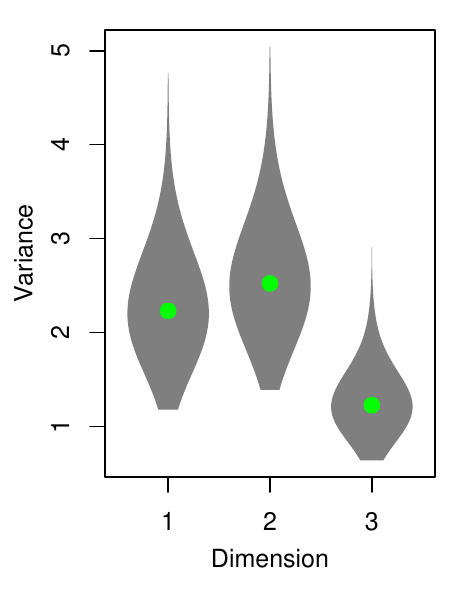}
         \caption{Simulated network}
         \label{fig:mgp_prior_sim}
     \end{subfigure}
     \hfill
     \begin{subfigure}[b]{0.243\linewidth}
         \centering
         \includegraphics[width=\linewidth]{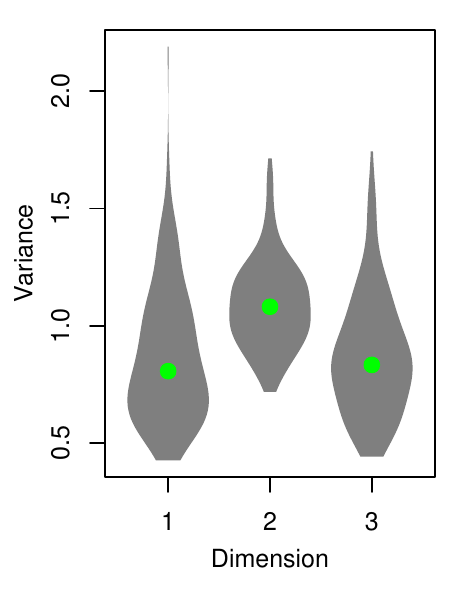}
         \caption{Cat brain network}
         \label{fig:mgp_prior_cat}
     \end{subfigure}
          \begin{subfigure}[b]{0.243\linewidth}
         \centering
         \includegraphics[width=\linewidth]{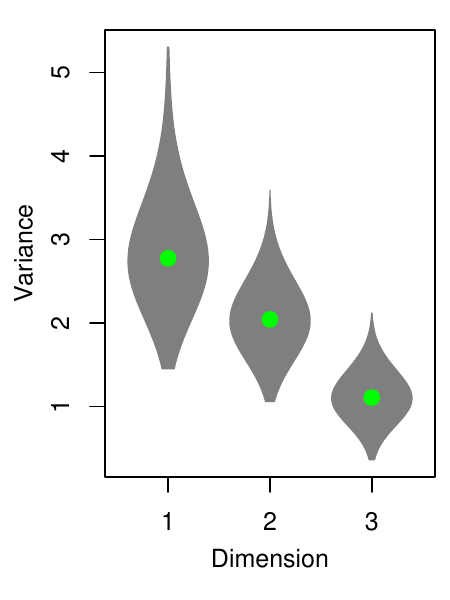}
         \caption{Simulated network}
         \label{fig:mtgp_prior_sim}
     \end{subfigure}
     \hfill
     \begin{subfigure}[b]{0.243\linewidth}
         \centering
         \includegraphics[width=\linewidth]{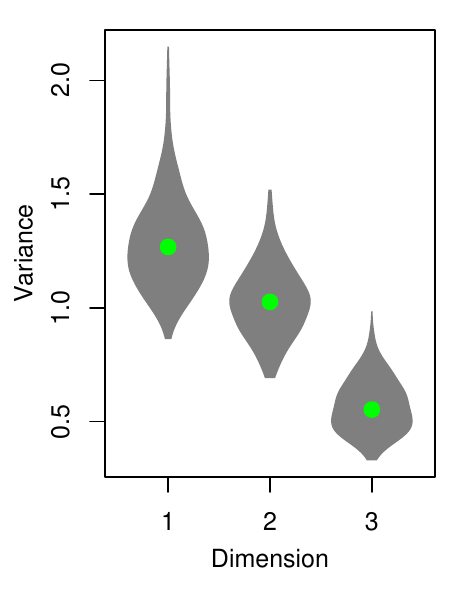}
         \caption{Cat brain network}
         \label{fig:mtgp_prior_cat}
     \end{subfigure}
        \caption{Posterior distributions of the variance parameter under the logistic LSPM with the MGP prior in (a) and (b), and with the MTGP prior in (c) and (d) fitted to a simulated network from Section 4.3 where $\alpha=5$, and to the cat brain network from Section 5.2. The green points indicate the posterior mean variance.}
\label{fig:mgp_prior}
\end{figure}

Figure \ref{fig:mgp_prior_d} and \ref{fig:mgp_prior} illustrate two specific examples that clearly highlight the difference under the MTGP and MGP priors.

\begin{table}[htb]
\caption{Performance of the logistic LSPM with the MGP prior. The posterior mode of the number of active dimensions, the proportion of the 30 simulated networks for which $p_m = p^*$, and the Procrustes correlation of the 30 predicted networks' positions and the LSPM posterior mean positions. The 95\% credible intervals are given in the brackets.}
\label{tab:trunc_nonT_dim_protest}
\begin{tabular}{|c|c|c|c|}
\hline
$p_0$ & \begin{tabular}[c]{@{}c@{}} $p_m$\end{tabular} & \begin{tabular}[c]{@{}c@{}}Proportion where\\ $p_m = p^*$\end{tabular} & \begin{tabular}[c]{@{}c@{}}Procrustes correlation\\ \end{tabular} \\ \hline
auto  & 5 (4, 6)                                                     & 0.43                                                                & 0.968 (0.912, 0.986)                                                                     \\ \hline
2     & 3 (3, 5)                                                     & 0.30                                                                & 0.788 (0.722, 0.959)                                                                     \\ \hline
4     & 4 (3, 5)                                                     & 0.67                                                                & 0.961 (0.911, 0.985)                                                                     \\ \hline
10    & 4 (3, 6)                                                     & 0.37                                                                & 0.979 (0.934, 0.985)                                                                     \\ \hline
\end{tabular}
\end{table}

% \begin{figure}
% \includegraphics[width=\linewidth]{figures/mgp_prior.pdf} 
% \caption{Posterior distributions of variance under the logistic LSPM with MGP prior fitted on datasets used in the paper. Green dot indicates the posterior mean variance.}
% \label{fig:mgp_prior}
% \end{figure}

\clearpage
\section{Simulation studies' additional posterior predictive checks.}
\label{app:simstudies}

The LSPM performance is assessed via the posterior distributions of the dimension, variance and shrinkage strength parameters, and Procrustes correlations between inferred and true locations. Posterior predictive checks are employed to assess model fit. Included here are plots to assist in assessing the performance of the LSPM, to supplement those provided in the paper.

% TRUNCATION SIMULATION STUDIES
% \subsubsection{Different initial truncation levels}

\begin{figure}[htb]
\includegraphics[width=.925\linewidth]{figures/trunc_pmv.pdf} 
\includegraphics[width=.925\linewidth]{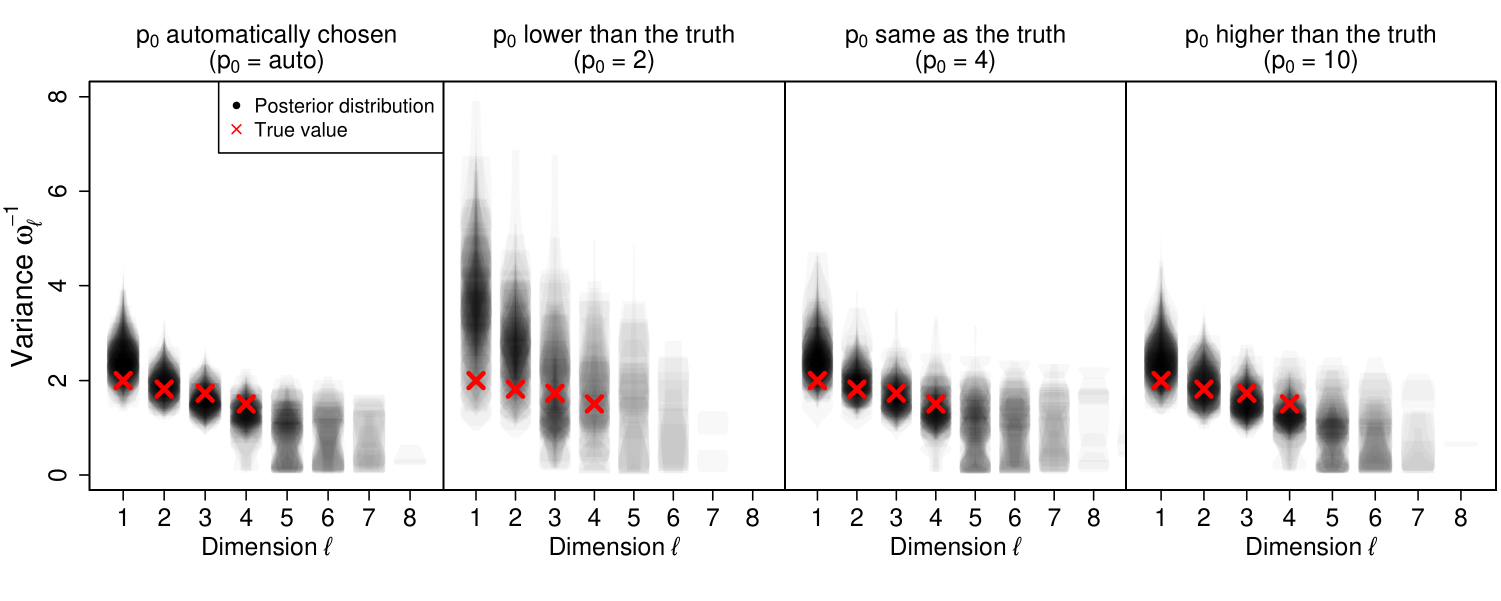} 
\caption{Simulation study 1. Posterior distributions of the variance parameters across dimensions for different initial truncation levels, $p_0$ for the logistic LSPM for binary networks (top) and for the Poisson LSPM for count networks (bottom).}
\label{fig:pmv_trunc}
\end{figure}

% NETWORK SIZE SIMULATION STUDIES
% \subsubsection{Different network sizes

\begin{figure}[!ht]
\includegraphics[width=\linewidth]{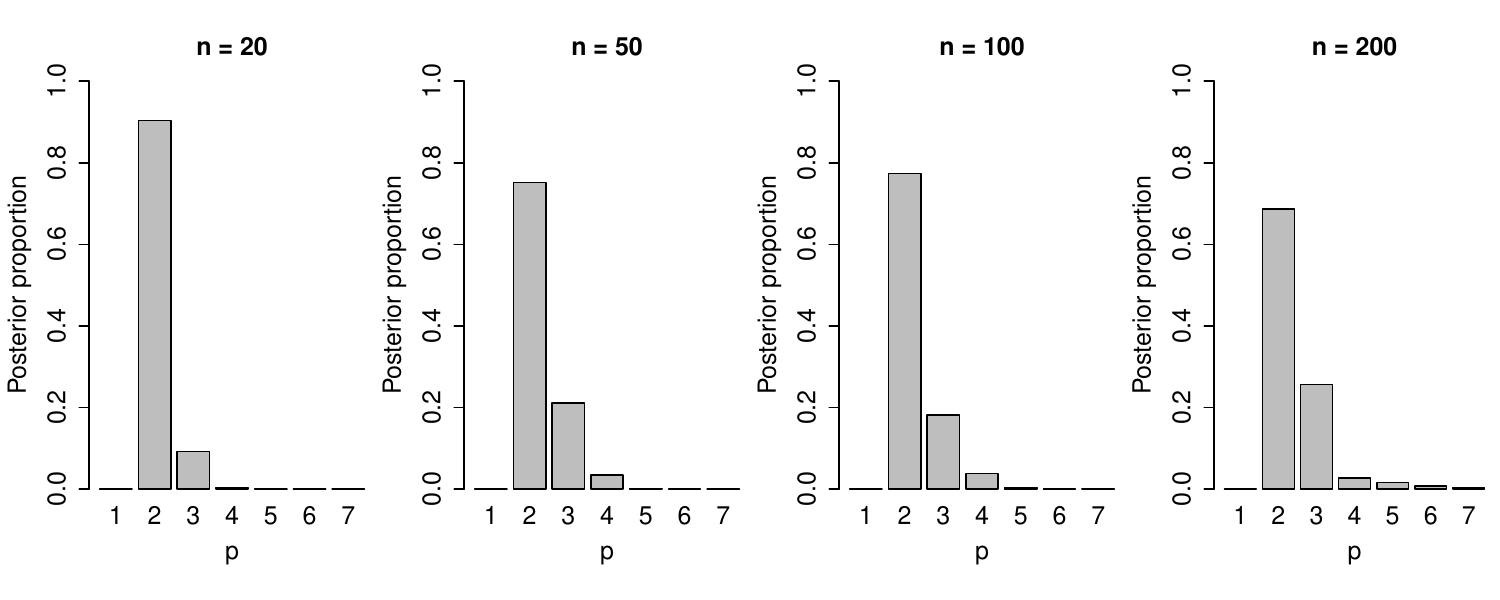}
\includegraphics[width=\linewidth]{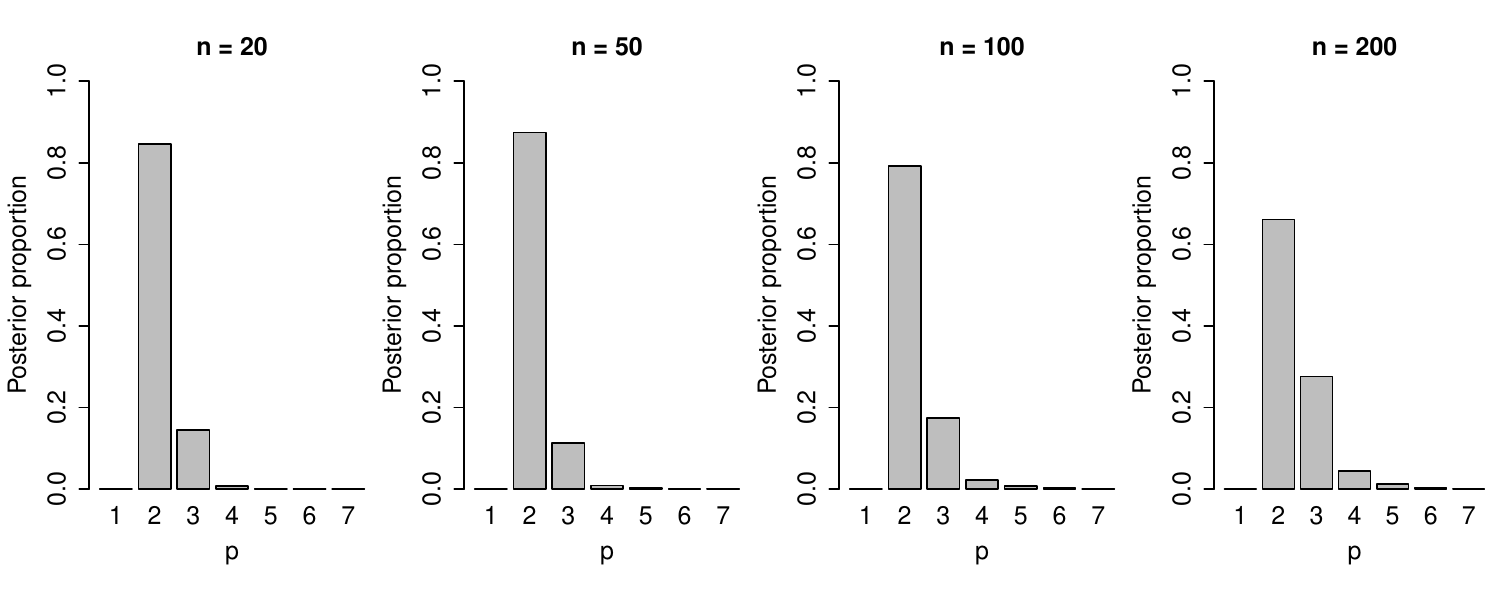}
        \caption{Simulation study 2. For different network sizes, $n$, the posterior distribution of $p$ in 30 simulated networks for the logistic LSPM for binary networks (top) and for the Poisson LSPM for count networks (bottom).}
        \label{dim_size}
\end{figure}

\begin{figure}[tb]
\includegraphics[width=.925\linewidth]{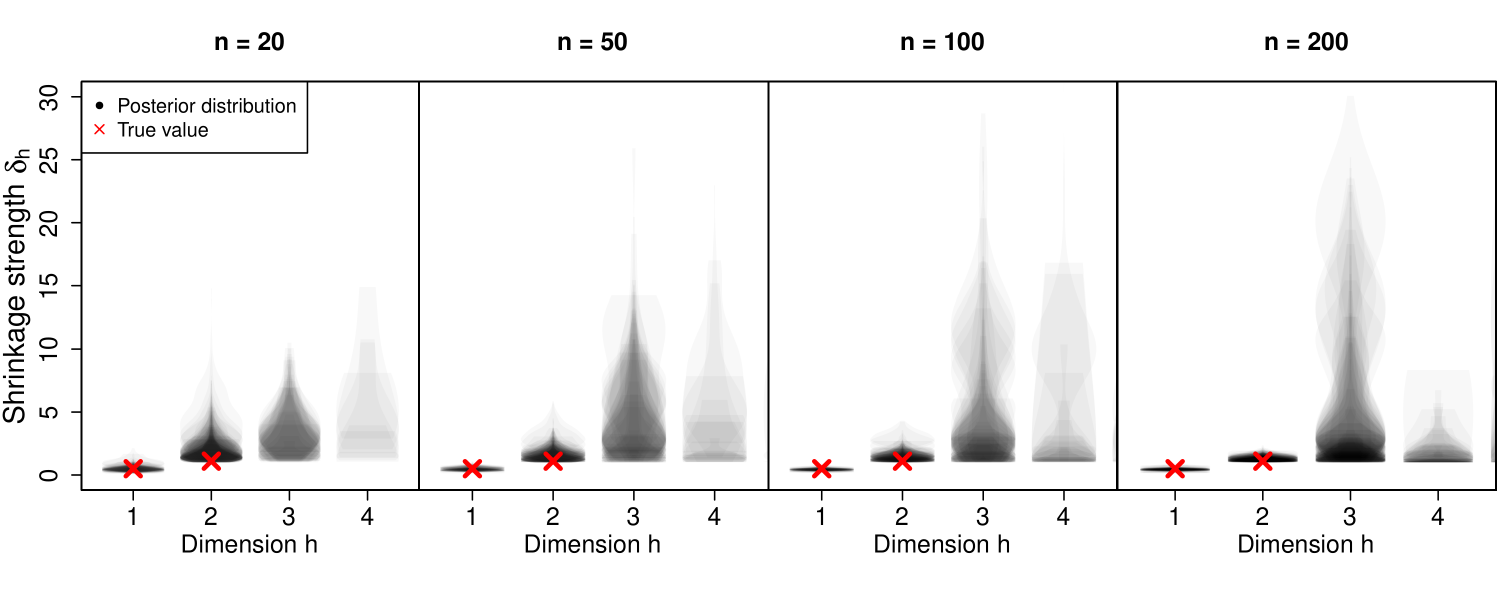} 
\includegraphics[width=.925\linewidth]{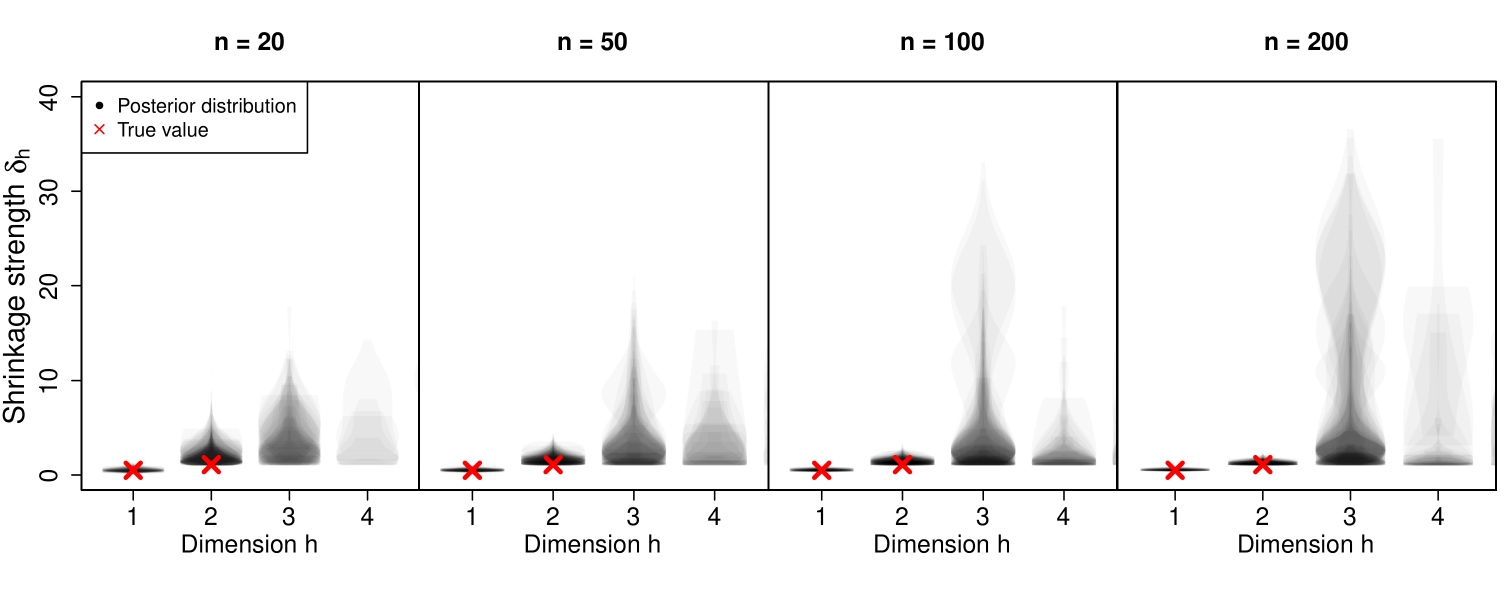} 
\caption{Simulation study 2. Posterior distributions of shrinkage strength parameters across dimensions for different network sizes, $n$ for the logistic LSPM for binary networks (top) and for the Poisson LSPM for count networks (bottom).}
\label{fig:pmd_size}
\end{figure}

\begin{figure}[tb]
\includegraphics[width=.925\linewidth]{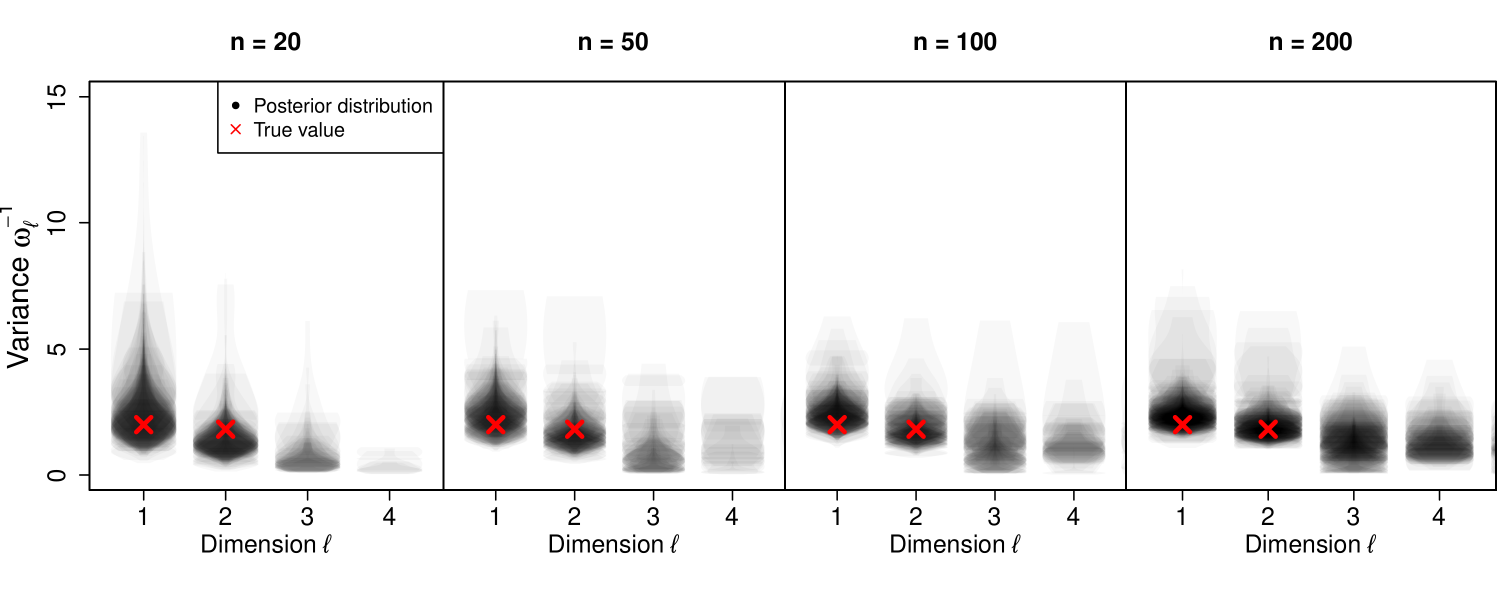} 
\includegraphics[width=.925\linewidth]{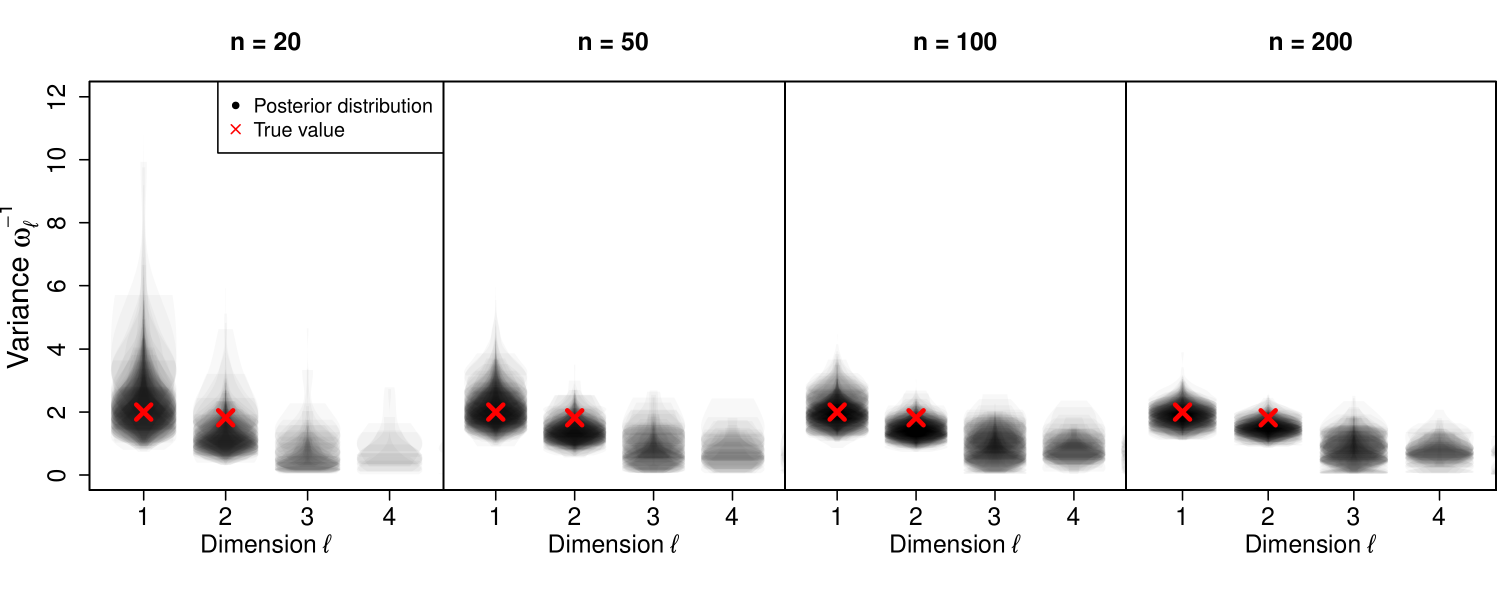} 
\caption{Simulation study 2. Posterior distributions of variance parameters across dimensions for different network sizes, $n$ for the logistic LSPM for binary networks (top) and for the Poisson LSPM for count networks (bottom).}
\label{fig:pmv_size}
\end{figure}

\begin{figure}[htb]
     \centering
     \begin{subfigure}[b]{0.495\linewidth}
         \centering
         \includegraphics[width=\linewidth]{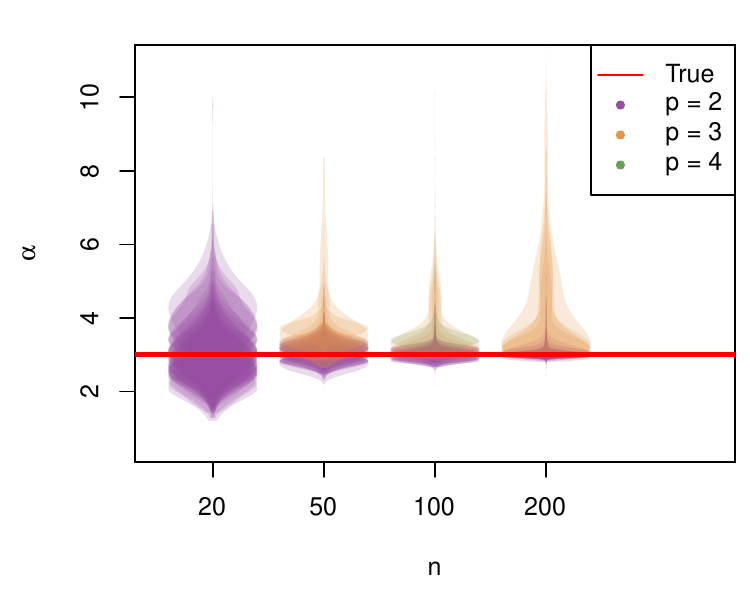}
         \caption{}
         \label{fig:size_alpha}
     \end{subfigure}
     \hfill
     \begin{subfigure}[b]{0.495\linewidth}
         \centering
         \includegraphics[width=\linewidth]{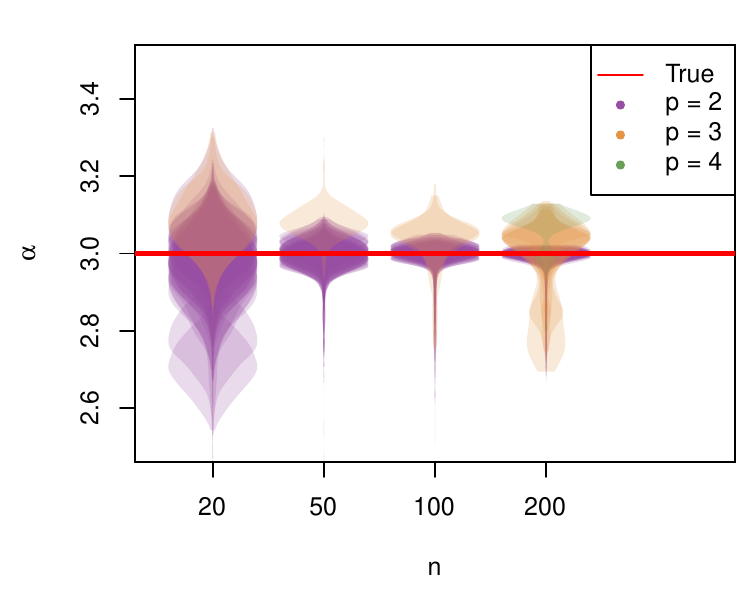}
         \caption{}
         \label{fig:size_alpha_c}
     \end{subfigure}
        \caption{Simulation study 2. Posterior distributions of $\alpha$ for different network sizes $n$ for (a) the logistic LSPM for binary networks and (b) the Poisson LSPM for count networks.}
        \label{fig:alpha_size}
\end{figure}

\begin{figure}[htb]
\includegraphics[width=\linewidth]{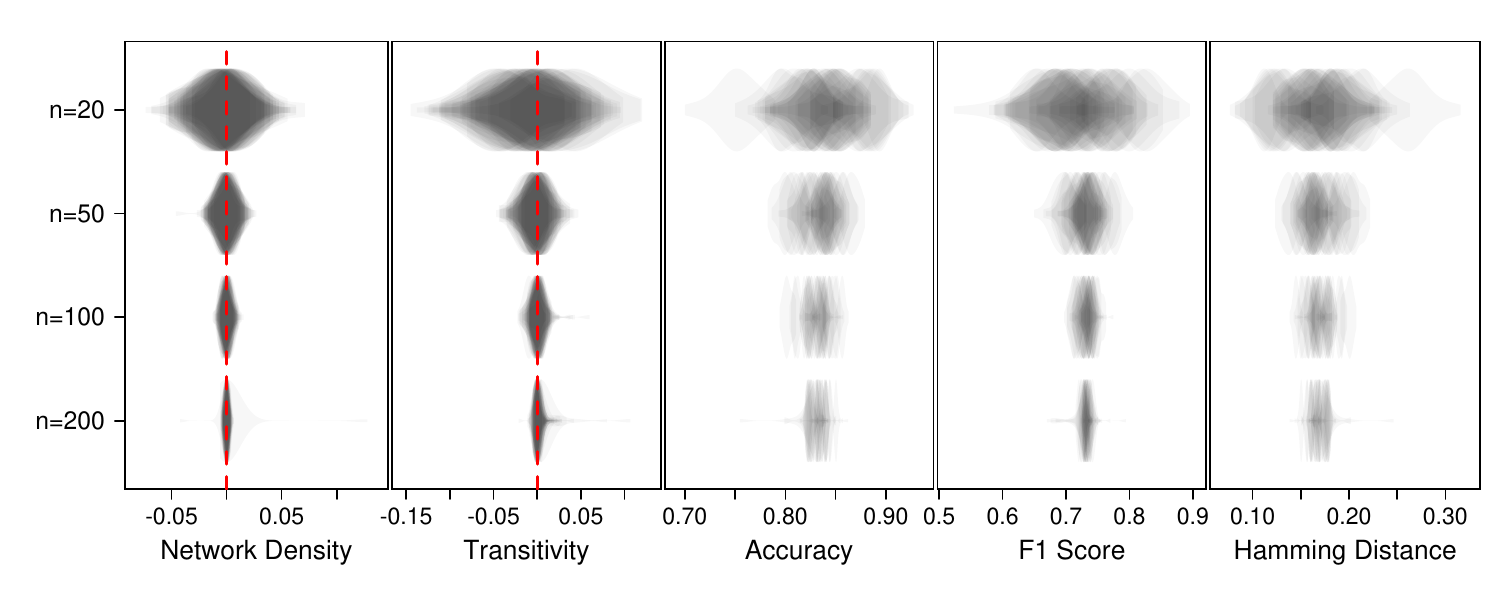} 
\caption{Simulation study 2. Posterior predictive checks from fitting the logistic LSPM conditioned on $p_m$ across different network sizes $n$ to simulated binary networks. The violin plots indicate the metrics' distributions from the predicted networks. The network density and transitivity plots illustrate the differences between the posterior predictive networks and the observed networks.}
\label{fig:pred_size}
\end{figure}

\begin{figure}[!ht]
%\nextfloat
\hfill
          \begin{subfigure}[b]{0.49\linewidth}
         \centering
         \includegraphics[width=\linewidth]{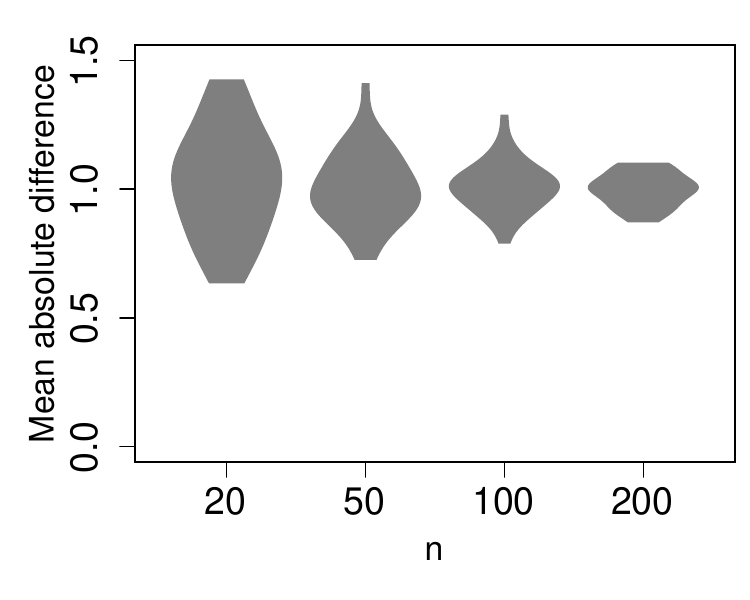}
         \caption{}
         \label{fig:size_abs}
     \end{subfigure}
          \begin{subfigure}[b]{0.49\linewidth}
         \centering
         \includegraphics[width=\linewidth]{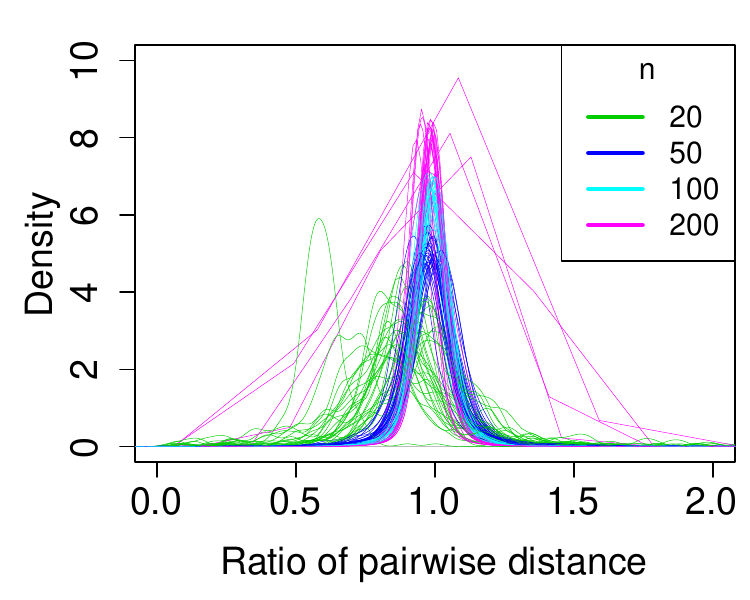}
         \caption{}
         \label{fig:size_ratio}
     \end{subfigure}

    \begin{subfigure}[b]{0.98\linewidth}
         \centering
         \includegraphics[width=\linewidth]{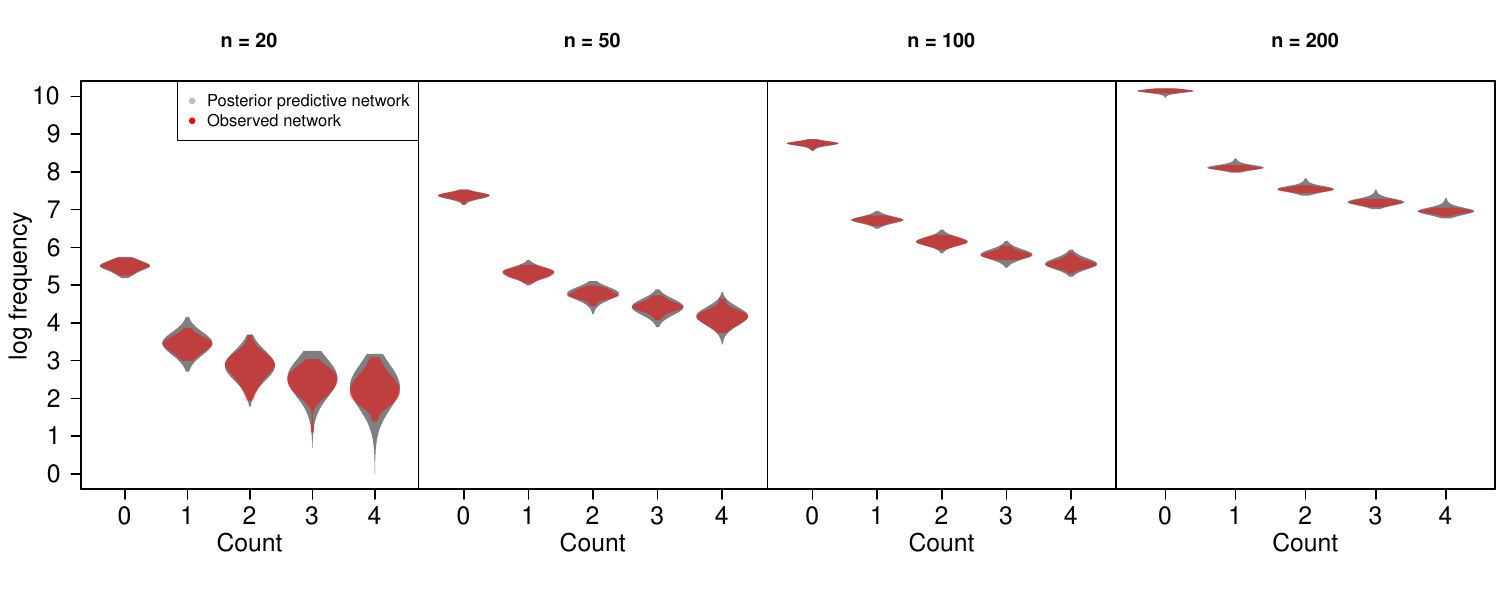}
         \caption{}
         \label{fig:size_logcount}
     \end{subfigure}
        \caption{Simulation study 2. Assessment of model fit for the Poisson LSPM via (a) the mean absolute difference between replicate and observed network counts, (b) the ratios of the pairwise distances between inferred and true node locations, and (c) the log frequency between replicate and observed network counts. }
        \label{fig:size_trunc_c}
\end{figure} 

% NETWORK DENSITIES SIMULATION STUDIES
% \subsubsection{Different network densities}

\begin{figure}[!ht]
\includegraphics[width=\linewidth]{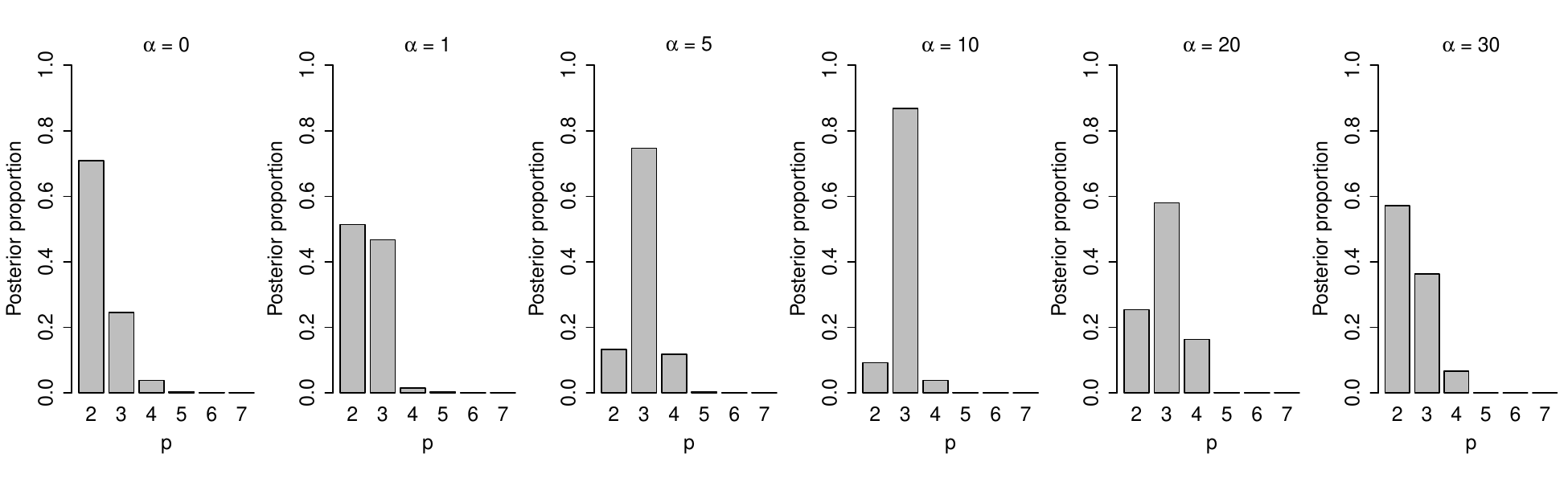}
        \caption{Simulation study 3. For different network densities, the posterior distribution of $p$ across the 30 simulated networks.}
        \label{dim_dens}
\end{figure}

\begin{figure}[tb]
\includegraphics[width=.95\linewidth]{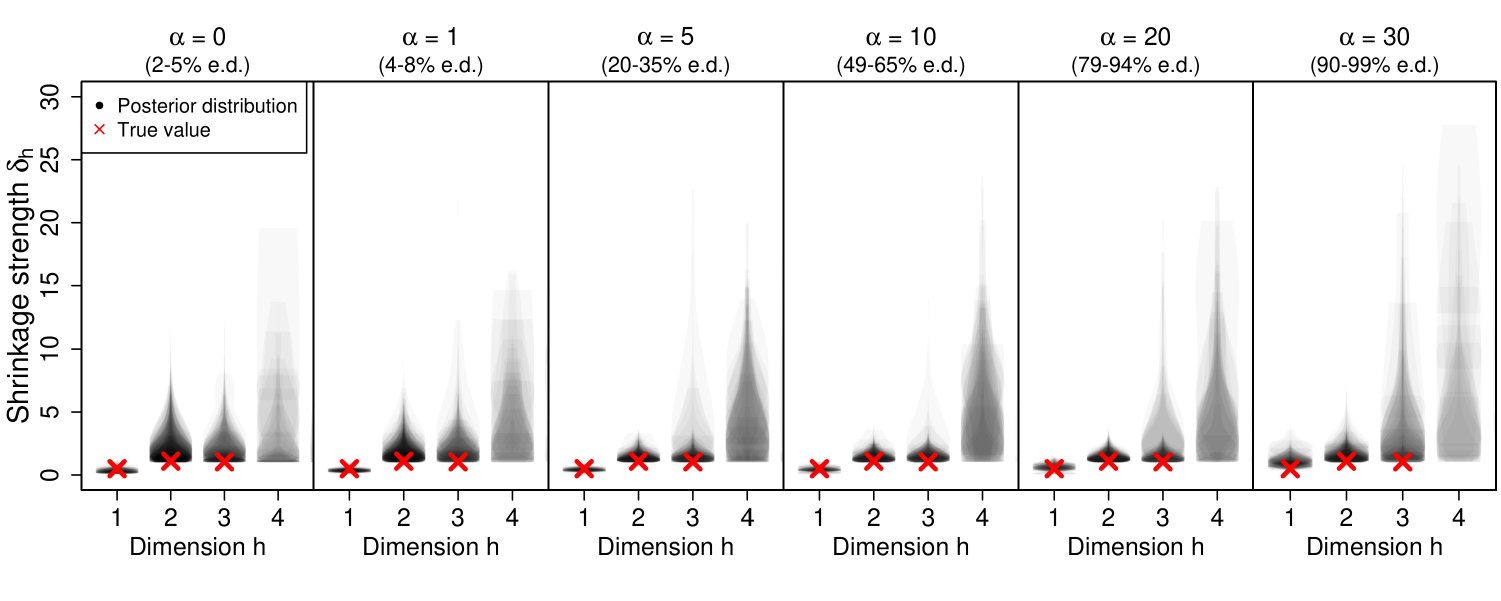} 
\caption{Simulation study 3. Posterior distributions of shrinkage strength parameters across dimensions for different network densities.}
\label{fig:pmd_dens}
\end{figure}

\begin{figure}[tb]
\includegraphics[width=.95\linewidth]{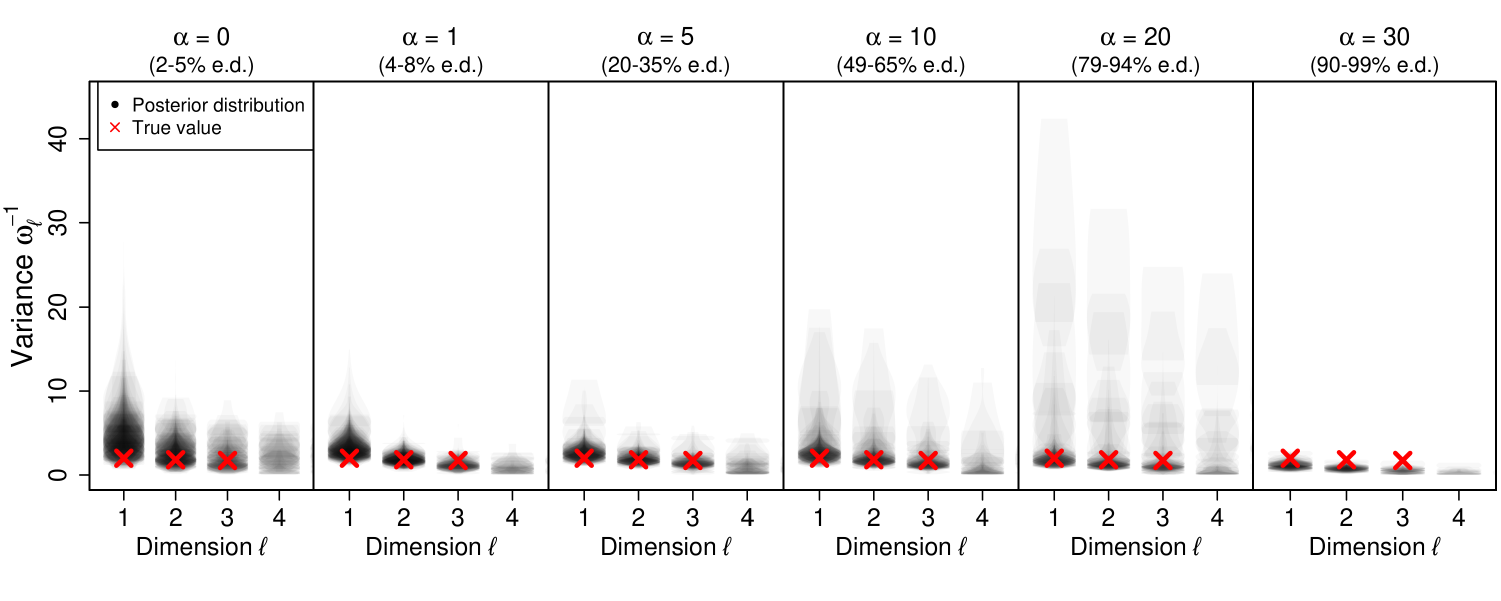} 
\caption{Simulation study 3. Posterior distributions of variance parameters across dimensions for different network densities.}
\label{fig:pmv_dens}
\end{figure}

\begin{figure}[htb]
\includegraphics[width=\linewidth]{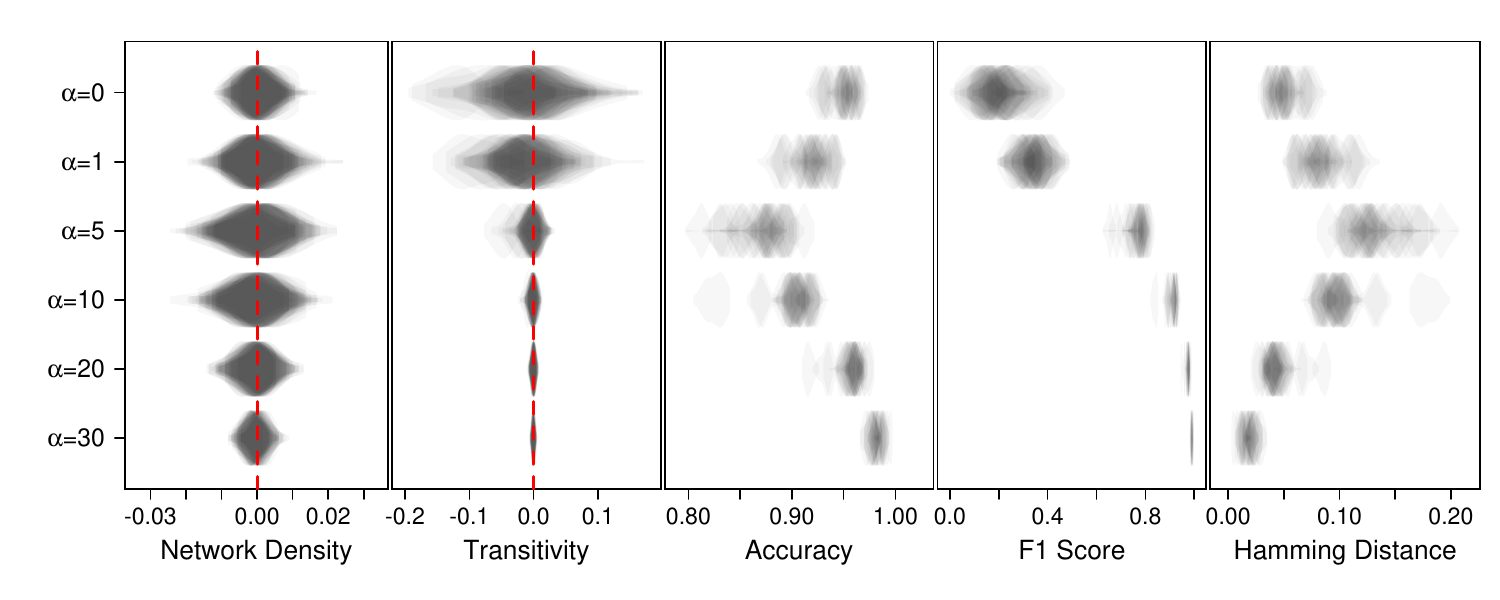} 
\caption{Simulation study 3. Posterior predictive checks under the logistic LSPM conditioned on $p_m$ across different network densities levels corresponding to different true $\alpha$ for simulated binary networks. The violin plots indicate the metrics' distributions from the replicate networks. The network density and transitivity plots illustrate the differences between the posterior predictive networks and the observed networks.}
\label{fig:pred_dens}
\end{figure}

% NETWORK OVERDISPERSION SIMULATION STUDIES
% \subsubsection{Different levels of overdispersion networks}

\begin{figure}[!ht]
\includegraphics[width=\linewidth]{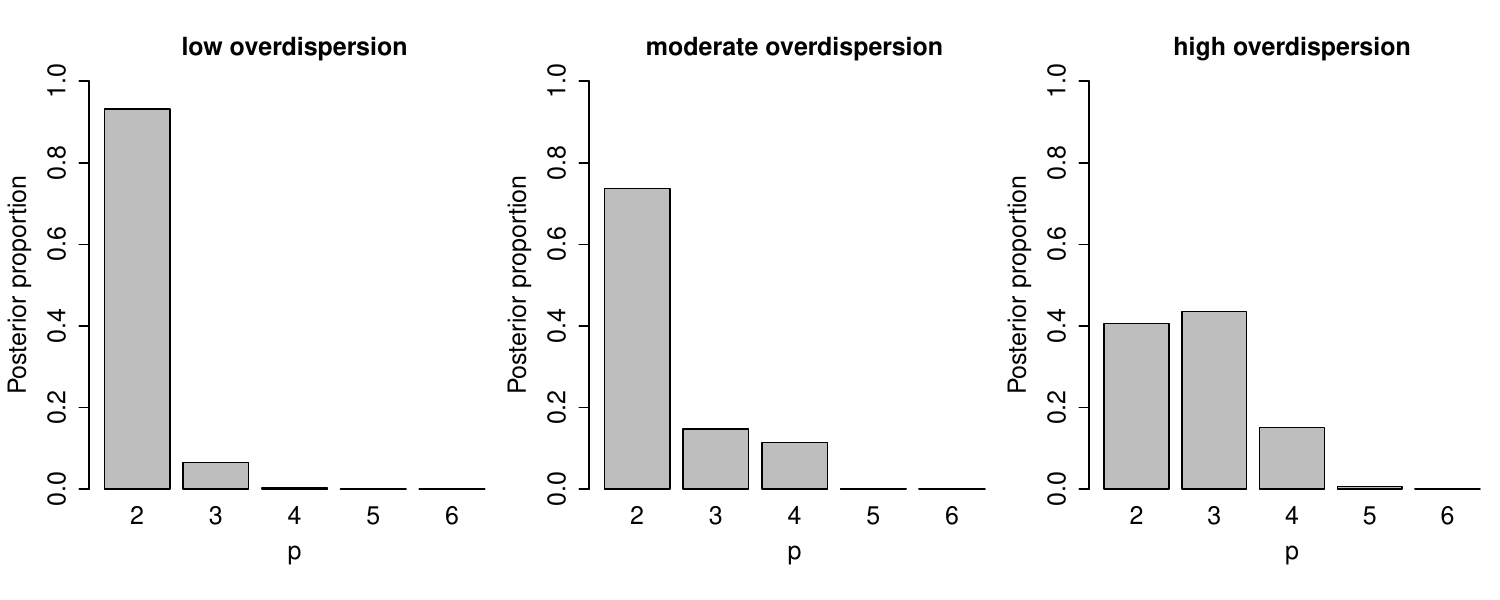}
        \caption{Simulation study 4. For different levels of overdispersion, the posterior distribution of $p$ in 30 simulated networks.}
        \label{dim_disp}
\end{figure}

\begin{figure}[tb]
\includegraphics[width=.95\linewidth]{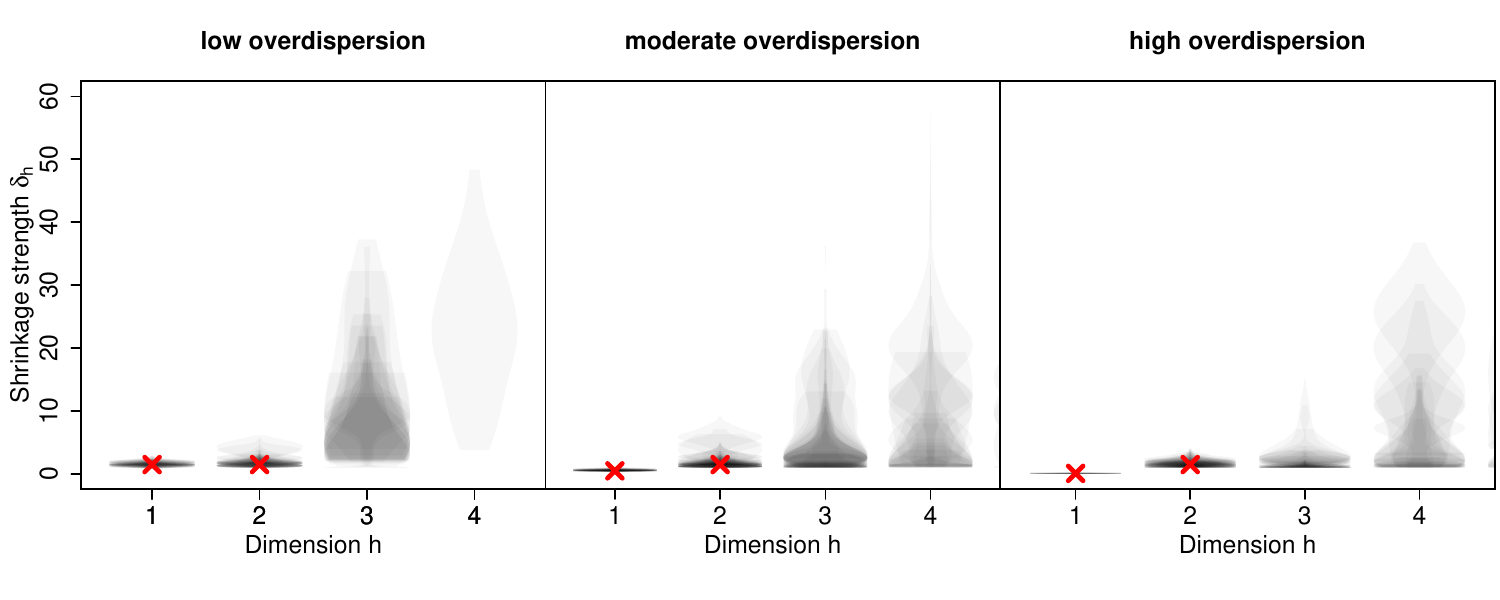} 
\caption{Simulation study 4. Posterior distributions of shrinkage strength parameters across dimensions for different levels of overdispersion.}
\label{fig:pmd_disp}
\end{figure}

\begin{figure}[htb]
     \centering
     \begin{subfigure}[b]{0.495\linewidth}
         \centering
         \includegraphics[width=\linewidth]{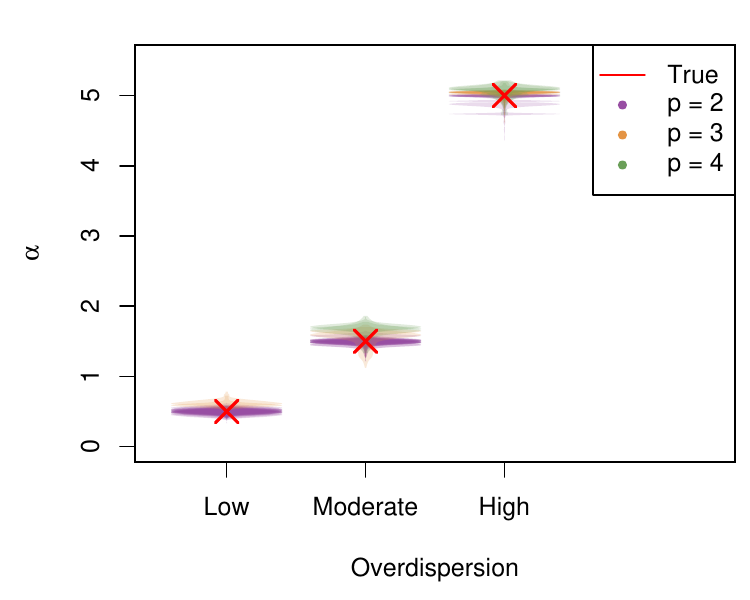}
         \caption{}
         \label{fig:disp_alpha}
     \end{subfigure}
     \hfill
        \caption{Simulation study 4. Posterior distributions of $\alpha$ for different levels of overdispersion.}
        \label{fig:alpha_disp}
\end{figure}

\begin{figure}[!ht]
%\nextfloat
\hfill
          \begin{subfigure}[b]{0.49\linewidth}
         \centering
         \includegraphics[width=\linewidth]{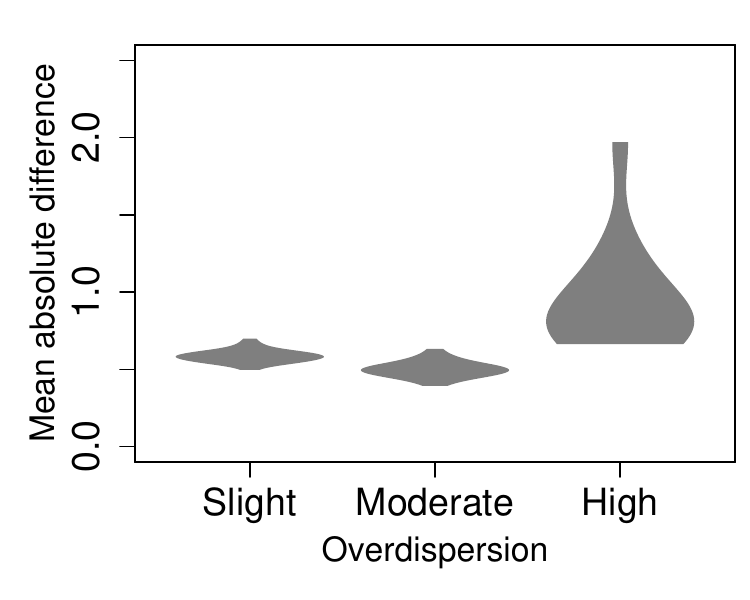}
         \caption{}
         \label{fig:disp_abs}
     \end{subfigure}
          \begin{subfigure}[b]{0.49\linewidth}
         \centering
         \includegraphics[width=\linewidth]{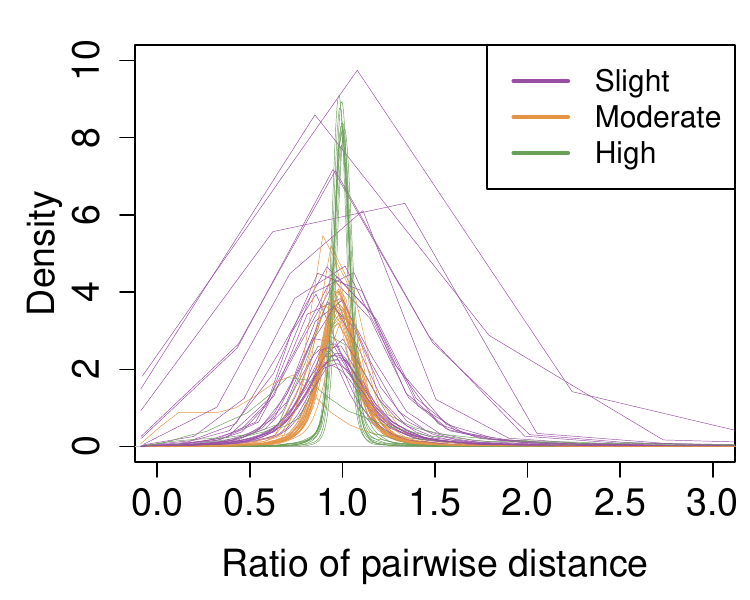}
         \caption{}
         \label{fig:disp_ratio}
     \end{subfigure}

    \begin{subfigure}[b]{0.98\linewidth}
         \centering
         \includegraphics[width=\linewidth]{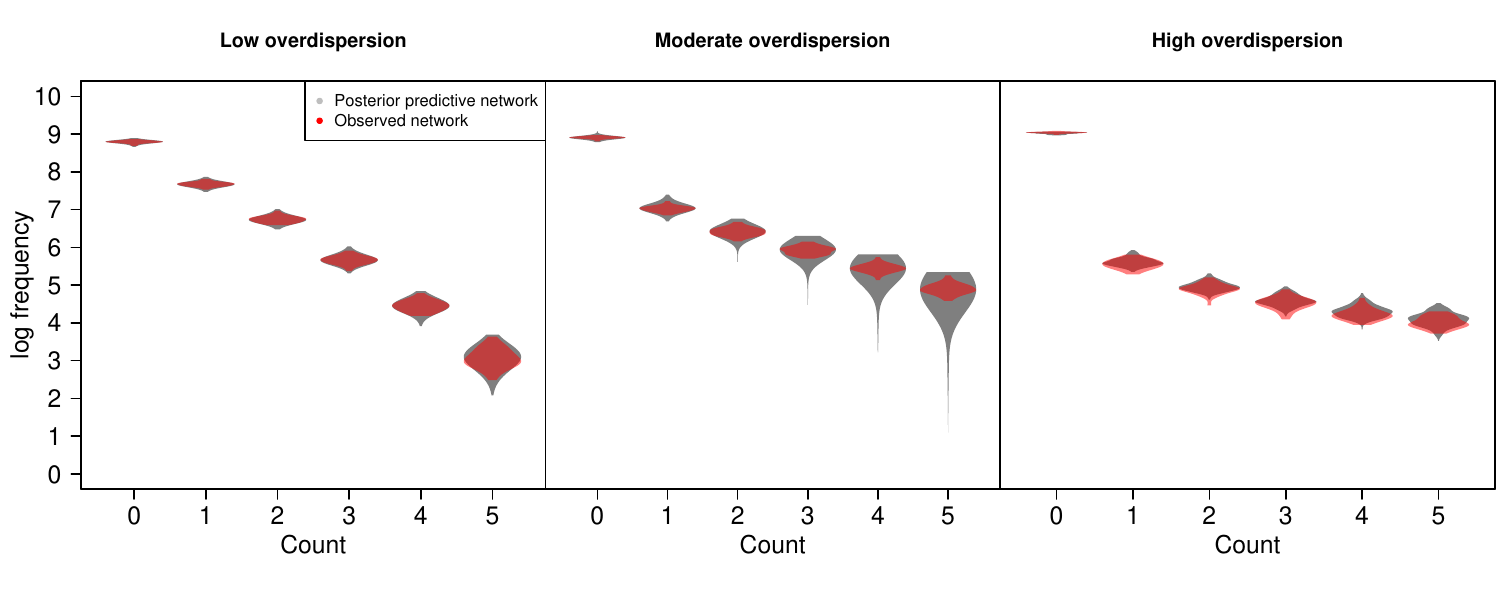}
         \caption{}
         \label{fig:disp_logcount}
     \end{subfigure}
        \caption{Simulation study 4. Assessment of model fit for the Poisson LSPM via (a) the mean absolute difference between replicate and observed network counts, (b) the ratios of the pairwise distances between inferred and true node locations, and (c) the log frequency between replicate and observed network counts. }
        \label{fig:disp_trunc_c}
\end{figure}

\clearpage

\newpage

\section{Additional posterior predictive checks for illustrative data sets} \label{app:application}

Further results obtained from applying the LSPM on the application networks can be found in this appendix, including the posterior distributions of the dimension, variance and shrinkage strength parameters, along with posterior predictive checks, posterior mean latent positions, and trace plots.

% ZACHARY APPLICATION

\begin{figure}[htb]
     \centering
     \begin{subfigure}[b]{0.495\linewidth}
         \centering
         \includegraphics[width=.95\linewidth]{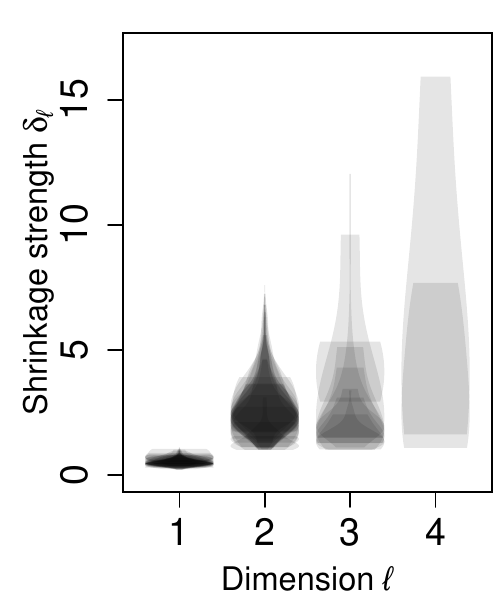} 
         \caption{}
         \label{fig:zach_pmd}
     \end{subfigure}
    \begin{subfigure}[b]{0.495\linewidth}
         \centering
         \includegraphics[width=.95\linewidth]{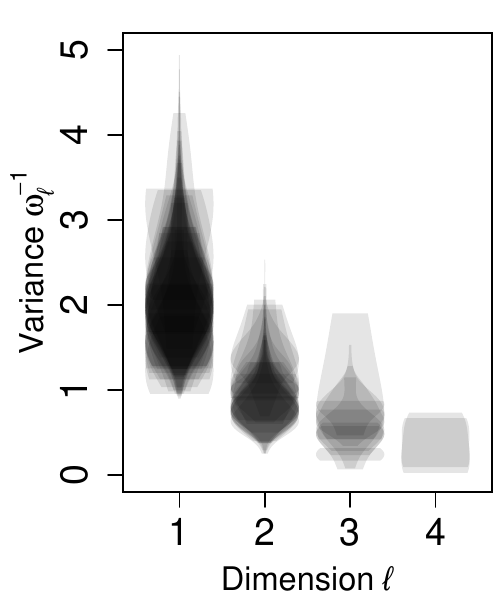} 
         \caption{}
         \label{fig:zach_pmv}
     \end{subfigure}
     \hfill
        \caption{For the Zachary karate club network (a) the posterior distributions of shrinkage strength across dimensions and (b) the posterior distributions of the variance parameters across dimensions.}
        \label{fig:pmd_pmv_zach}
\end{figure}

\begin{figure}[htb]
     \centering
     \begin{subfigure}[b]{0.45\linewidth}
         \centering
         \includegraphics[width=.95\linewidth]{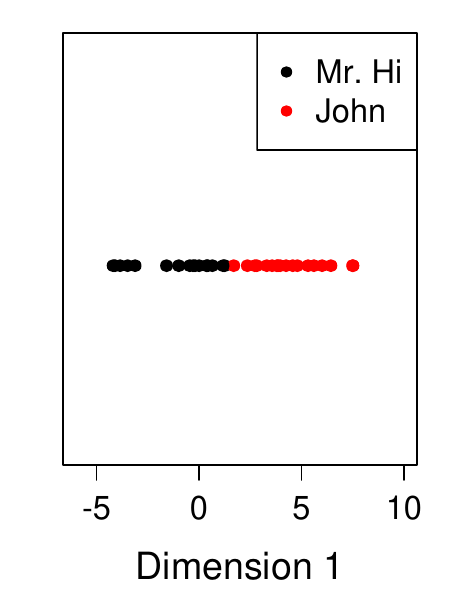} 
         \caption{}
         \label{fig:zach_lpm}
     \end{subfigure}
    \begin{subfigure}[b]{0.535\linewidth}
         \centering
         \includegraphics[width=.95\linewidth]{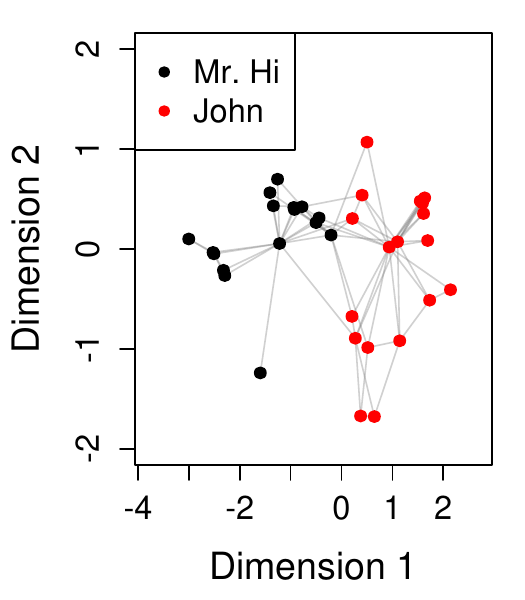} 
         \caption{}
         \label{fig:zach_lspm}
     \end{subfigure}
     \hfill
        \caption{For the Zachary karate club network (a) the LPM $p=1$ positions and (b) the LSPM $p_m=2$ posterior mean positions. The node colour indicated the faction membership of the actors after the karate club was split into two separate clubs, one led by Mr.Hi, the other led by John. }
        \label{fig:pos_zach}
\end{figure}

% \begin{figure}[!t]
% \includegraphics[width=\linewidth]{figures/zach_pred.pdf} 
% \caption{Zachary karate club network. Posterior predictive checks for the logistic LSPM with 2 active dimensions and LPM with 2 dimensions.}
% \label{fig:pred_zach}
% \end{figure}

% CAT APPLICATION

\begin{figure}[htb]
     \centering
     \begin{subfigure}[b]{0.495\linewidth}
         \centering
         \includegraphics[width=.95\linewidth]{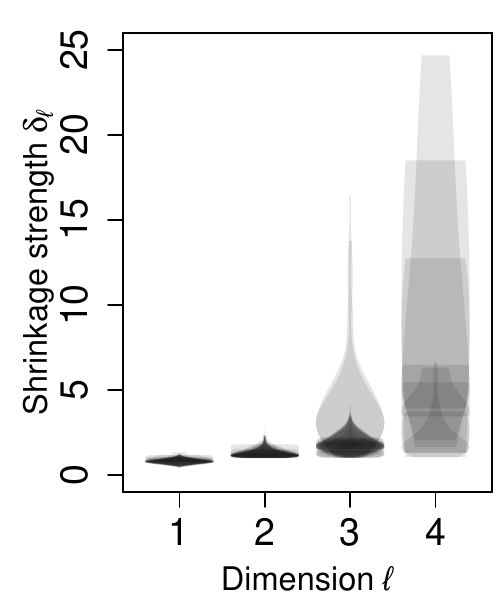} 
         \caption{}
         \label{fig:cat_pmd}
     \end{subfigure}
    \begin{subfigure}[b]{0.495\linewidth}
         \centering
         \includegraphics[width=.95\linewidth]{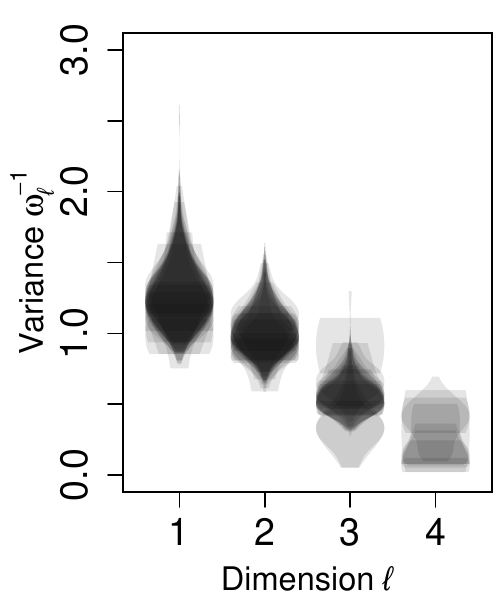} 
         \caption{}
         \label{fig:cat_pmv}
     \end{subfigure}
     \hfill
        \caption{For the cat brain connectivity network: (a) posterior distributions of shrinkage strength parameters across dimensions and (b) posterior distributions of variance parameters across dimensions.}
        \label{fig:pmd_pmv_cat}
\end{figure}

\begin{figure}[!t]
\includegraphics[width=\linewidth]{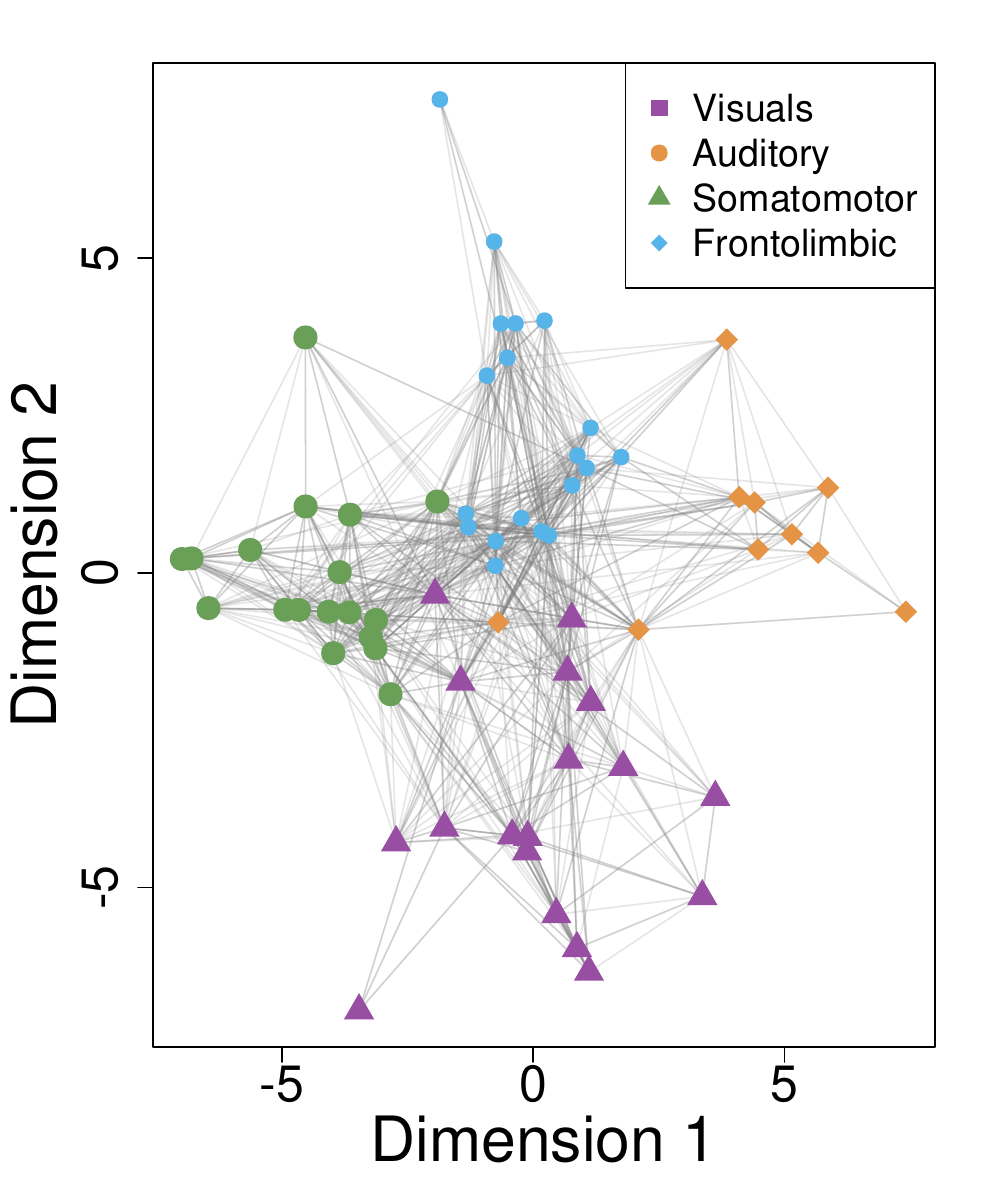} 
\caption{Posterior mean latent positions for the $p =2$ LPM on the cat brain connectivity data.}
\label{fig:lpm_cat}
\end{figure}

\begin{figure}[!t]
\includegraphics[width=\linewidth]{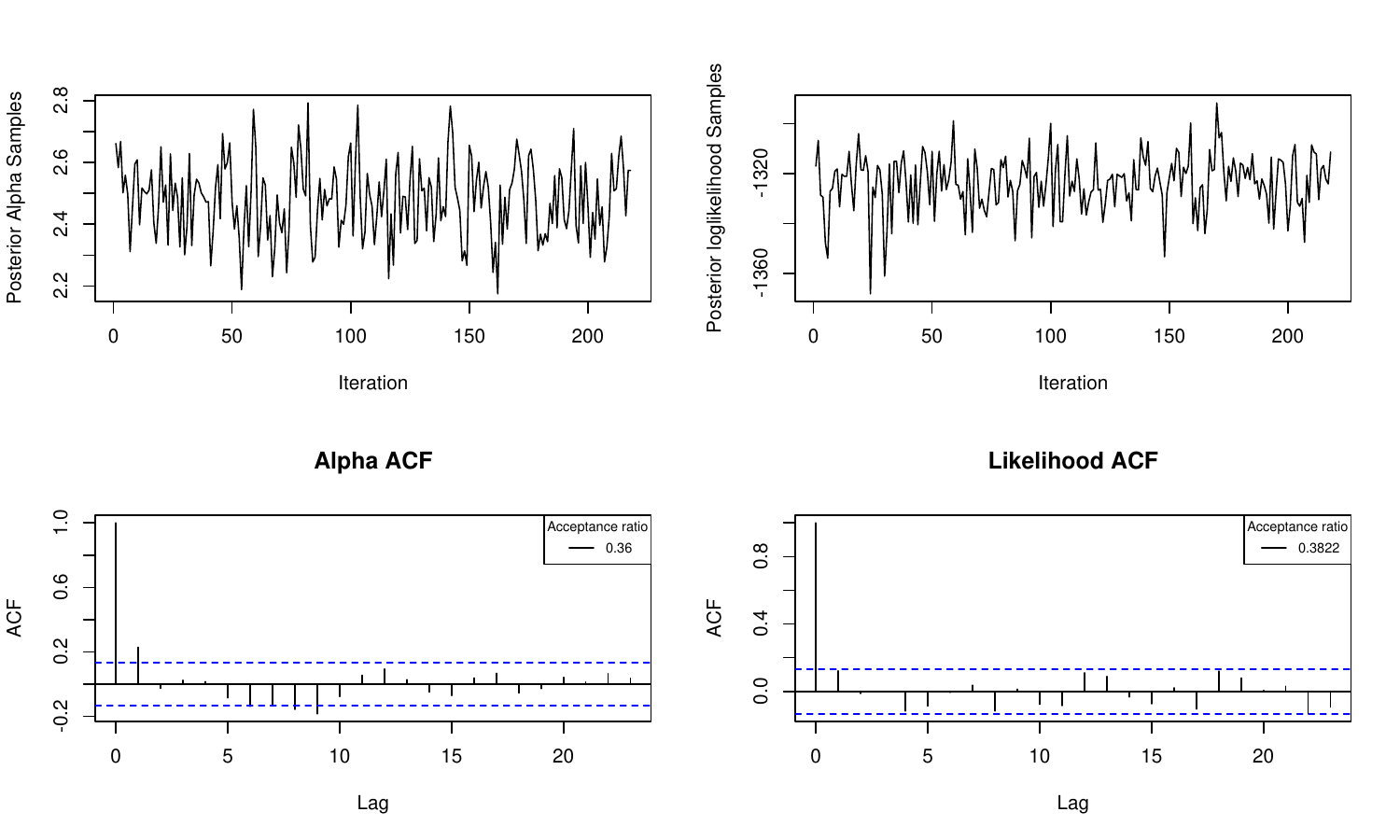} 
\caption{Trace plots (top) for $\alpha$ and the log likelihood for the cat brain connectivity data, with associated autocorrelation plots (bottom).}
\label{fig:trace_cat}
\end{figure}

% WORM APPLICATION

\begin{figure}[htb]
     \centering
     \begin{subfigure}[b]{0.495\linewidth}
         \centering
         \includegraphics[width=.95\linewidth]{figures/worm_pmd.pdf} 
         \caption{}
         \label{fig:worm_pmd_supp}
     \end{subfigure}
    \begin{subfigure}[b]{0.495\linewidth}
         \centering
         \includegraphics[width=.95\linewidth]{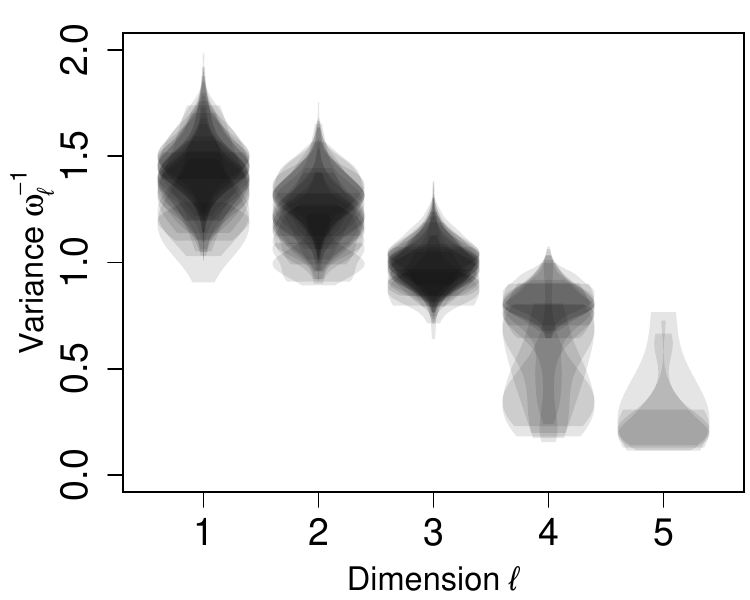} 
         \caption{}
         \label{fig:worm_pmv}
     \end{subfigure}
     \hfill
        \caption{For the worm nervous system network: (a) posterior distributions of shrinkage strength parameters across dimensions and (b) posterior distributions of variance parameters across dimensions.}
        \label{fig:pmd_pmv_worm}
\end{figure}

% PHONES APPLICATION

\begin{figure}[htb]
     \centering
     \begin{subfigure}[b]{0.495\linewidth}
         \centering
         \includegraphics[width=.95\linewidth]{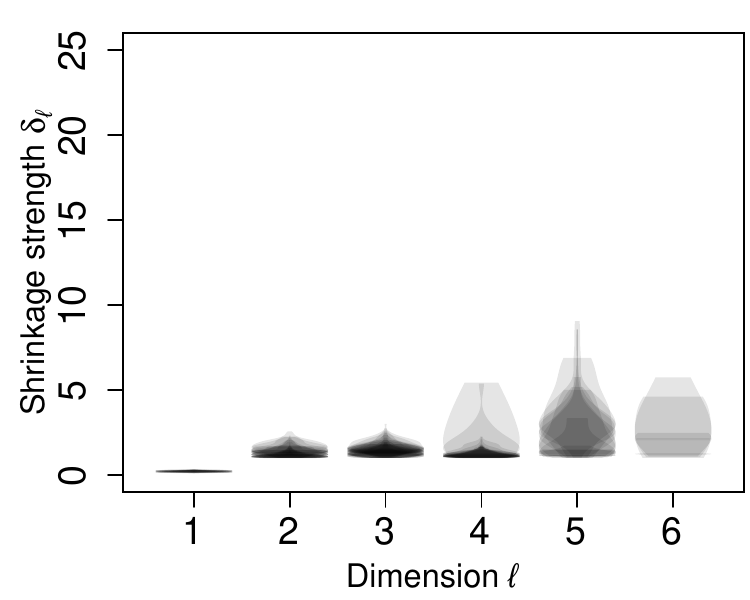} 
         \caption{}
         \label{fig:phones_pmd}
     \end{subfigure}
    \begin{subfigure}[b]{0.495\linewidth}
         \centering
         \includegraphics[width=.95\linewidth]{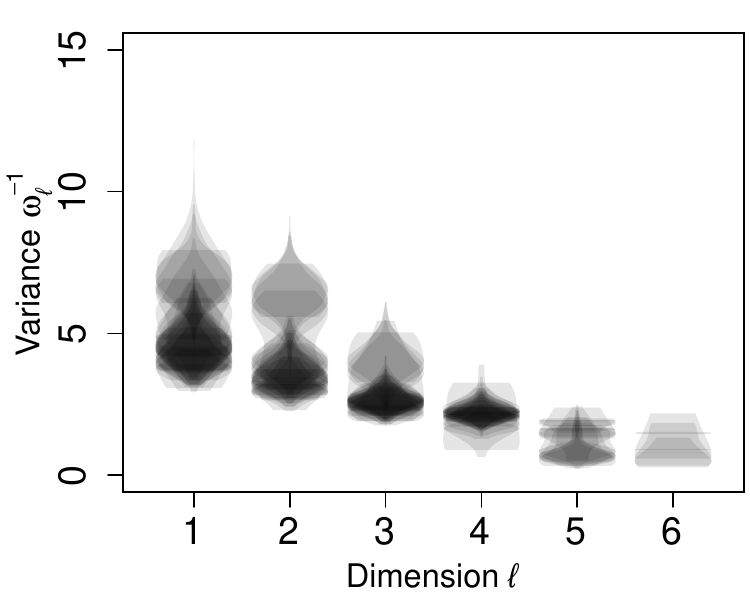} 
         \caption{}
         \label{fig:phones_pmv}
     \end{subfigure}
     \hfill
        \caption{For the phone calls count network: (a) posterior distributions of shrinkage strength parameters across dimensions and (b) posterior distributions of variance parameters across dimensions.}
        \label{fig:pmd_pmv_phones}
\end{figure}

\begin{figure}[!t]
\includegraphics[width=\linewidth]{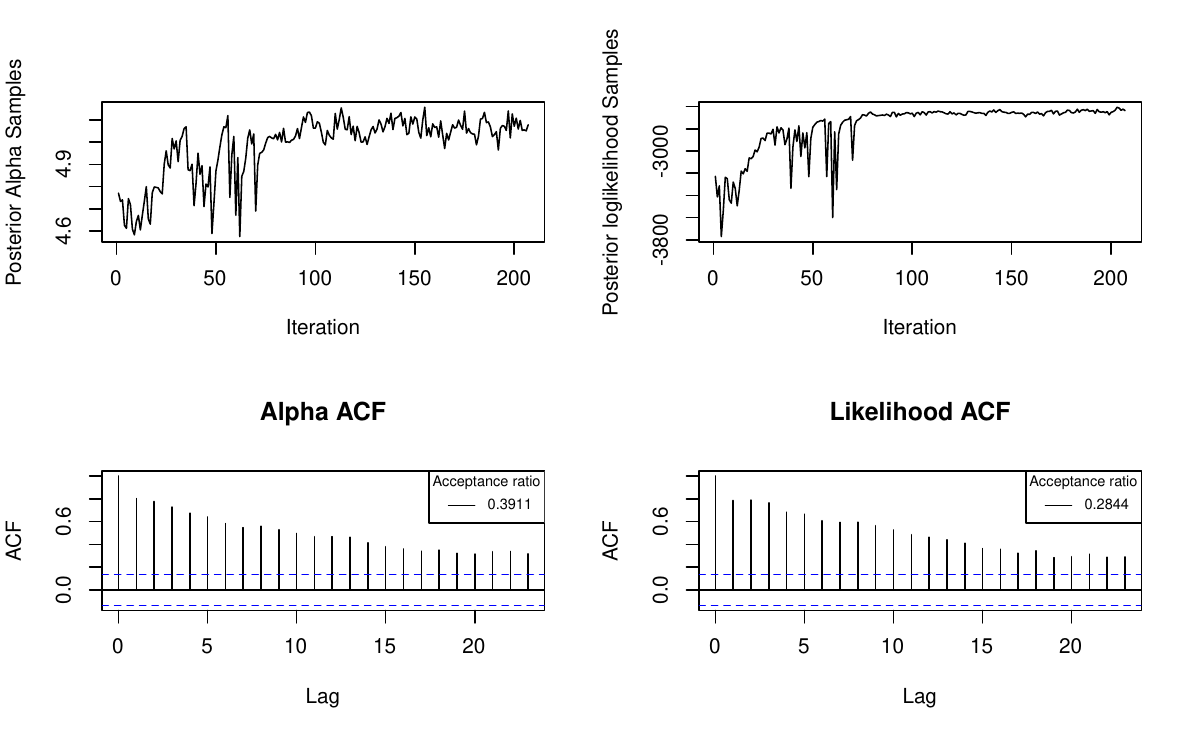} 
\caption{Trace plots (top) for $\alpha$ and the log likelihood for the phone calls count data, with associated autocorrelation plots (bottom).}
\label{fig:phone_trace}
\end{figure}
\clearpage
% \bibliographystyle{authordate1}
\bibliographystyle{plainnat}
\bibliography{LSPM-ArXiv.bib}{}   % name of .bib file

\end{document}